\documentclass[trackchanges,twocolumn,twocolappendix]{aastex701}

\begin{document}

\title{The solar-like latitudinal distribution of flaring activities revealed by TESS, APOGEE and GALAH}

\author[orcid=0000-0003-4917-7221,sname='Yang']{Huiqin Yang}

\affiliation{Key Laboratory of Optical Astronomy, National Astronomical Observatories, Chinese Academy of Sciences}
\affiliation{Institute for Frontiers in Astronomy and Astrophysics, Beijing Normal University, Beijing 102206, China}
\email[show]{yhq@nao.cas.cn}  

\author[sname='Liu']{Shuai Liu}
\affiliation{Key Laboratory of Optical Astronomy, National Astronomical Observatories, Chinese Academy of Sciences}
\affiliation{School of Astronomy and Space Sciences, University of Chinese Academy of Sciences}
\email{liushuai@nao.cas.cn}

\author[sname='Huang']{Yang Huang}
\affiliation{Key Laboratory of Optical Astronomy, National Astronomical Observatories, Chinese Academy of Sciences}
\affiliation{School of Astronomy and Space Sciences, University of Chinese Academy of Sciences}
\email{liushuai@nao.cas.cn}

\author[sname='Zhang']{Bowen Zhang}
\affiliation{Key Laboratory of Optical Astronomy, National Astronomical Observatories, Chinese Academy of Sciences}
\affiliation{School of Astronomy and Space Sciences, University of Chinese Academy of Sciences}
\email{liushuai@nao.cas.cn}

\author[sname='Liu']{Jifeng Liu} 

\affiliation{Key Laboratory of Optical Astronomy, National Astronomical Observatories, Chinese Academy of Sciences}

\affiliation{School of Astronomy and Space Sciences, University of Chinese Academy of Sciences}
\email{jfliu@nao.cas.cn}

\begin{abstract}

Flare flux reflect contribution from active regions rather than the whole hemisphere of a star. Unlike the amplitude of light-curves caused by starspots, the flare detection is independent of inclination. The two valuable properties of flares can be used to reveal the latitudinal distribution of active regions (LaDAR) given that LaDAR is coupled with inclination and location information in spatially unresolved stars. We detected $\sim 27000$ flares of 1510 flaring stars in the TESS mission with the corresponding inclinations obtained. The detection rate of flaring stars shows that flares are hard to detect on stars with low inclination, indicating that flares occur mainly at low latitudes. Further investigation of the relationship between the apparent flaring activity and inclination along with the rotation period finds that as the rotation period increases from a solar-like rotation to an ultra-fast rotation, the mean latitude of active regions increases from $\theta \approx 15^{\circ}$ to $\theta \approx 27^{\circ}$, whose trend is in line with the rotation--activity relationship. The LaDAR indicates that flares are attributed to small-scale fields that are formed at low latitudes, while polar spots that are associated with large-scale fields are inactive and are difficult to trigger flares.

\end{abstract}



\section{Introduction} 

The most prominent feature of the solar dynamo is that sunspots, active regions, and flares only appear at low latitudes, which to date has not been fully understood \citep{Charb2020,Nandy2002}. This crucial feature presents strong constraints and implications for the dynamo simulation and the internal magnetic process of the Sun. Although the Sun is an extraordinary laboratory, the input parameters cannot be changed for this laboratory to validate the dynamo theory and simulation. Therefore, it is necessary to investigate the latitudinal distribution of active regions (LaDAR) of other stars as an alternative choice to change the input parameters. 

The current inversion techniques on recovering the surface distribution of stars include the (Zeeman) Doppler Imaging (ZDI) \citep{Donati1997}, light-curve inversion \citep{Ikuta2020} and interferometric imaging \citep{Roett2016}. However, the ZDI technique tends to cancel out small-scale magnetic fields and does not reflect the distribution of magnetic fields of local spots \citep{Kochu2020}. It is also sensitive to several factors such as stellar activity \citep{Bruls1998}, microturbulence and blend lines \citep{Unruh1995,Berdy1998}, stellar parameters \citep{Unruh1996}, and differential rotation \citep{Hackman2001}, which may result in opposite results or artifacts. Moreover, it may loss information on mid-latitude activity \citep{Senav2021}, and its capacity is seriously weakened near the equator \citep{Rice2002,Berdy2005}. The light-curve inversion typically reveals only weak relative latitude information for active regions and cannot reflect active regions at high latitudes. The interferometric imaging requires a very high observation condition, which cannot be applied to big data. The disadvantages of those techniques make our understanding of the LaDAR of stars still ambiguous to date.

With the advent of the big data era, we recently proposed that the LaDAR of stars can be revealed from a statistical point of view by studying the variation of the apparent flaring activity along with the inclination for stars with similar rotation periods \citep{Yang2025}. Unlike other activity proxies such as chromospheric activity $R'_{\rm HK}$ and coronal activity $R_{\rm X}$, a flare only reflects the contribution from local active regions rather than the whole hemisphere. Meanwhile, the detection of a flare is independent of inclination and rotation, whereas other proxies such as the light-curve modulation caused by starspots depend on them. Moreover, the variation of the location from the edge to the center of the disk could yield enough variation of the apparent flare energy that is at least 360\% \citep{Yang2025}. By contrast, the variation of the S-index (the proxy for chromospheric activity) observed from pole-on to equator-on is $\sim 0.008$ \citep{Vanden2025}, which is lower than the uncertainty, excluding its application under the current observational condition. These valuable characters of flares are important for decoupling the location information from spatially unresolved stars by investigating the relation between the activity and inclination.

In this study, we combined the data from the Transiting Exoplanet Survey Satellite mission \citep[TESS;][]{Ricker2015}  with the Apache Point Observatory Galactic Evolution Experiment-2 spectrograph \citep[APOGEE;][]{Abdu2022} and the Galactic Archaeology with HERMES survey \citep[GALAH;][]{Buder2024} to study the relationship between the flaring activity and inclination, which can reveal the LaDAR of those flaring stars.

\section{Data and Method}

We cross-matched the data of the TESS mission with the APOGEE DR17 and the GALAH DR4 and obtained about 32000 stars, of which 9418 stars have detectable rotation periods \citep{Mc2014,Santos2021,Reinhold2020,Colman2024,Fetherolf2023,Kounkel2022} and 1510 stars have detectable flares. The details of the sample selection, the measurement of rotation periods, and the flare detection are presented in Appendix~\ref{sample_select}, \ref{sec:rotation}, and \ref{sec:flaredect}, respectively.  We calculated the inclination of those rotating stars through the following equation:

\begin{equation}\label{eq_inclination}
{\rm sin}i = \frac{v {\rm sin} i \cdot P}{2\pi R},
\end{equation}
where $v$sin$i$ is the projected rotational velocity measured from APOGEE \citep{Abdu2022} or GALAH \citep{Buder2024}, $R$ is the stellar radius, which can be estimated from the isochrone fitting \citep{Somers2020}, and $P$ is the rotation period measured from the light curve modulation of the TESS mission \citep{Mc2014,Santos2021,Claytor2024,Reinhold2020,Colman2024,Fetherolf2023,Kounkel2022}. For each star, the apparent flaring activity $R_{\rm flare}$ is defined as the ratio the total energy of detectable flares to the total energy a star emitted during the observation \citep{Yang2017,Yang2018,Yang2019}:

\begin{equation}\label{eq_fa}
 R_{\rm flare}=\frac{\sum E_{\rm flare}}{\int L_{\rm bol}dt} = \frac{\tilde{L}_{\rm flare}}{L_{\rm bol}}.
\end{equation}

Here, $\sum E_{\rm flare}$ is the sum of all detectable flare
energies during the whole observation and $L_{\rm bol}$ is the bolometric luminosity. The definition of $R_{\rm flare}$ is similar to the proxy of chromospheric activity $R'_{\rm HK}$ \citep{Noyes1984} and coronal activity $R_{\rm X}$ \citep{Wright2011} and is a normalized quantity that has removed the influence of stellar luminosity. 
\begin{figure*}
\includegraphics[width=0.8\textwidth]{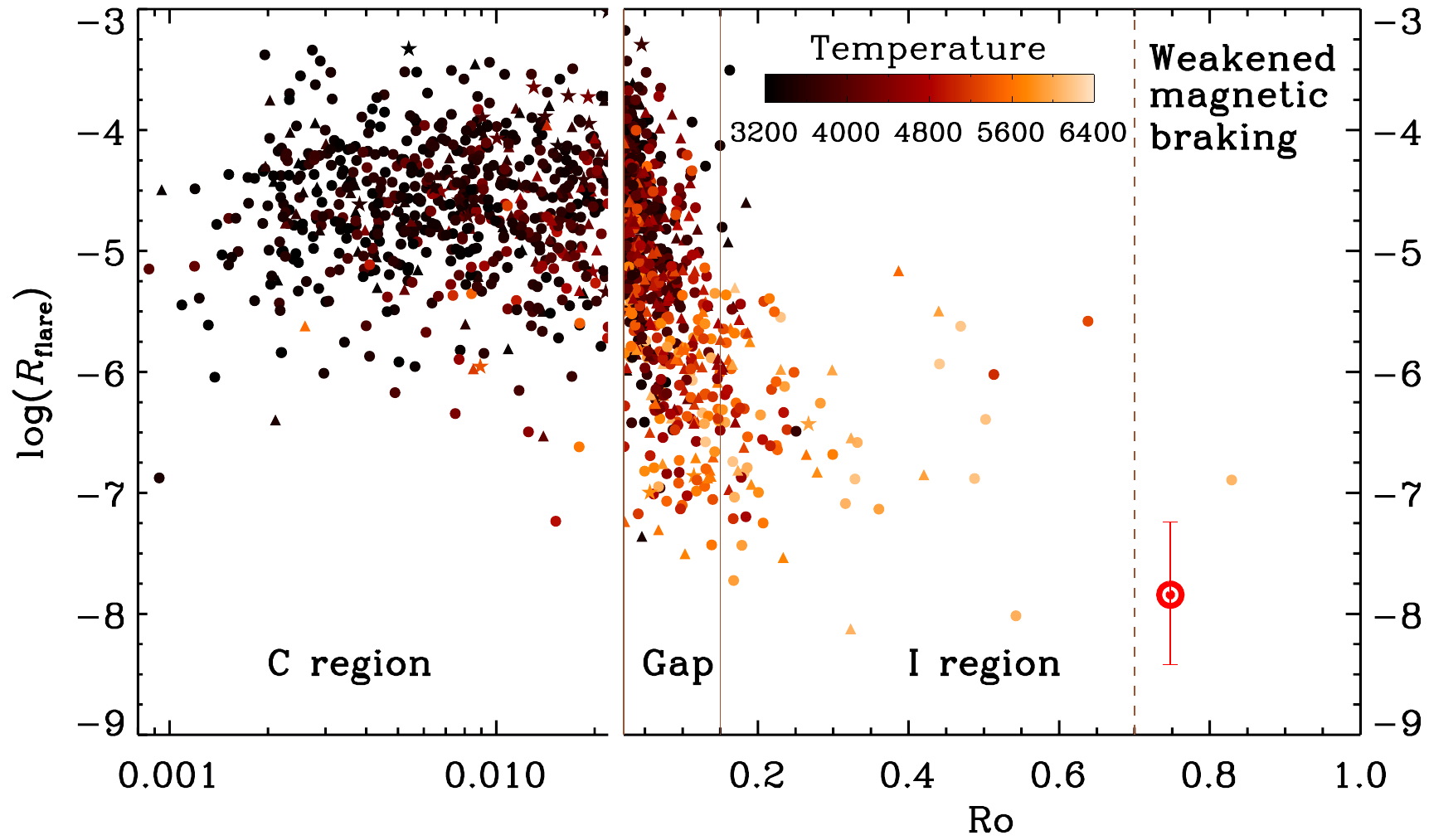}
\includegraphics[width=0.8\textwidth]{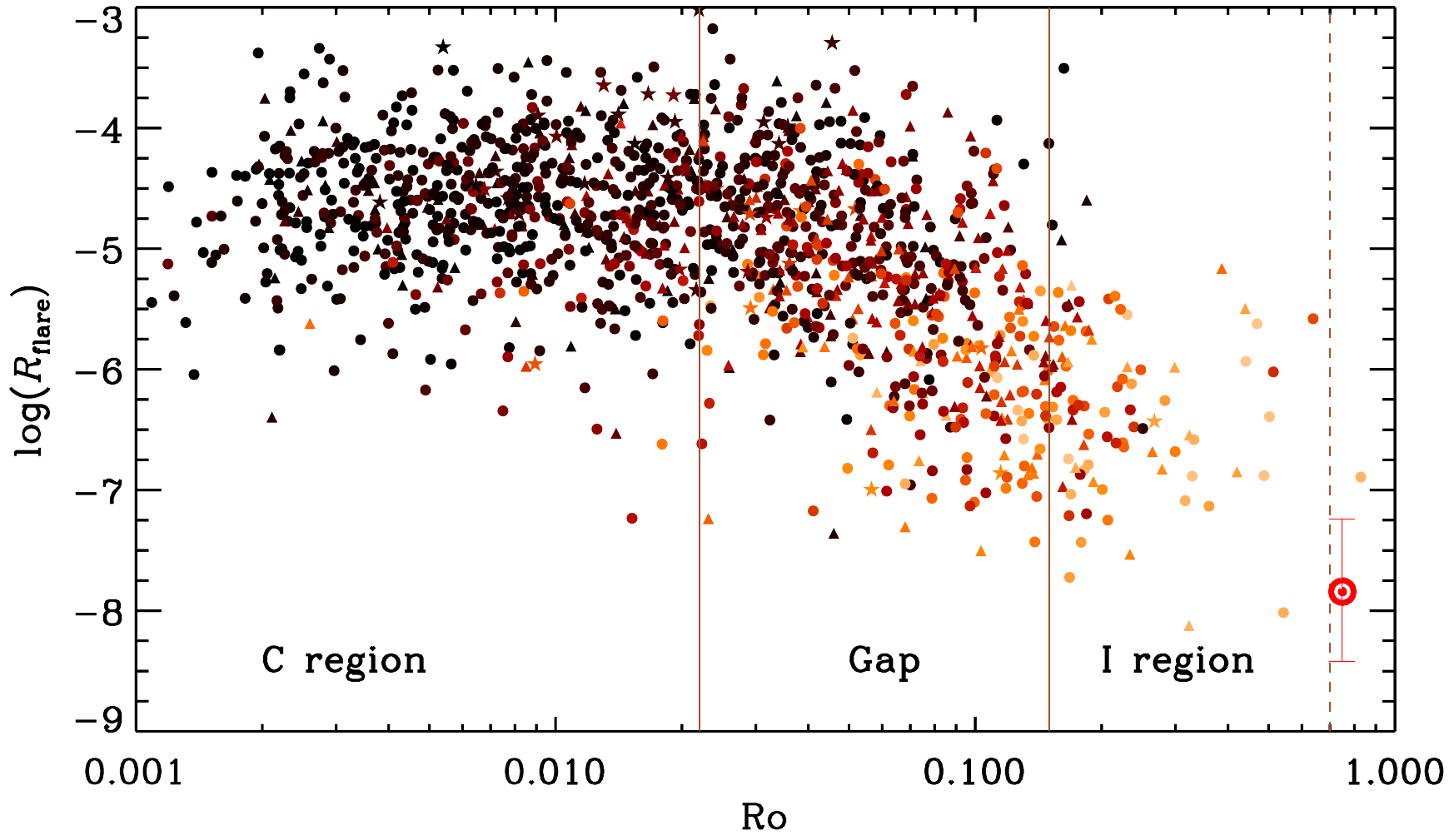}
\caption{Top panel: the rotation--flaring activity relationship is shown, which is separated into four regions according to the CgIW scenario. The C region is shown is a log--log plane for clarity, and the other regions are shown in a linear--log plane. The $x$ axis is the Rossby number (Ro; the ratio of the rotation period to the global convective turnover time). The vertical lines are Ro = 0.022, 0.15 and 0.70, respectively. The shapes of symbols refer to the following star types: circle = dwarf; triangle =  binary; and five-pointed star = subgiant. The Sun is marked as an $\odot$ symbol. Bottom panel: the Same as the top panel, but is shown in a log--log plane for the whole sample.}
\label{fig_ro_fa}
\end{figure*}

We recently proposed that a star undergoes the convective phase (C), gap (g), interface phase (I), and weakened magnetic braking phase (W) as it ages \citep[the CgIW scenario;][]{Yang2025b}. In each phase, stars have distinct activity dependence on rotation and angular momentum loss rate, indicating that each phase has a different stellar dynamo and stellar structure (coupled or decoupled of the core-envelope state). Specifically, at the beginning, a star is in the C phase. Its radiative core and convective envelope are decoupled. It thus has a convective dynamo, resulting in stellar activity being independent of rotation (the saturated regime); As a star ages, its core and envelope begin to recouple and this transition phase corresponds to the gap; Finally, the core and envelope become coupled and have a solar-like dynamo (the I phase); When a star further ages, its surface magnetic geometry could change, resulting in less large-scale field and more small-scale field. The lack of large-scale field could significantly reduce the angular momentum loss rate while the increase of small-scale field could rekindle stellar activity. This phase corresponds to the W phase, where the Sun is located. This scenario unifies two paradigms on stellar evolution: the rotation--activity relationship and gyrochonology, by showing the one-to-one mapping of the two paradigms in terms of the four phases.  Figure~\ref{fig_ro_fa} shows the rotation--flaring activity relationship for our sample in terms of the CgIW scenario, which are grouped by four regions. Stars in each region have different stellar structures, activity dependence on rotation. The rarity of stars in the I region and the W region is due to the capacity of TESS on detecting long periods of stars and the detection limit of flares. Many studies on this relationship with various proxies have demonstrated that stars with a similar Rossby number (Ro) have a similar intrinsic activity level \citep{Noyes1984,Wright2011,Yang2017,Yang2019}. This phenomenon is most significant for fast rotating stars that have reached the saturated activity level (the C region) and are independent of rotation.

\begin{figure*}
\includegraphics[width=0.31\textwidth]{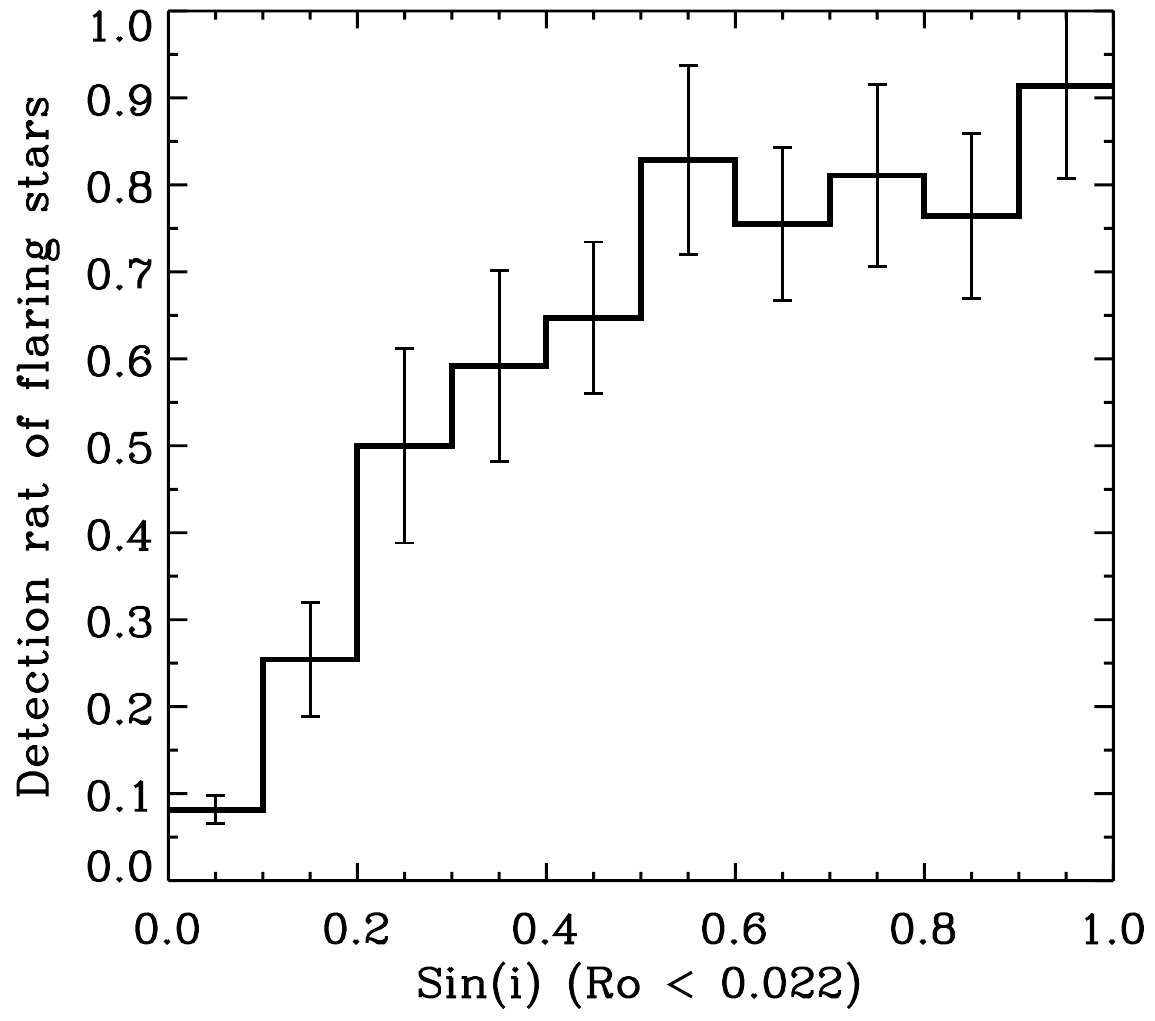}
\includegraphics[width=0.31\textwidth]{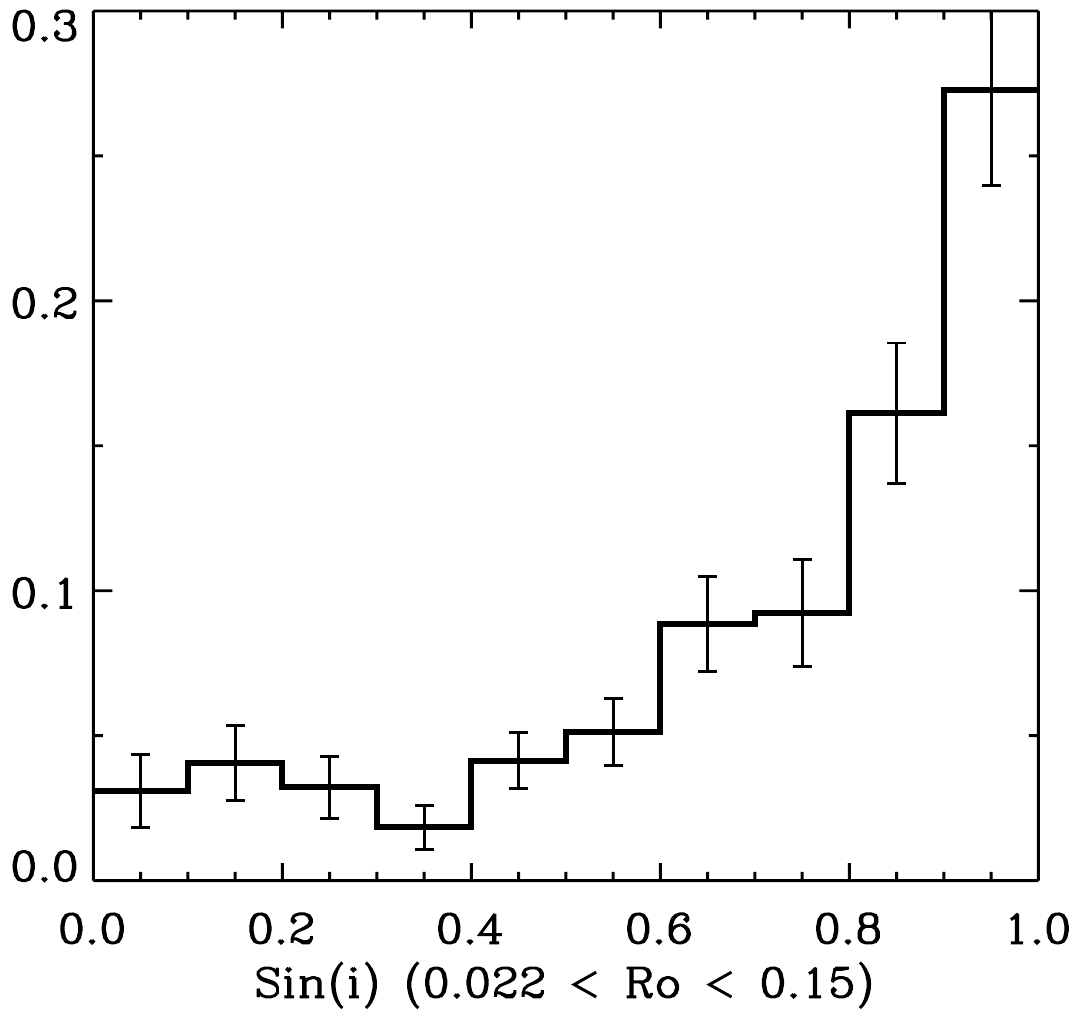}
\includegraphics[width=0.31\textwidth]{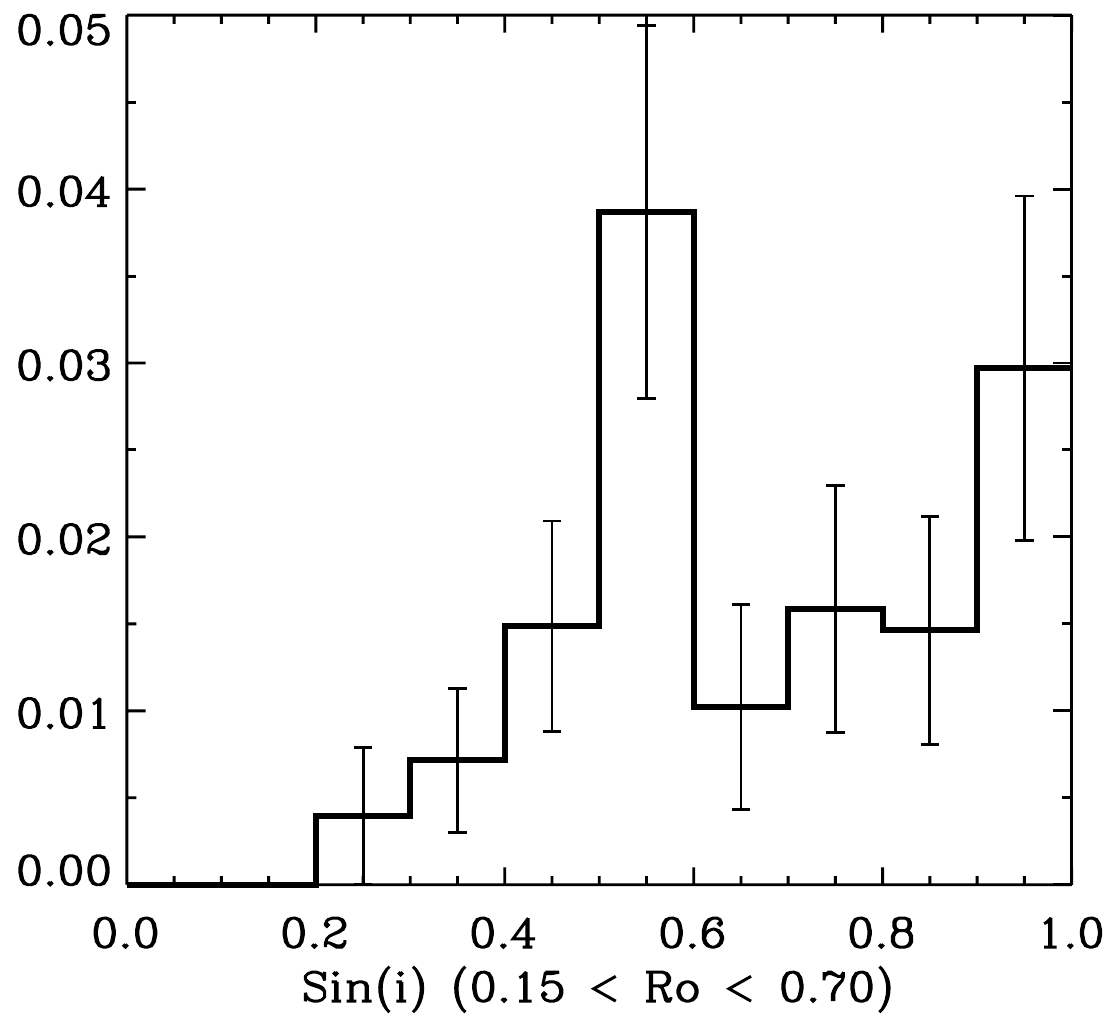}
\caption{The detection rates of flaring stars along with inclination in the C, g, and I phases. The detection rate is the ratio between the number of flaring stars and the number of rotating stars in a sin$i$ bin. Error bars of each sin$i$ bin were estimated using a square root of the number of flaring star in each bin.}
\label{fig_detect_rate}
\end{figure*}

\begin{figure*}
\includegraphics[width=0.5\textwidth]{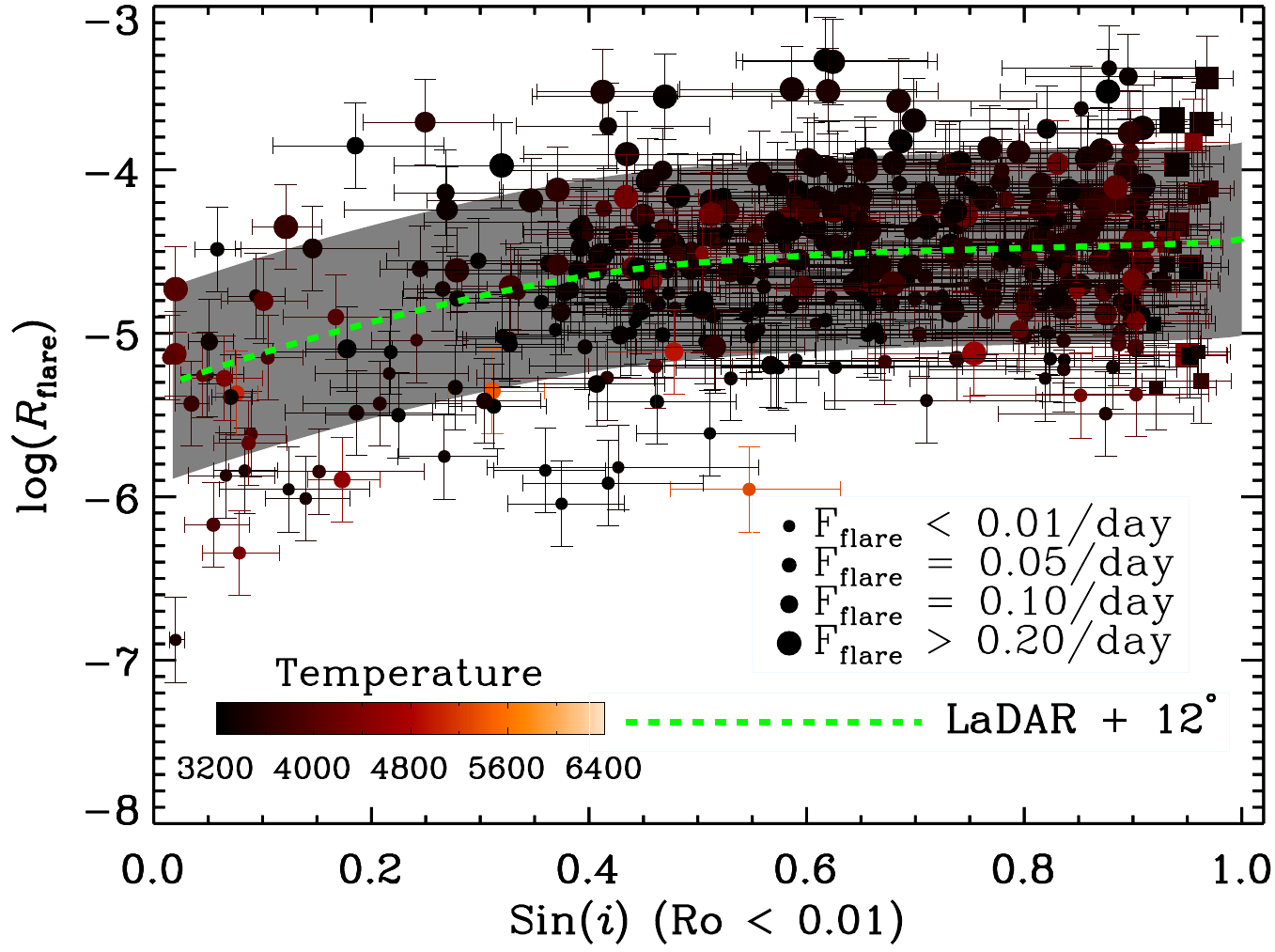}
\includegraphics[width=0.5\textwidth]{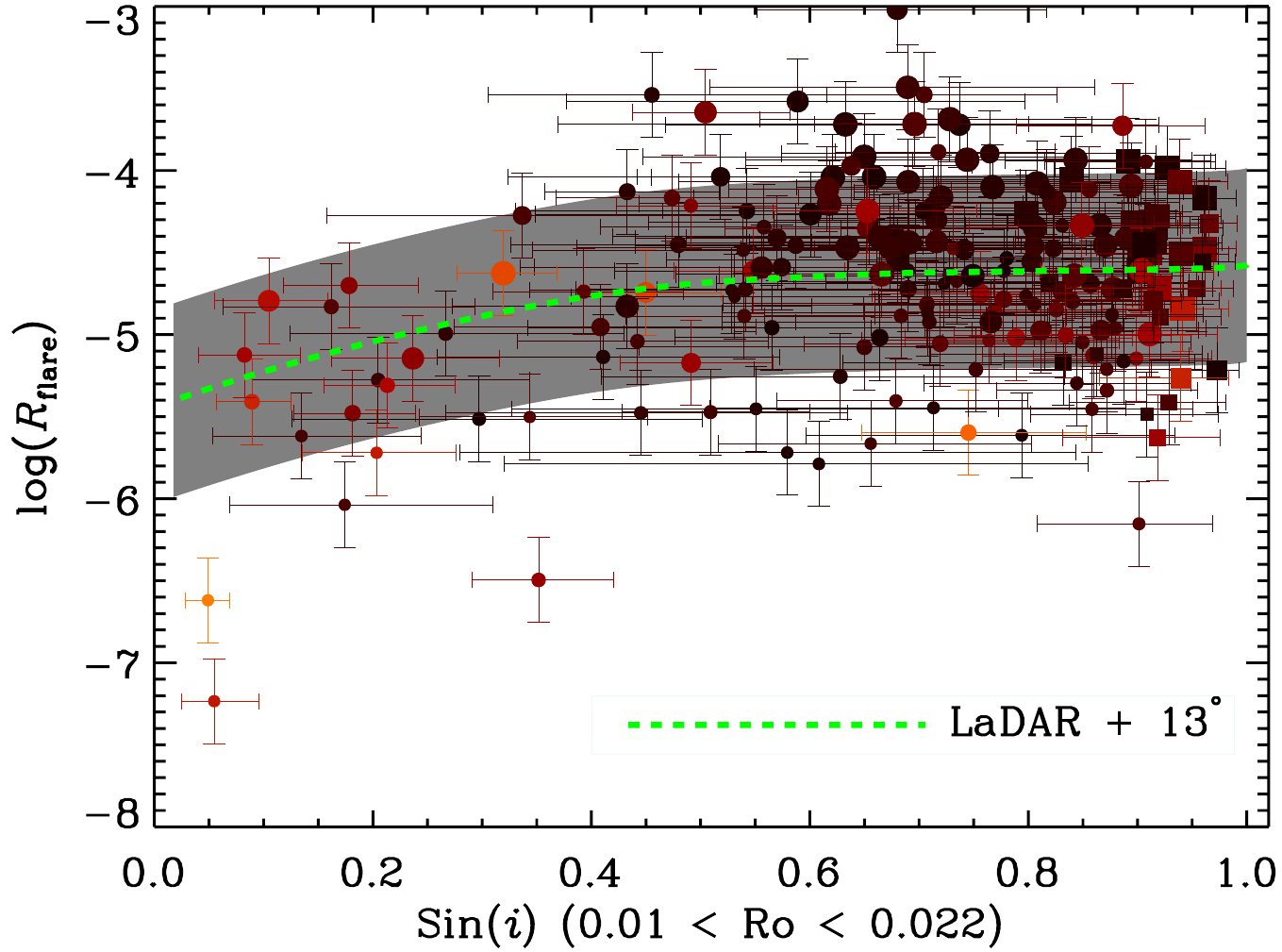}
\includegraphics[width=0.5\textwidth]{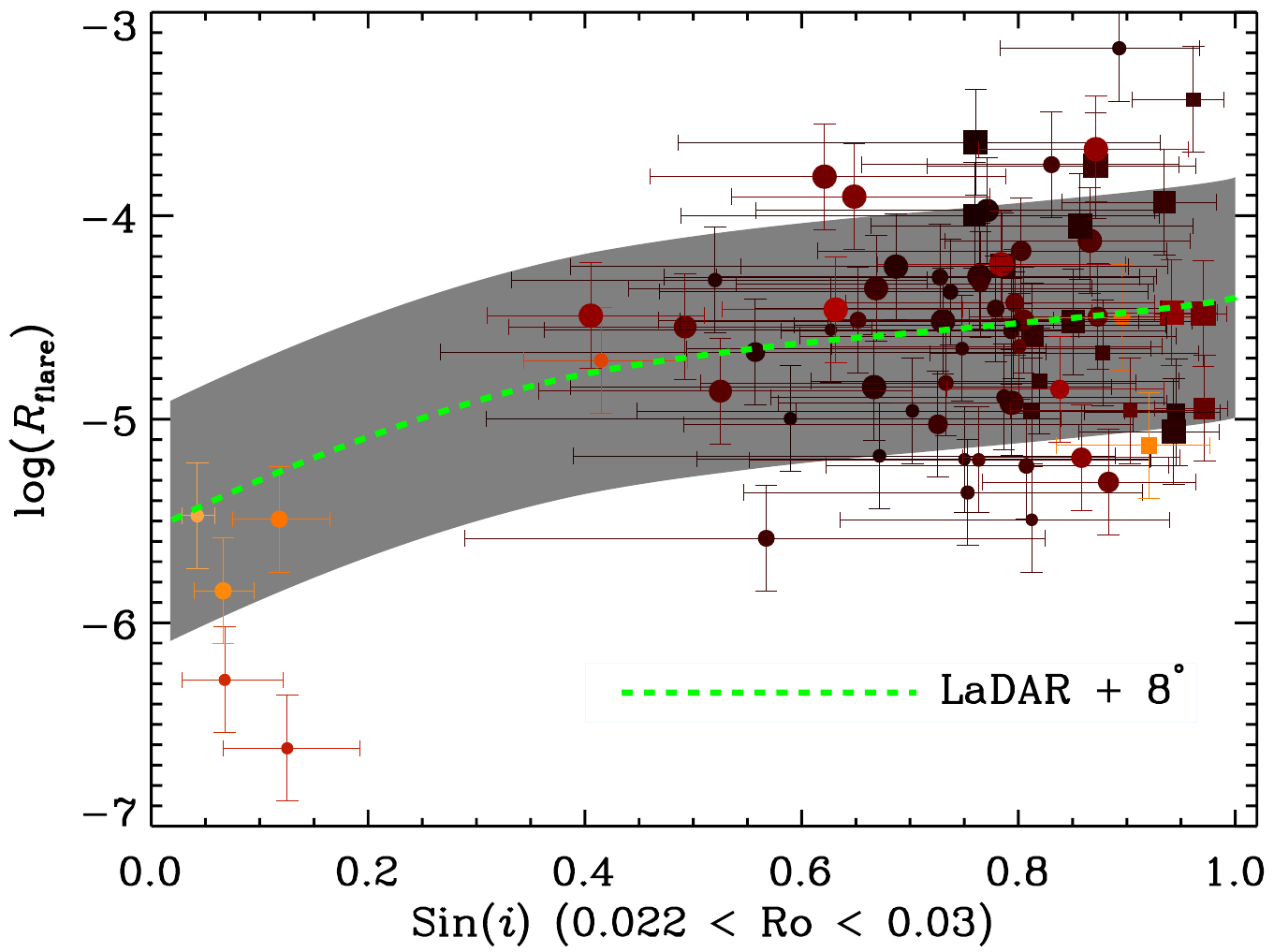}
\includegraphics[width=0.5\textwidth]{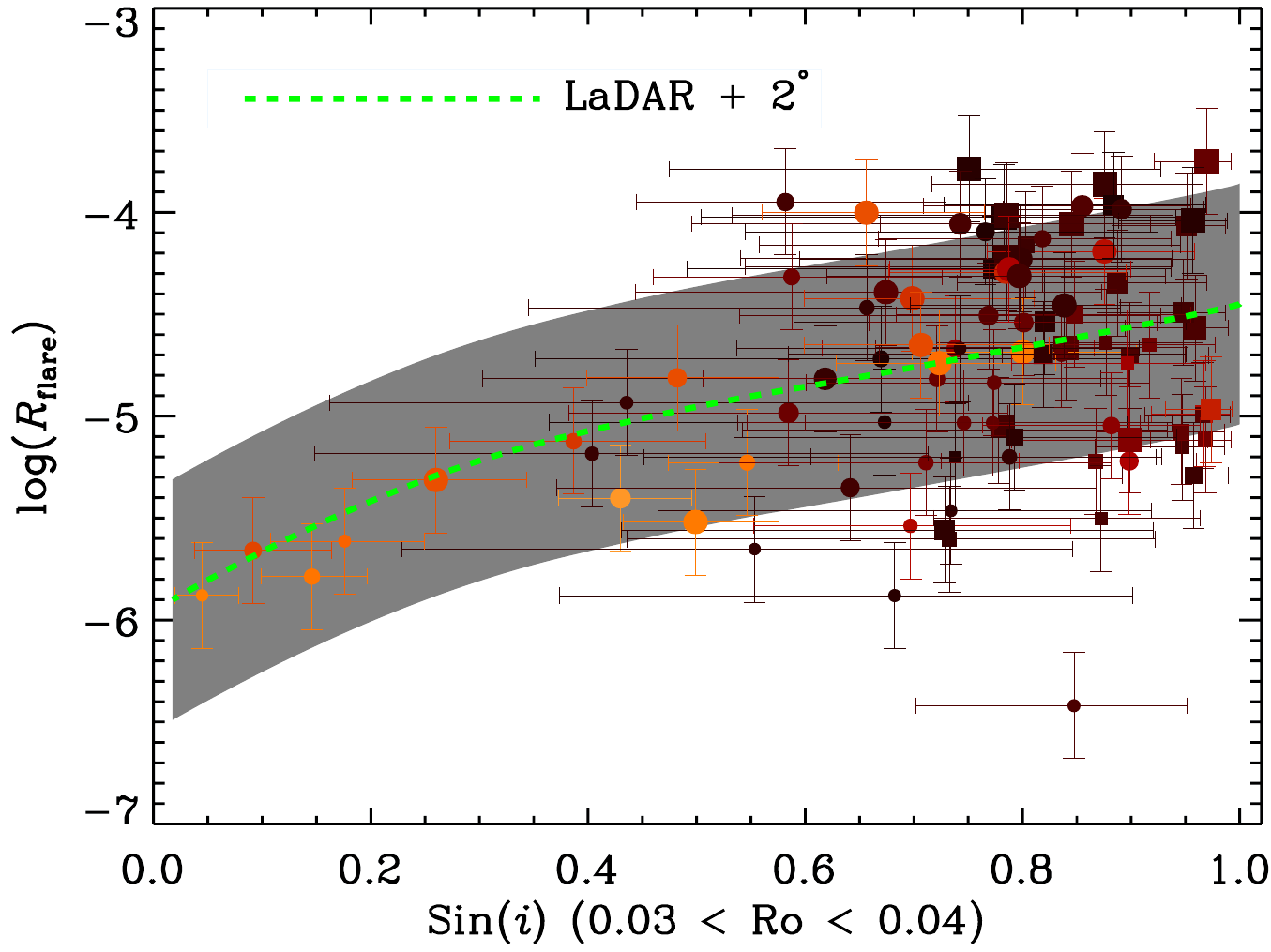}
\caption{ Inclination sin$i$ vs. the flaring activity in different Ro bins. The size of the circle represents the flaring frequency of a star. The green dashed line represents the best match of the stellar LaDAR. The shaded region represents the uncertainty caused by the maximum and minimum of the solar cycle. The uncertainty of log$R_{\rm flare}$ is 0.26 dex that is from the error propagation of the flare energy. The uncertainty of sin$i$ is from 16th and 84th percentiles of the posterior probability distribution of sin$i$.}
\label{fig_sini_fa}
\end{figure*}
\begin{figure}
\includegraphics[width=0.5\textwidth]{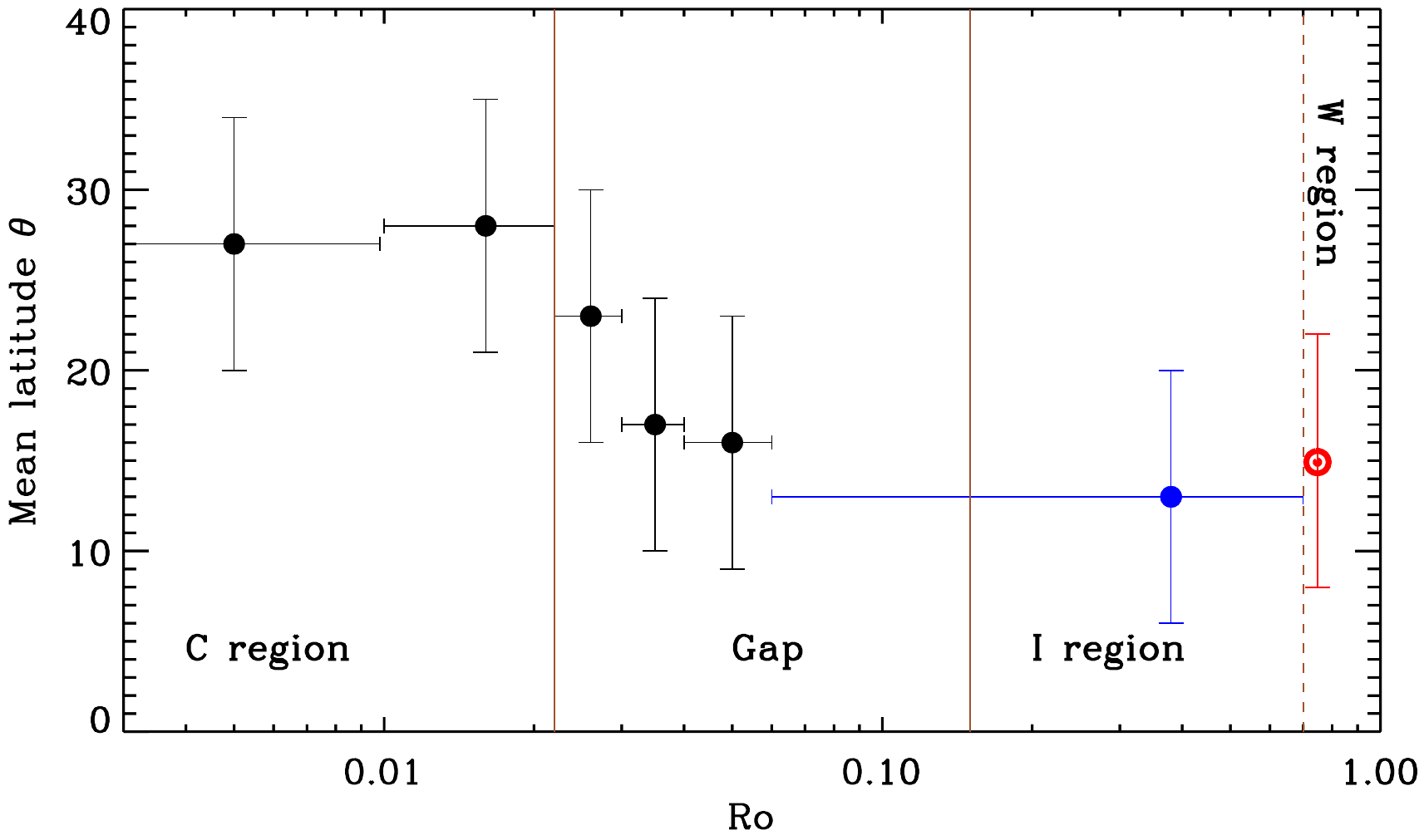}
\caption{Rossby number vs. the mean latitude of active regions. Each black point represents the best match between the observation and the simulated LaDAR in a Ro bin (the green dashed line in Figure~\ref{fig_sini_fa}). The blue point is obtained by the upper envelope of the  sin$i$--flaring activity relationship in  Figure~\ref{fig_sini_fa_ext}. The Sun is marked with an $\odot$ symbol. The uncertainty of the mean latitude is $\pm 7^\circ$ that is the same as the Sun. The uncertainty of Ro represents the range of each Ro bin used to fit the sin$i$--flaring activity relationship.}
\label{fig_ro_ladar}
\end{figure}

\section{result and discussion}
\subsection{The relation between  the apparent flaring activity and inclination } \label{sec:relation}

As the flare detection is independent of the inclination, the detection rates of flaring stars with different inclinations should be similar if flares occur uniformly on a hemisphere. However, the detection rate--inclination relationship in the C, g, and I phase (Figure~\ref{fig_detect_rate}) shows that the detection rate has a strong positive correlation with the inclination. Although the detection rate in the gap and I phase is influenced by the detection limit of the telescope and the decreasing rotation, it is very difficult to detect white-light flares for stars with a low inclination. This indicates that flares occur mainly at low latitudes because the limb darkening effect seriously weakens the flare detection when stars are near pole-on.

We have proposed that the variation of the apparent flaring activity along with the inclination can reveal the LaDAR of stars \citep{Yang2025}.  To illustrate the variation for different LaDARs, we take the Sun as a benchmark to carry out a simulation as follows: we have collected solar flare data for over 40 years \citep{Yang2025}, and found that their mean latitude is $\approx (15^{+7}_{-7})^{\circ}$ (Figure~\ref{fig_solar_ladar}). We created a series of stellar LaDAR by taking the equator as the axis of symmetry and decreasing/increasing the latitudes of flares from the solar LaDAR - $15^\circ$ to the solar LaDAR + $70^\circ$ by steps of $1^\circ$. Those mock data represent different latitudinal distributions of flares from the equator-concentrated to the pole-concentrated. We then simulated an observation of those distributions by the TESS telescope from pole-on ($i \approx 0^\circ$) to equator-on ($i \approx 90^\circ$) \citep{Yang2025}. The variation of the inclination determines whether and where each flare is observed, resulting in the relationships between the apparent flaring activity and the inclination for different LaDARs (Figure~\ref{fig_simu}). For example, when we observe the Sun as a star from the pole-on, all observable flares are near the limb whose apparent amplitudes and durations are seriously weakened by the limb darkening effect in the white-light band, resulting in the apparent flaring energy and observable flare number dramatically decreasing. Meanwhile, we show the relation between the apparent flaring activity and the inclination for groups of stars in separate Ro bins in Figure~\ref{fig_sini_fa}. As the same rotation corresponds to the same intrinsic flaring activity, the variation of the apparent flaring activity results from the LaDAR of stars in a Ro bin. The apparent flaring activity increases with increasing inclination in all Ro bins from the C region to the gap, indicating that their LaDARs are solar-like. We then vertically shifted the simulated relations of Figure~\ref{fig_simu} from log($R_{\rm flare}$) = -7 to log($R_{\rm flare}$) = -3 to compare with the data of Figure~\ref{fig_sini_fa} in each Ro bin, and took the best match with the minimum $\chi^2$ as the LaDAR in each Ro bin (the green dashed lines in Figure~\ref{fig_sini_fa}). 

It should be noted that the uncertainty of the sin$i$--flaring activity relationship increases with increasing Ro because of the telescope's capacity to determine $v$sin$i$ of slow rotators and the detection limit of flares (Figure~\ref{fig_sini_fa_ext}). The rotation period could also cause uncertainty in the unsaturated regime (Ro $> 0.022 $) as the intrinsic flaring activity decreases along with the rotation period. However, we argue that the upper envelope of this relationship (The bottom right panel of Figure~\ref{fig_sini_fa_ext}) can reveal the LaDAR of those slow rotators because it minimizes or reduces the influence from the detection limit of flares and rotation. Our sample have $\sim 25\%$ stars with $v$sin$i$/$v >1$ because of the uncertainty of $P_{\rm rot}$, $R$, and $v$sin$i$. Since they have high amplitudes of lightcurve modulations and relatively large $v$sin$i$, they are probably stars near equator-on. We note that the sin$i$--flaring activity relationship does not change when excluding stars of $v$sin$i$/$v >1$ (Fig.~\ref{fig_sini_fa_ext_exceed1}). The sin$i$--flaring activity relationship of fast rotators in three temperature bins are basically consistent (Fig.~\ref{fig_sini_fa_teffbin}), indicating that the influence of temperature on the LaDAR could be small. Since our sample does not have enough stars in each temperature bin for slow rotators, a precise analysis on the influence of temperature is needed in the future.

Figure~\ref{fig_ro_ladar} shows the evolution of stellar LaDAR along with Ro, whose mean latitudes are from the best match of Figure~\ref{fig_sini_fa} and Figure~\ref{fig_sini_fa_ext}. The mean latitudes are $\theta \approx 27^\circ$ in the C phase (saturated regime). Then they rapidly decrease in the gap towards the equator and reach a solar-like LaDAR in the I phase ($\theta \approx 13^\circ$). Finally, the transition from the I phase to the W phase makes the LaDAR slightly increase. This variation trend is consistent with the CgIW scenario that shows not only a saturated activity in the C phase, a rapid decline of stellar activity in the gap and a relatively slow decline of stellar activity in the I phase, but also a slight re-kindling of stellar activity from the I phase to the W phase \citep{Yang2025b}. Thus, the stellar LaDAR evolves in lockstep with stellar activity, which is supported by solar observations that the average latitudes of sunspots are strongly correlated with the strength of the stellar cycle \citep{Solanki2008,Jiang2011}. Specifically, for a solar mass star, the duration from $\theta \approx 27^\circ$ to $\theta \approx 13^\circ$ is very short, and it will have a solar-like LaDAR throughout its whole main sequence era ( Figure~\ref{fig_period_ladar_ind}; $P_{\rm rot} > 1$ day).

Our interpretation on the sin$i$--flaring activity relationship is based on the assumption that white-light flares originate from layers at or below the temperature minimum region so that the limb darkening effect could work \citep{Kuhar2016,Name2017}. However, the origin of white-light flares has not been fully understood \citep{Kowalski2024}. If white-light emission is primarily chromospheric in origin, it would not be subject to limb darkening. Whether the observed sin$i$--flaring activity relationship is caused by the limb darkening effect could be verified by studying X-ray data for this relationship, because the origin of X-ray flares is in or above the chromosphere. Unfortunately, there are few X-ray data for stellar flares, which prevents a statistical result.

\begin{figure*}
\includegraphics[width=0.8\textwidth]{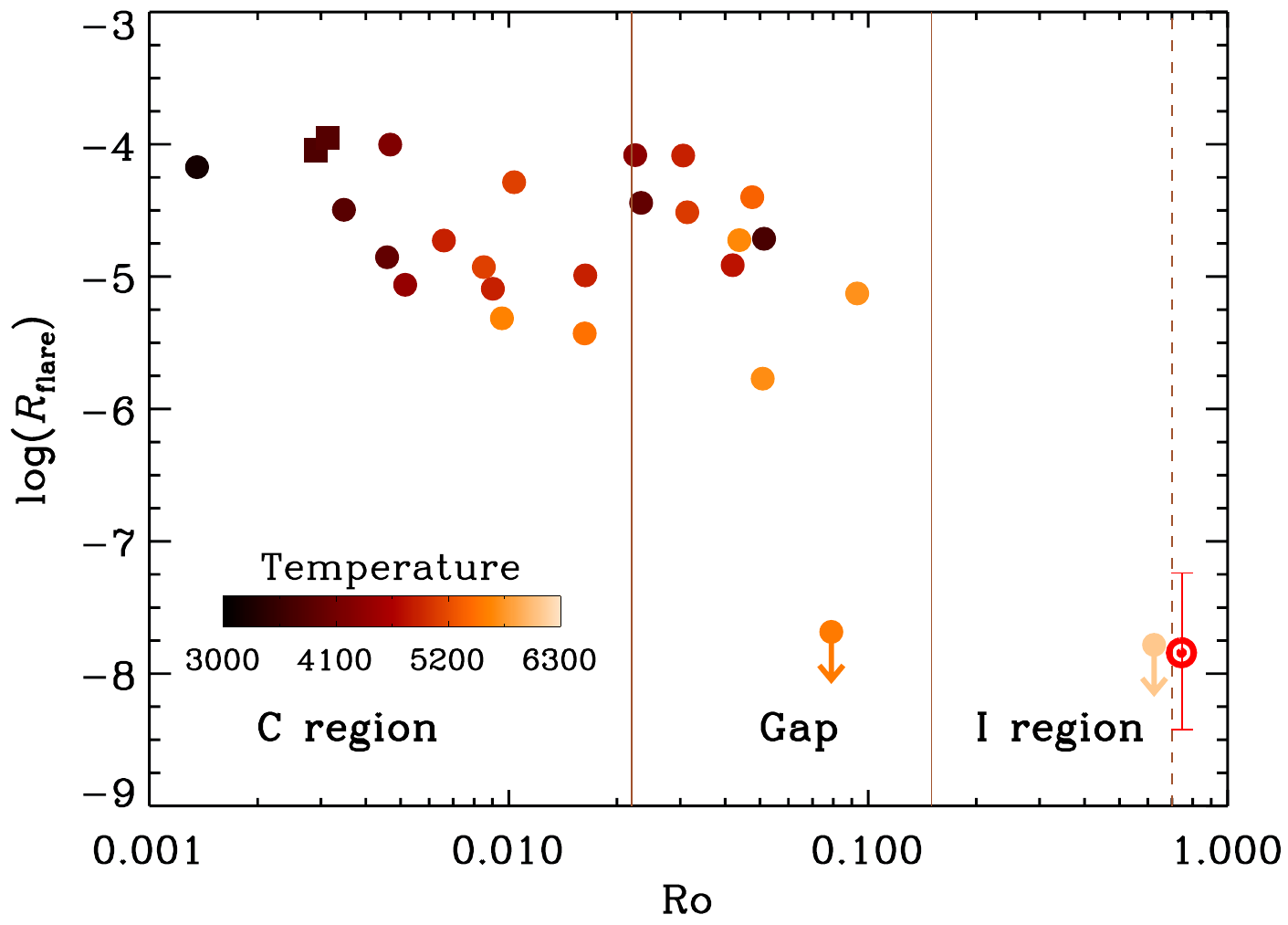}
\includegraphics[width=0.4\textwidth]{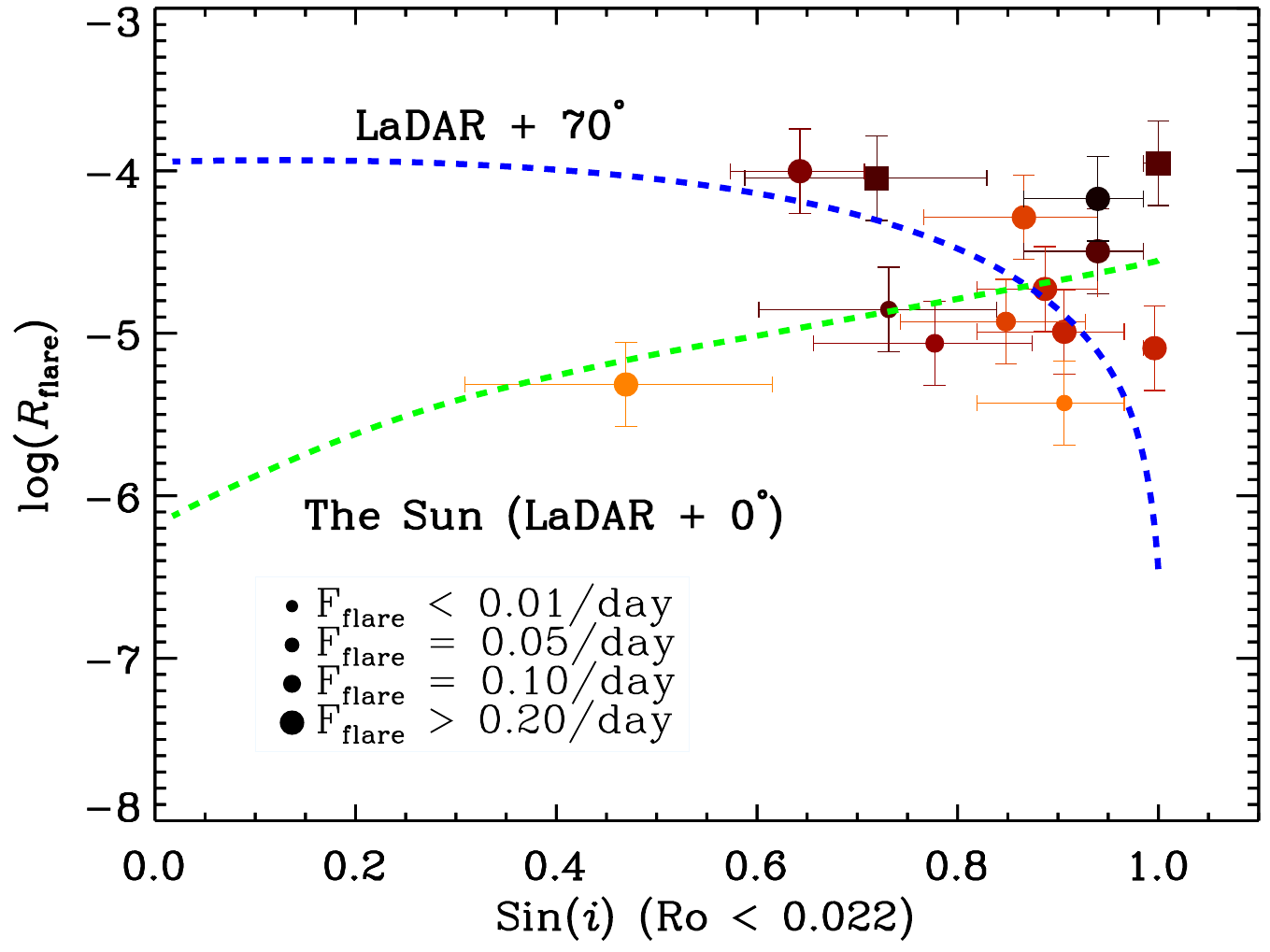}
\includegraphics[width=0.4\textwidth]{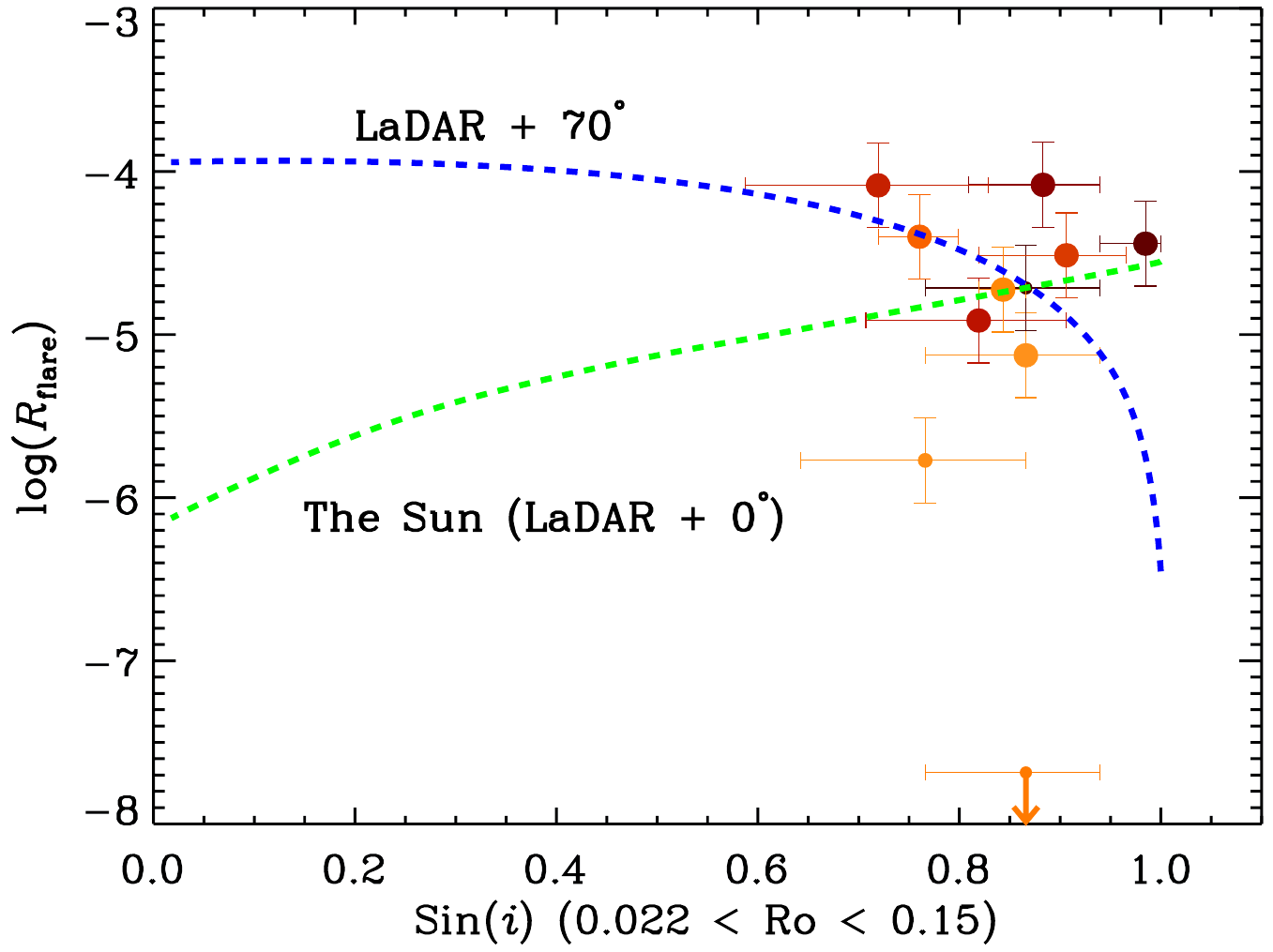}
\caption{ Top panel: the rotation--flaring activity relationship of 26 stars that have the TESS observations and (Z)DI measurements. The symbol of circle represents stars that are found to be high-latitude by the (Z)DI. The symbol of square represents stars that are found to be without high-latitude spots by the (Z)DI. Stars without detectable flares are plotted with downward arrows. Bottom panel: same as Figure~\ref{fig_ro_ladar} but for stars in the top panel. The green and blue dashed line denote the solar-like and polar latitudinal distribution, respectively.}
\label{fig_polarcap}
\end{figure*}
\subsection{The discrepancy of the polar spot and the latitudinal distribution of flaring activities} 

In the past 30 years, theoretical understanding and numerical simulation predicted the formation of the polar spot \citep{Schuessler1992}. Although our results show the predicted trend that faster rotators have higher LaDAR, the LaDAR is still in low latitudes and is far from the pole. We note that, in spite of that many observations have claimed polar spots on fast rotators through the (Z)DI, most of them also reported the presence of low-latitude spots at the same (Z)DI. It is not clear about the fraction of high-latitude spots and whether they are active. If active regions of fast rotators are mainly in the vicinity of the pole, it contradicts our results. We cross-matched stars with (Z)DI \citep{Strass2009} and the TESS mission. We found that 26 stars have the TESS observation, of which 24 stars are claimed to have high-latitude or polar spots by (Z)DI (Table~\ref{table_polarcap}). We calculated their flaring activities and compared them with their inclination in Figure~\ref{fig_polarcap}. Unfortunately, there are no stars with low inclination (sin$i < 0.4$), preventing the comparison from a definite result. However, given that the latitudinal distribution of polar spots will result in a significant drop when sin$i > 0.8$ (the blue dashed line in Figure~\ref{fig_polarcap}), it is very unlikely that those stars only have polar spots. 

In order to interpret the discrepancy between our results and the polar spots, we propose that, on one hand, the (Z)DI has three defects, resulting in a preference for polar spots: (1) it only reconstructs large-scale structures that can distort the line profiles and the small-scale magnetic fields that represent local spots tend to be canceled out \citep{Senav2021,Kochu2020,Namekata2024}. As high-latitude spots can be associated with a dipole that is a large-scale magnetic field \citep{Morin2010}, and low-latitude spots are attributed to small-scale fields, the results of the (Z)DI might be biased. (2) As we have mentioned above, the (Z)DI is sensitive to several factors that may result in opposite results or artifacts. For example, the young solar-like star EK Dra has been investigated by many studies through the (Z)DI \citep{Jarvi2009,Rosen2016,Waite2017,Jarvi2018,Senav2021,Namekata2024}. However, even for a similar observation date, different studies may obtain opposite results for the latitudinal distribution of spots. Given that their frequent and irregular changes, it is unlikely that the opposite results are due to the stellar cycle. (3) The capacity of (Z)DI near the equator is severely weakened and only the strongest features can be recovered \citep{Rice1989,Rice2002,Lee2025}. This can also bias the results of (Z)DI. On the other hand, an appealing explanation is that high-latitude spots are inactive, so high-latitude flares are rare. For example, a numerical simulation shows that the eruption of an active region will be suppressed by an overlying large-scale structure \citep{Alva2018}. It indicates that a polar spot (if it really exists) overlaid by a dipole is very stable and is difficult to trigger flares and coronal mass ejections. This is supported by the recent observation of EK Dra that a flare originates from low-latitude spots of $\theta \approx 25^\circ$ rather than polar spots \citep{Namekata2024}, although the polar spots have stronger magnetic fields. This is also supported by the TESS flare detection that finds high-latitude flares are rare \citep{Ilin2021,Bicz2024}. In this scenario, flares occur mainly at low latitudes, where they correspond to small-scale fields. 

The dynamo simulation of the magnetic flux emergence shows that, as rotation increases, the latitudinal distribution of magnetic flux emergence is split into multiple peaks, one of which is at low latitudes ( Figure~\ref{fig_isik_hist}) \citep{Isik2018,Isik2024}, corresponding to low-latitude spots and small-scale fields. The others are at mid- and high-latitudes with larger tilt angles, which account for the formation of a dipole, given that larger tilt angles lead to a larger contribution to the global axial dipole field and less flux cancels out between opposite-polarity patches of bipolar magnetic regions \citep{Wang1991}. The split of the magnetic flux emergence is supported by the finding that starspots show a bimodality of latitude distribution \citep{Netto2020,Lehtinen2022}. We separate the lower component of the distribution (the vertical dashed line in  Figure~\ref{fig_isik_hist}) and find its mean latitude increases from $\theta \approx 15^\circ$ to $\theta \approx 25^\circ$, as the rotation period increases from a solar-like rotation to $P_{\rm rot} = 3.125$ days. This rotation--mean latitude relationship is basically in line with our results of Figure~\ref{fig_ro_ladar} in the gap and I phase. Further simulation of ultra-fast rotators is necessary to verify this relationship in the C phase. 

We present the relation between $R_{\rm flare}$ and the proxy of light-curve modulation $S_{\rm ph}$ ( Figure~\ref{fig_sini_sph}). Their strong positive correlation can also prove that even if polar spots are widespread, they are stable and inactive. Given that $S_{\rm ph}$ depends on the inclination, if most flares occur near the pole, there should be a negative correlation between $R_{\rm flare}$ and $S_{\rm ph}$. If flares occur uniformly on the sphere, there should not be a correlation. Only if most flares occur at low latitudes, there should be a positive correlation. Another way to verify the flaring polar spots is that there could be some flaring polar-spot stars with very low inclinations. In this case, stars could have a high level of $R_{\rm flare}$ but their rotations are undetectable. Our sample has 130 flaring stars without detectable rotation periods, implying that they may belong to this case. We compare their distributions of $R_{\rm flare}$, flaring amplitude, and maximum flaring luminosity with that of fast rotators ( Figure~\ref{fig_np_hist}). The fact that their distributions are closer to a lower activity level suggests that they are fast rotators with low-latitude spots or slow rotators, excluding the possibility that they are fast rotators with flaring polar spots.

\begin{figure*}
\includegraphics[width=0.5\textwidth]{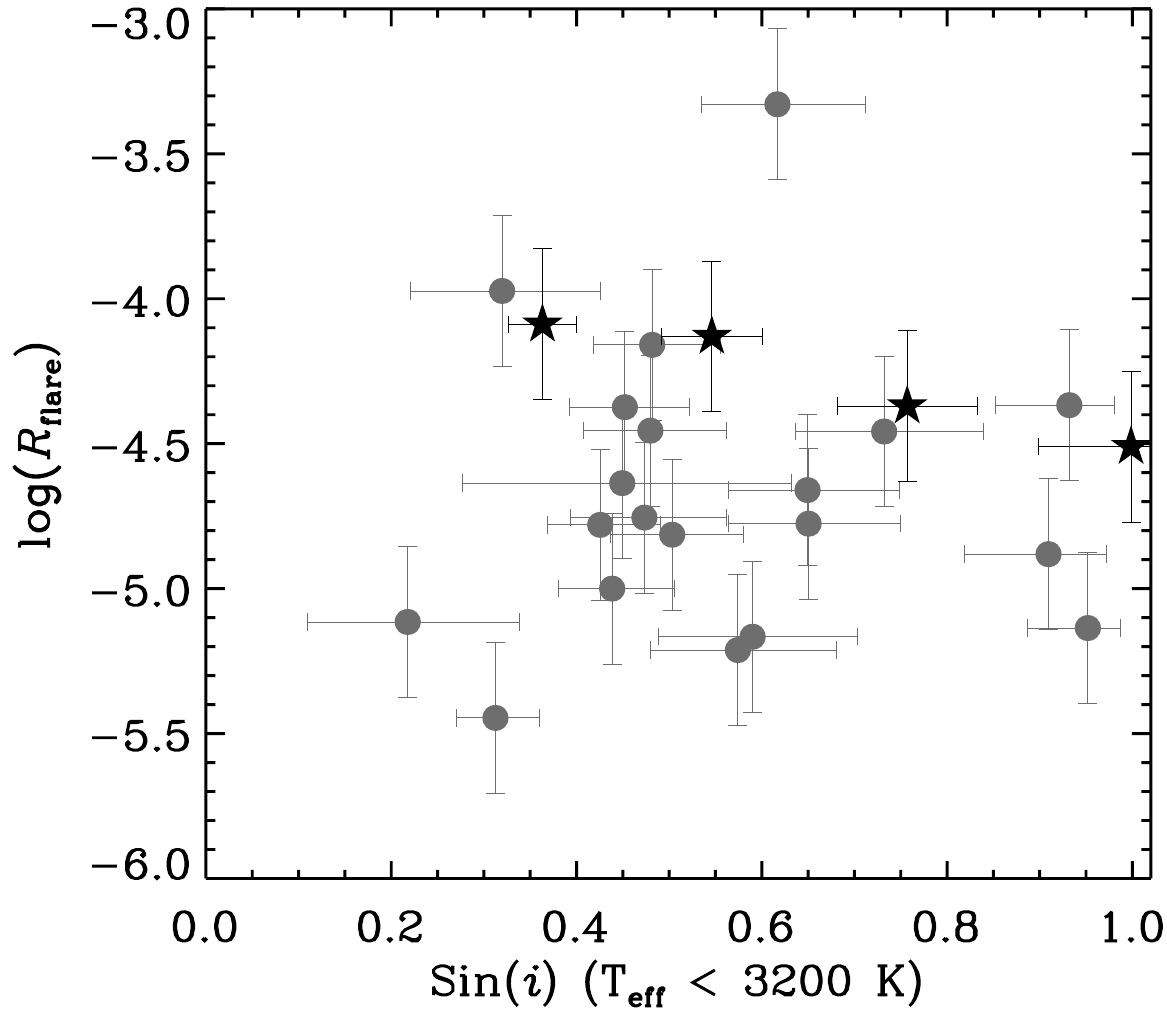}
\includegraphics[width=0.5\textwidth]{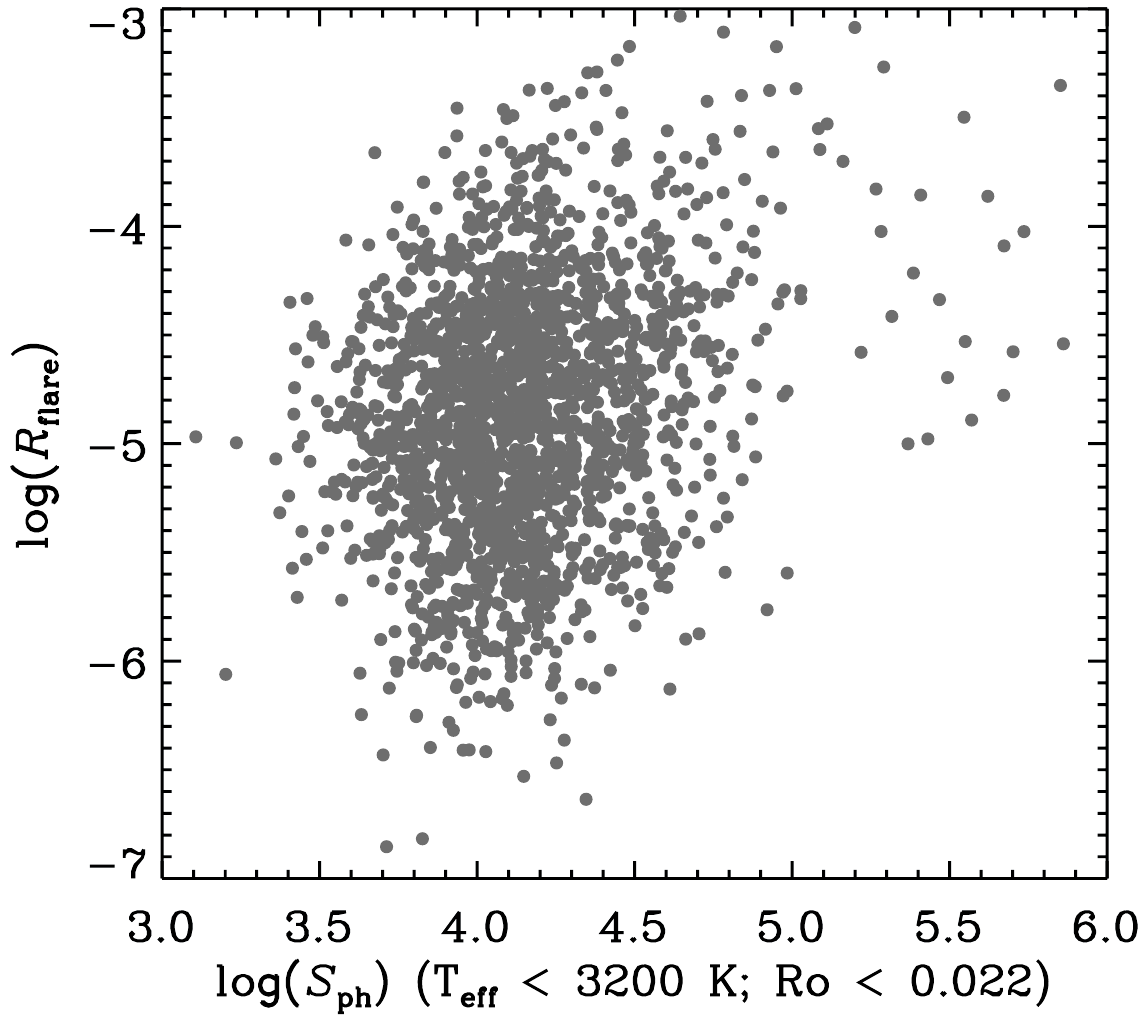}
\caption{Left panel: the sin$i$--flaring activity relationship for fully convective stars ($T_{\rm eff} < 3200$ K). The symbol of circle denotes flaring stars in our sample and the symbol of five-point star denotes four stars in \cite{Ilin2021}, where high-latitude flares have been identified. All stars are fast rotators and in the C region. Right panel: the $S_{\rm ph}$--flaring activity relationship for fully convective stars in the C region. The Pearson correlation coefficient is $r \approx 0.23$. }
\label{fig_fc}
\end{figure*}

\subsection{The LaDAR of fully convective stars}\label{sec4}

\citet{Ilin2021} found high-latitude flares in fully convective (FC) stars ($T_{\rm eff} < 3200$ K; M4), which is different from partially convective (PC) stars of our sample. The coolest temperature of our sample is 3120 K and FC stars are few (Fig.~\ref{fig_fc}), preventing us from validating the LaDAR of FC stars in our sample. However, compared to partially convective stars that have a strong positive correlation between $S_{\rm ph}$ and $R_{\rm flare}$, FC stars show a weaker correlation (Fig.~\ref{fig_fc}), indicating that the flare amplitudes of FC stars are similar regardless of the amplitude of starspots (i.e., even if a star is near pole-on, its apparent flare could be very strong). This weak correlation implies that the LaDAR of FC stars may be higher than that of partially convective (PC) stars or may be uniformly distributed in the hemisphere, which is consistent with that high-latitude flares have been found in FC stars \citep{Ilin2021}. It also reveals the dynamo discrepancy between PC and FC stars, given that a dynamo simulation of FC stars shows that uniformly distributed small-scale fields that carry most of the magnetic energies can be formed through turbulent convection \citep{Yadav2015}. 

To verify the inclination--flaring activity relationship of FC stars, it requires enough FC stars with a low inclination to obtain a direct comparison. However, FC stars are small and faint, making it difficult to observe them with a high-resolution telescope and get an accurate $v$sin$i$ especially for stars with low inclinations. In the future, a definite inclination--flaring activity relationship for FC stars is necessary.

\begin{acknowledgments}
We sincerely thank the referee for helpful comments. We sincerely thank J. Jiang and K. Namekata for the discussion. We sincerely thank E.Isik for the discussion and kindly provide simulation data.
\end{acknowledgments}





%
\facilities{TESS \citep{Ricker2015}, APOGEE \citep{Abdu2022}, GALAH \citep{Buder2024}}
\software{astropy \citep{2013A&A...558A..33A,2018AJ....156..123A,2022ApJ...935..167A},  
          Interactive Data Language \citep{Landsman1993} , 
          TESS Data Integration Platform \citep{Yang2019}
          }


\appendix

\restartappendixnumbering

\section{The sample selection}\label{sample_select}

We cross-matched the TESS mission with the APOGEE DR17 \citep{Abdu2022} and removed stars with the following APOGEE quality flags:  (ASPCAPFLAGS): VSINI\_BAD,  SN\_BAD, STAR\_BAD,  CHI2\_BAD,  VMICRO\_BAD;  (STARFLAGS): BAD\_PIXELS, LOW\_SNR, or VERY\_BRIGHT\_NEIGHBOR.  We cross-matched the TESS mission with the GALAH DR4 \citep{Buder2024} and removed stars with the following GALAH quality flags: flag\_sp $> 1$; flag\_red $ \neq 0$. We also removed stars with the following parameters : $T_{\rm eff} > 6400$, log$g < 3.5$ (giants whose spectrum broadening are dominated by macroturbulent velocity \citep{Holtzman2018}). In total, we obtained 31735 stars, in which 22803 stars are from APOGEE.


\section{The measurement of rotation periods} \label{sec:rotation}

\begin{table*}
\caption{\label{table_rot}The priorities of references for rotation periods in this study.}
\centering
\begin{tabular}{cccc}
\hline\hline\\
Reference&Priority&Number of rotation stars & Mission\\[1ex]
&&in this study&\\

\hline\\
\citet{Mc2014}           &1&223&Kepler \\[1ex]
\citet{Santos2021}           &2 &292&Kepler\\[1ex]
\citet{Colman2024}          &3&1349&TESS\\[1ex]
\citet{Fetherolf2023}       & 4&3931&TESS \\[1ex]
\citet{Kounkel2022}          &5&173&TESS \\[1ex]
This work        & 6&2746&TESS  \\[1ex]
\citet{Reinhold2020}        & 7&704&K2\\[1ex]
\hline\\
total      &--    &9418&Kepler, K2, TESS \\[1ex]
\hline
\end{tabular}
\tablecomments{The priority of the reference for rotation periods increases with the decreasing priority number.}
\end{table*}

\begin{table*}
\caption{\label{table_var}An example on the parameters of 9418 rotating stars in this study}
\centering
\begin{tabular}{cccccccc}
\hline\hline\\
TIC&$T_{\rm eff}$&log$g$&$P_{\rm rot}$ & Ref.&flare flag &$S_{\rm ph}$\\[1ex]
&(K)&($\rm cm/s^2$)&(Day)&&&(ppm)\\[1ex]

\hline\\
25078924  &5238  &4.22  &0.91 &Colman et al.2024      & 0  &   26959.1\\[1ex]
25081005  &5683  &4.48  &5.73 &Colman et al.2024   &0    &    1532.7\\[1ex]
25081173  &6302  &4.13  &2.83 &Fetherolf et al.2023 &  0  &     1183.7\\[1ex]
25116563  &4749  &4.60  &10.09 &Colman et al.2024  & 0   &     2076.8\\[1ex]
25117741  &5346  &4.51  &10.77 &Fetherolf et al.2023  & 0   &     20237.8\\[1ex]
\hline
\end{tabular}
\tablecomments{The entire table is available online.}
\end{table*}

Firstly, we collected references for rotation periods \citep{Mc2014,Santos2021,Colman2024,Fetherolf2023,Kounkel2022,Reinhold2020} in the Kepler, K2 and TESS missions. As the capacity of TESS on detecting rotation period is weaker than that of the Kepler mission \citep{Boyle2025} and it is difficult to recover stars whose $P_{\rm rot} > 13.7$ days \citep{Claytor2024}, we preferentially adopted rotation periods from the Kepler mission. The priorities of the references are shown in Table~\ref{table_rot}.

Secondly, for stars without rotation periods in references, we carried out the rotation detection as follows:
1). We calculated the Lomb-Scargle (LS) periodogram for each sector of each star and obtained the normalized power peak of each LS.
2). If more than half of the power peaks of a star are greater than 0.05, the rotation period of a star is calculated from the median of those power peaks.
3). We only adopted rotation periods for stars whose $P_{\rm rot} < 13$ days.

Thirdly, the uncertainties of rotation period are mainly from the differential rotations and harmonic signals. We picked out stars with multiple periods(i.e. if a star has different rotation periods from different references or our LS periodogram, we take it as a star with multiple periods. Some references present the primary and secondary
periods, or present periods for individual sectors. If any of them are different, we take it as a
star with multiple periods). In total, multiple-period stars account for $\sim 28\%$ of our flaring sample, in which $\sim 16\%$ are stars with harmonic signals ($P_{\rm max} > 1.5 \times P_{\rm min}$) and $\sim 12\%$ are stars with differential rotation ($P_{\rm max} < 1.5 \times P_{\rm min}$). For stars with differential rotation, we adopted periods that could cause sin$i <$ 1. For stars with harmonic signals, we visually checked the lightcurves to make a decision (if their power are close, we preferred the periods that could make sin $i<$ 1; if the power of the primary signal is dominated and does not vary in each sector, we preferred the primary signal ). Note that the sin$i$--flaring activity relationship does not change when removing stars with multiple periods. We also checked stars with long periods in the TESS mission \citep{Claytor2024}. However, we found that those periods would make sin$i >> $ 1, indicating that TESS is not suitable for detecting periods of slow rotators.

In total, we obtained 9418 rotating stars. Table~\ref{table_var} shows an example on the parameters of them. In order to compare the flaring activity with other activity proxy, we also calculated the proxy of the light-curve amplitude $S_{\rm ph}$ \citep{Mathur2025} for each flaring star. We calculated $S_{\rm ph}$ in a duration of $P_{\rm rot} \times 5$ and took the mean of them as the final $S_{\rm ph}$ of a star (If a star do not have a period, we took a duration of 5 days to calculate $S_{\rm ph}$). In the calculation, we only used the quiet lightcurves by excluding the flare flux.

\section{The flare detection and contamination check } \label{sec:flaredect}

We used the light-curves of stars (2-min cadence; Pre-search Data Conditioning Simple Aperture Photometry from the Science Processing Operations Center pipeline) from Sector 1 to Sector 60 to detect flares in our cross-matched sample. We present our detection method step by step for clarity. 

1. Due to systematic issues such as the orbital period and the angular momentum dumps, the light-curve in each sector is often separated into several parts, resulting in discontinuities within a sector. We fit those discontinuities (when the discontinuity is longer than 6 hr) independently rather than taking a sector as a whole. 

2. In each part of a sector, a smoothing filter based on the spline algorithm is used to fit the baseline of the light-curve. The filter width is one-eighth of the most significant period that is given by the LS periodogram.  The upper limit of the filter width is set to be 24 hr \citep{Yang2017}. The degree of the spline fit is determined by the minimal Bayesian information criterion \citep{Liddle2007}. We perform an iterative $\sigma$-clipping approach to remove all outliers in the fitting process.

3. After detrending baselines, at least three continuous points higher than 3$\sigma$ are sorted out as a flare candidate, and all candidates must be no break points in the vicinity of the peak ($\pm 3$ points).  Its beginning and end are at the first continuous point and the last continuous point that are higher than 2$\sigma$, respectively.

4. If a flare candidate only has one peak, its rising phase should be shorter than the decay phase. If a flare candidate has multiple peaks, it will be divided into several sub-flares according to local minimum points of the flare profile. Points after the peak of the maximum sub-flare are taken as the decay phase. Its duration should be longer than the rising phase of the maximum sub-flare. Otherwise, the whole flare candidate will be discarded.

We have developed a visualization software, the TESS Data Integration Platform, which is developed from the framework of the Kepler Data Integration Platform \citep{Yang2019}. This software enabled us to visually and quickly check the candidate sample and removed stars including $\delta$ Scuti, $\gamma$ Doradus, RR lyrae and stars with asteroid crossing \citep{Yang2017,Yang2019,Seli2025}. The sample of flaring stars includes 203 binaries that are identified by the non-single star and variability catalog of $Gaia$ DR3 \citep{Gaiadr3}. We also found 21 potential close binaries whose APOGEE VSCATTER parameter is larger than 1 km/s. The spatial resolution of TESS is low ($\sim 21''$). Flaring stars might be polluted by bright neighboring stars\citep{Tu2020}. Based on $Gaia$ DR3, we find that 82 stars in our sample have neighboring stars (within $42''$) that are brighter than those flaring stars. We present all flaring stars and flares in Table~\ref{table_fa} and Table~\ref{table_flaregy}, respectively. In Table ~\ref{table_fa}, we present several flags to denote whether a star is binary or has a brighter neighboring star. In the sin$i$--flaring activity relationship, we do not include any binary or potentially contaminated stars.
\begin{table*}
\setlength{\tabcolsep}{1pt}
\caption{\label{table_fa} An example on parameters of 1510 flaring stars in this study.}
\centering
\begin{tabular}{ccccccccccccccccccccccccccccc}
\hline\hline\\
TIC&$T_{\rm eff}$&log$g$&$R/R_{\rm sun}$& $\tau_{g}$&$V_{\rm sini}$& $P_{\rm rot}$& $N_{\rm flare}$& log($E_{\rm max}$)& $t_{\rm obs}$&$R_{\rm flare}$&Sin$i$&$L_{\rm max}$&Vflag&Bflag&NS\\[1ex]
&(K)&($\rm cm/s^2$)&& (Day)&($\rm km/s$)& (Day) & & (erg)& (Day)&&&(erg/s)&\\[1ex]
\hline

25118964 &3347 & 4.63 &0.56 &159.6& 27.60 & 1.02 & 253  &34.75 &558.9393& -4.095 &0.9095  &31.4951&S&S&No\\[1ex]
25132694 &5170 & 4.54 &0.85 &55.9 &  1.52 & 9.90 & 2    &33.65 &572.1321& -7.213 &0.3305  &31.0493&S&S&No \\[1ex]
25132999 &4023 & 4.67 &0.63 &98.1 &  5.60 & 5.16 & 129  &34.97 &544.6092& -4.379 &0.9066  &31.8905&S&S&No\\[1ex]
29759435 &3733 & 4.64 &0.61 &140.8&  1.71 & -1.00&  2   &34.06 &515.3696& -5.953 & -1     &30.9733&S&S&No \\[1ex]
29779873 &3332 & 4.79 &0.45 &194.7&  12.28& 2.62 & 184  &34.65 &544.4557& -3.986 & 0.9270 &31.5757&S&S&No \\[1ex]

\hline
\end{tabular}
\tablecomments{The entire table is available online as some columns are not shown.}
\end{table*}

\begin{table*}
\setlength{\tabcolsep}{4pt}
\caption{\label{table_flaregy} An example on parameters of 26690 flares in this study.}
\centering
\begin{tabular}{ccccccccccccccccccccccccccccc}
\hline\hline\\
TIC&Sector&$t_{\rm start}$&$t_{\rm end}$&log($E_{\rm flare}$)&log($L_{\rm max}$)&$t_{\rm peak}$&$A_{\rm max}$\\[1ex]
&&(Day)& (Day) & (erg) & (erg/s) & (Day)& (ppm)\\[1ex]
\hline

25118964& S10 &1571.0947 &1571.1003  & 32.6563  & 30.1150 &1571.0947 &  30193.93 \\[1ex]
25118964& S10 &1573.7559 &1573.7725  & 32.9859  & 30.0820 &1573.7585 &  27983.22\\[1ex]
25118964& S10 &1575.0336 &1575.1323  & 34.1702  & 30.4429 &1575.0435 &  64251.05\\[1ex]
25118964& S10 &1576.4531 &1576.4698  & 33.2948  & 30.5088 &1576.4545 &  74769.70\\[1ex]
25118964& S10 &1578.2419 &1578.2502  & 32.8645  & 30.2343 &1578.2419 &  39743.95\\[1ex]

\hline
\end{tabular}
\tablecomments{The entire table is available online as some columns are not shown.}
\end{table*}

In total, we obtained 1510 flaring stars with 26690 flares. The examples of the flaring stars and flares are shown in Table~\ref{table_fa} and Table~\ref{table_flaregy}.

\begin{figure*}
\includegraphics[width=1\textwidth]{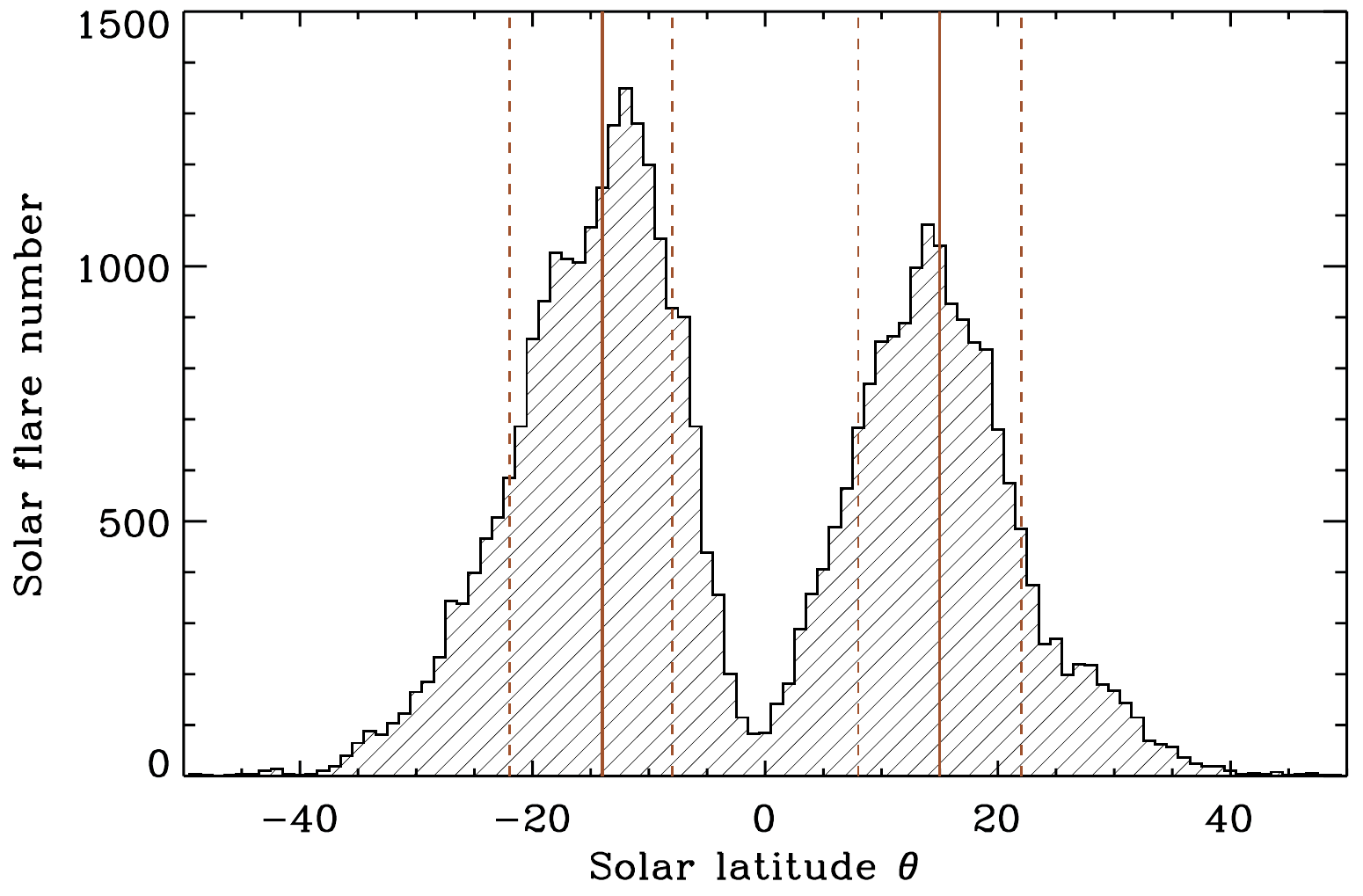}
\caption{The latitudinal distribution of solar flares from 1975 to 2017. In each hemisphere, the 50th percentile (the vertical solid lines) is taken as the mean latitude of solar flares, and the 16th and 84th percentiles (the vertical dashed lines) are taken as the uncertainties. Their vales are $\theta = (-14^{+6}_{-8})^{\circ}$ and $(15^{+7}_{-7})^{\circ}$, respectively.} 
\label{fig_solar_ladar}
\end{figure*}
\begin{figure*}
\includegraphics[width=1\textwidth]{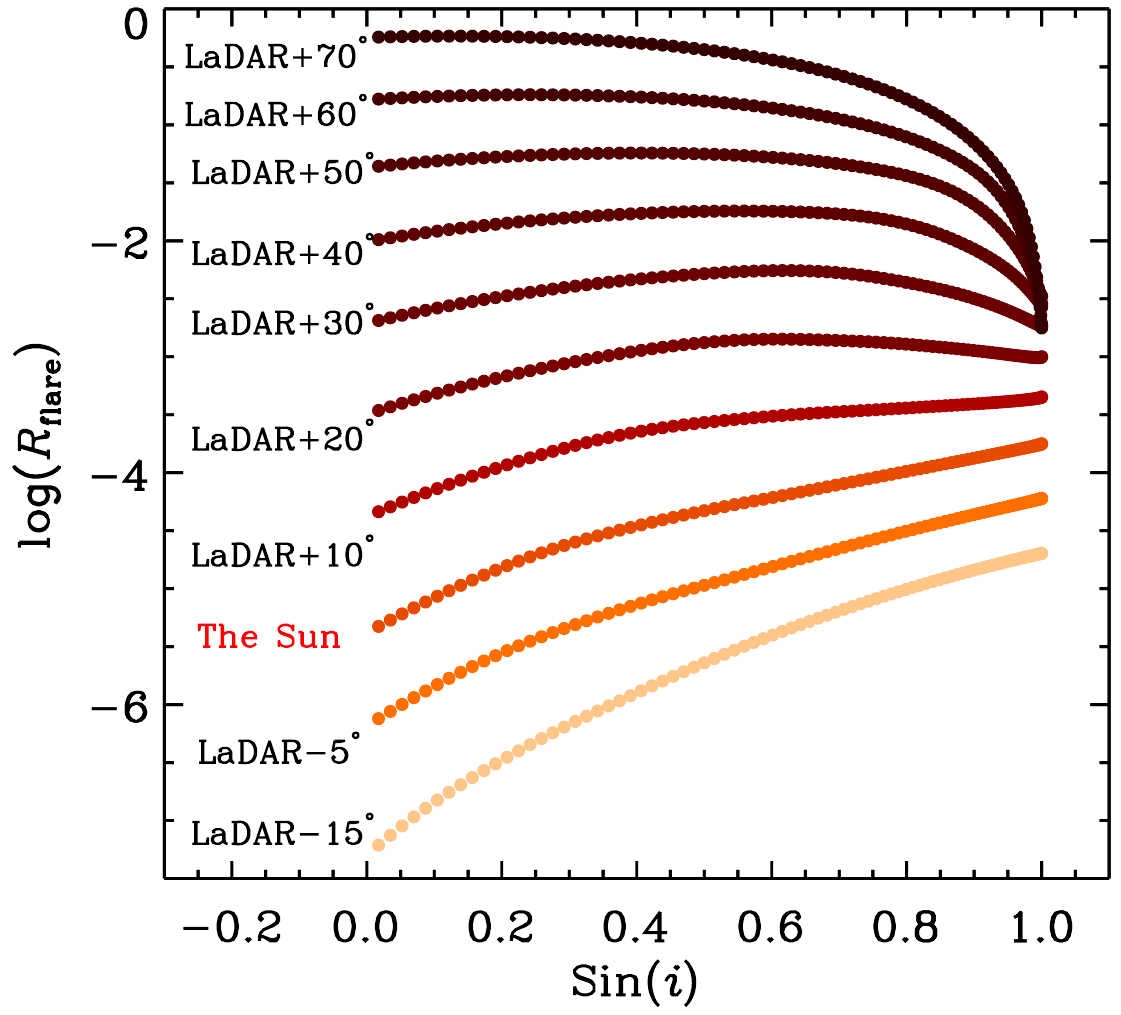}
\caption{ The simulated relationship between the inclination sin$i$ and the flaring activity for different LaDAR. Each relation represents a LaDAR that is created by increasing or decreasing the solar LaDAR for several degrees. Each relation is vertically shifted for clarity.}
\label{fig_simu}
\end{figure*}

\begin{figure*}
\includegraphics[width=0.5\textwidth]{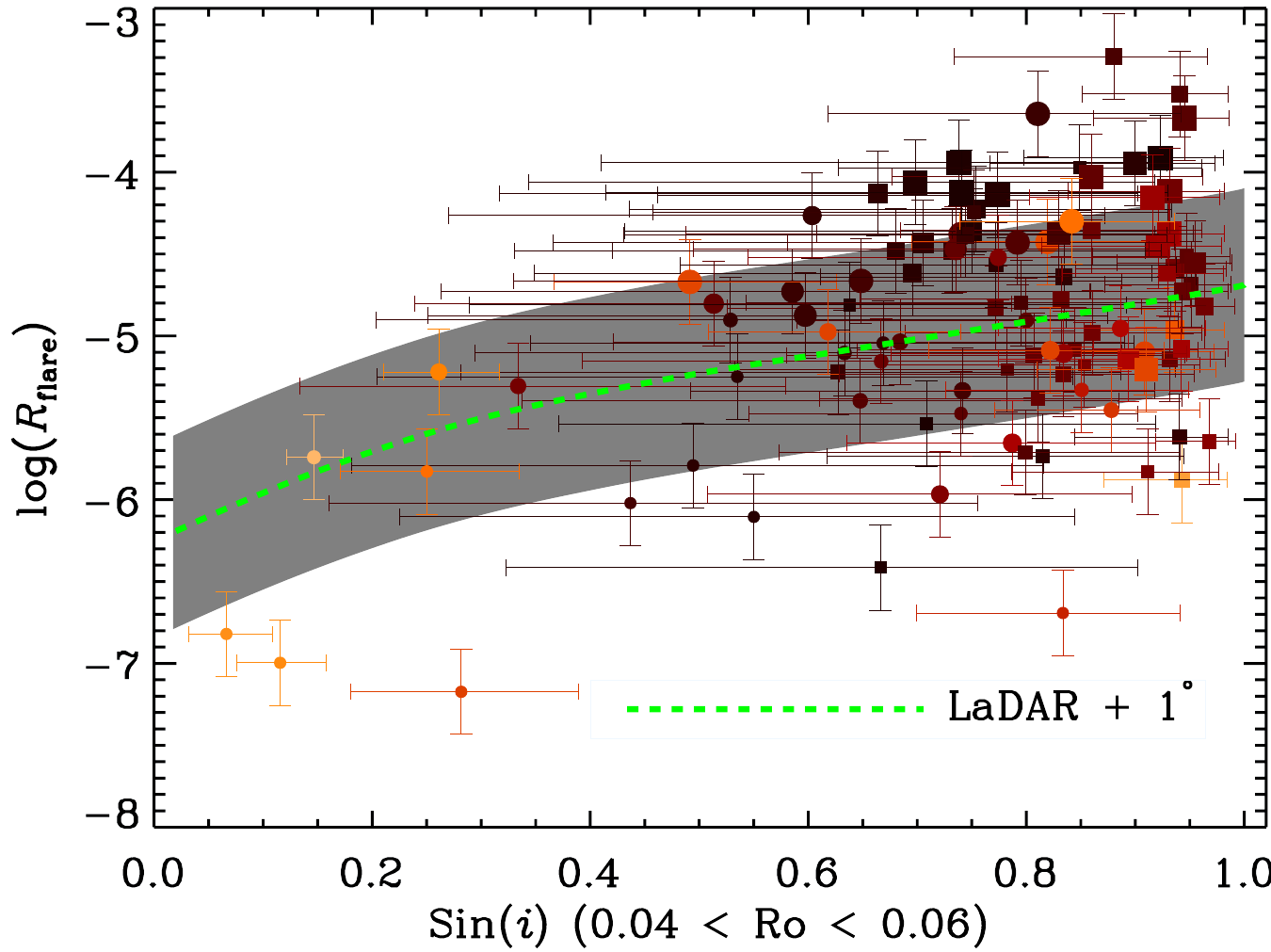}
\includegraphics[width=0.5\textwidth]{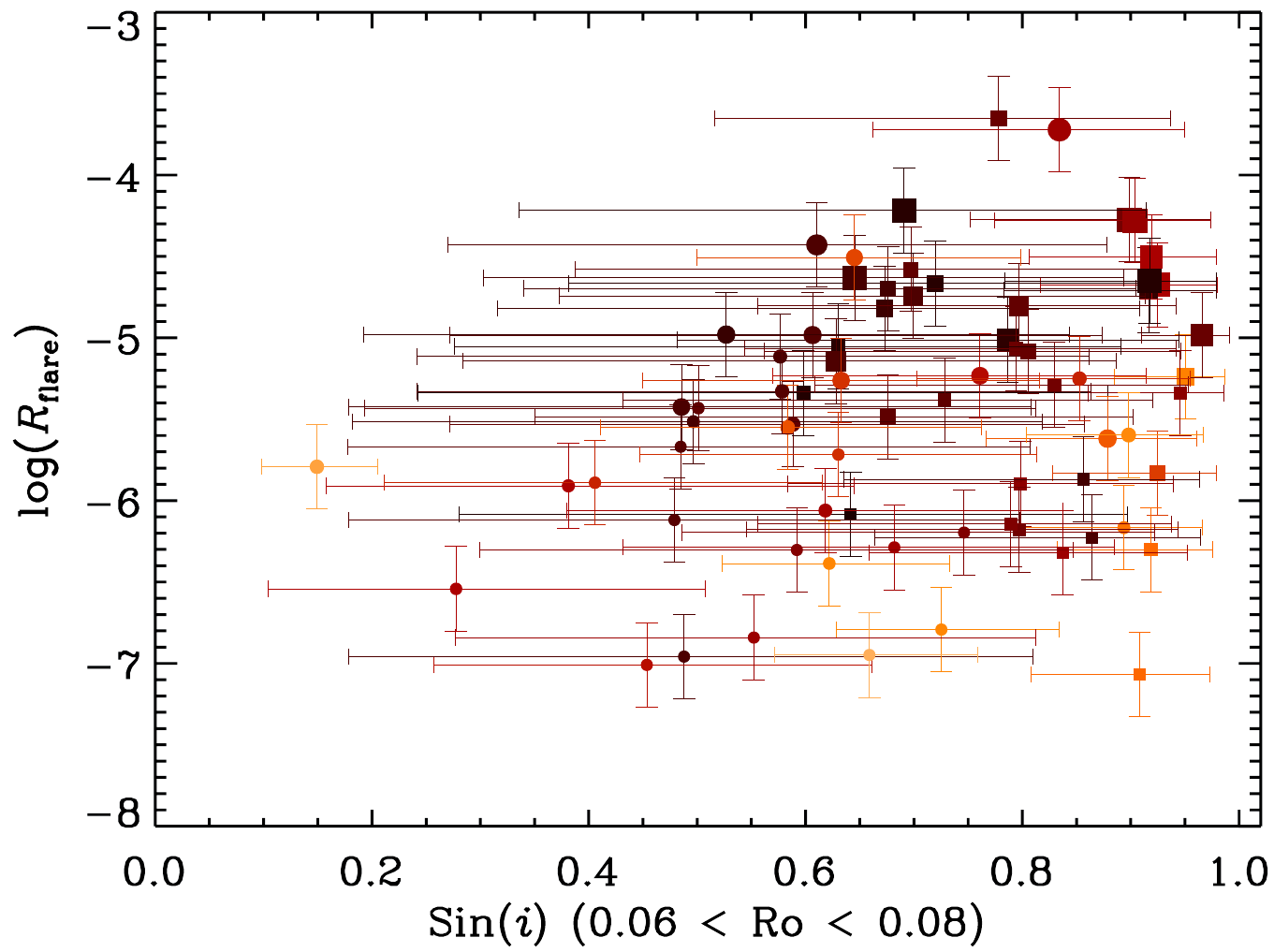}
\includegraphics[width=0.5\textwidth]{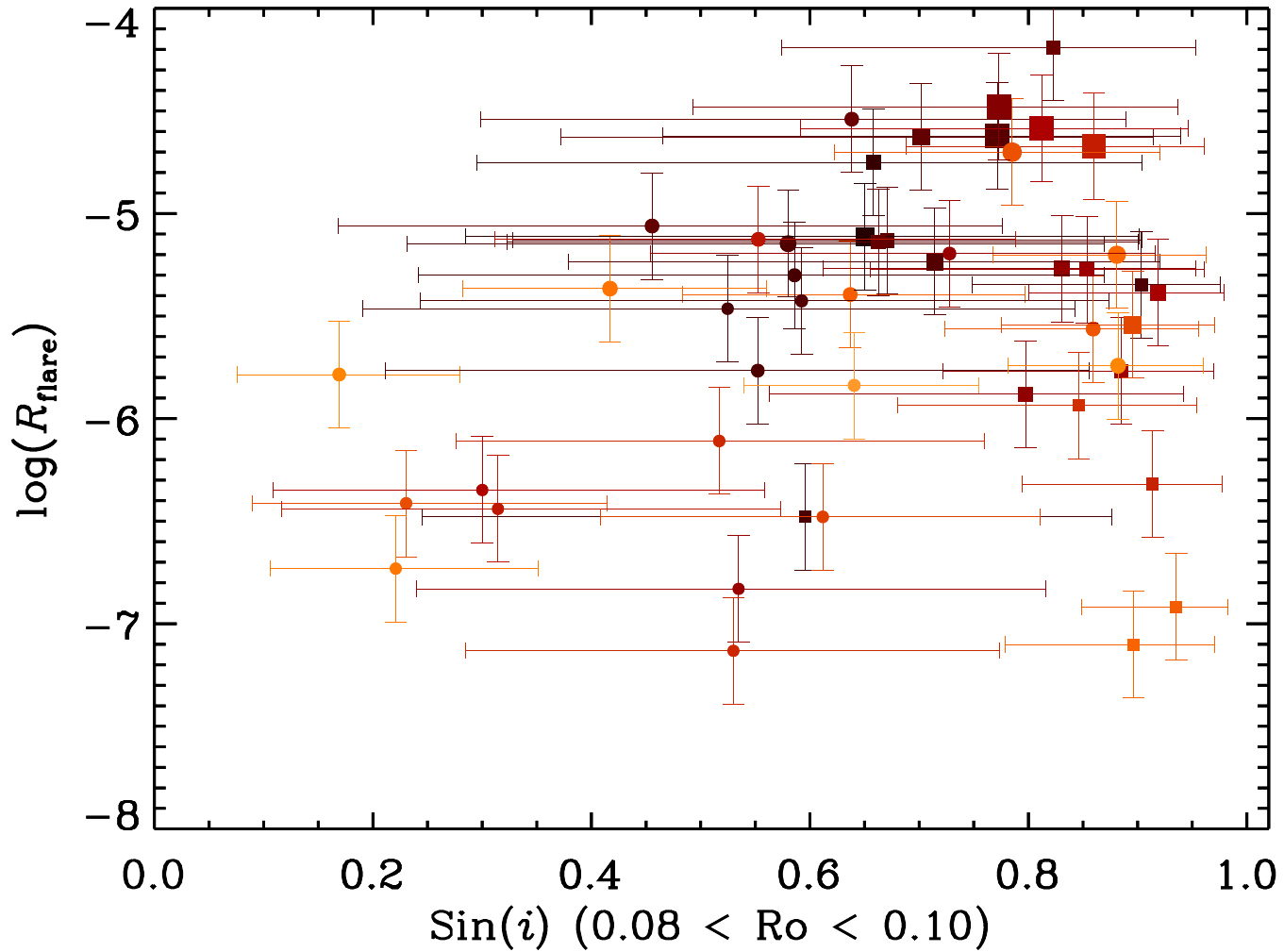}
\includegraphics[width=0.5\textwidth]{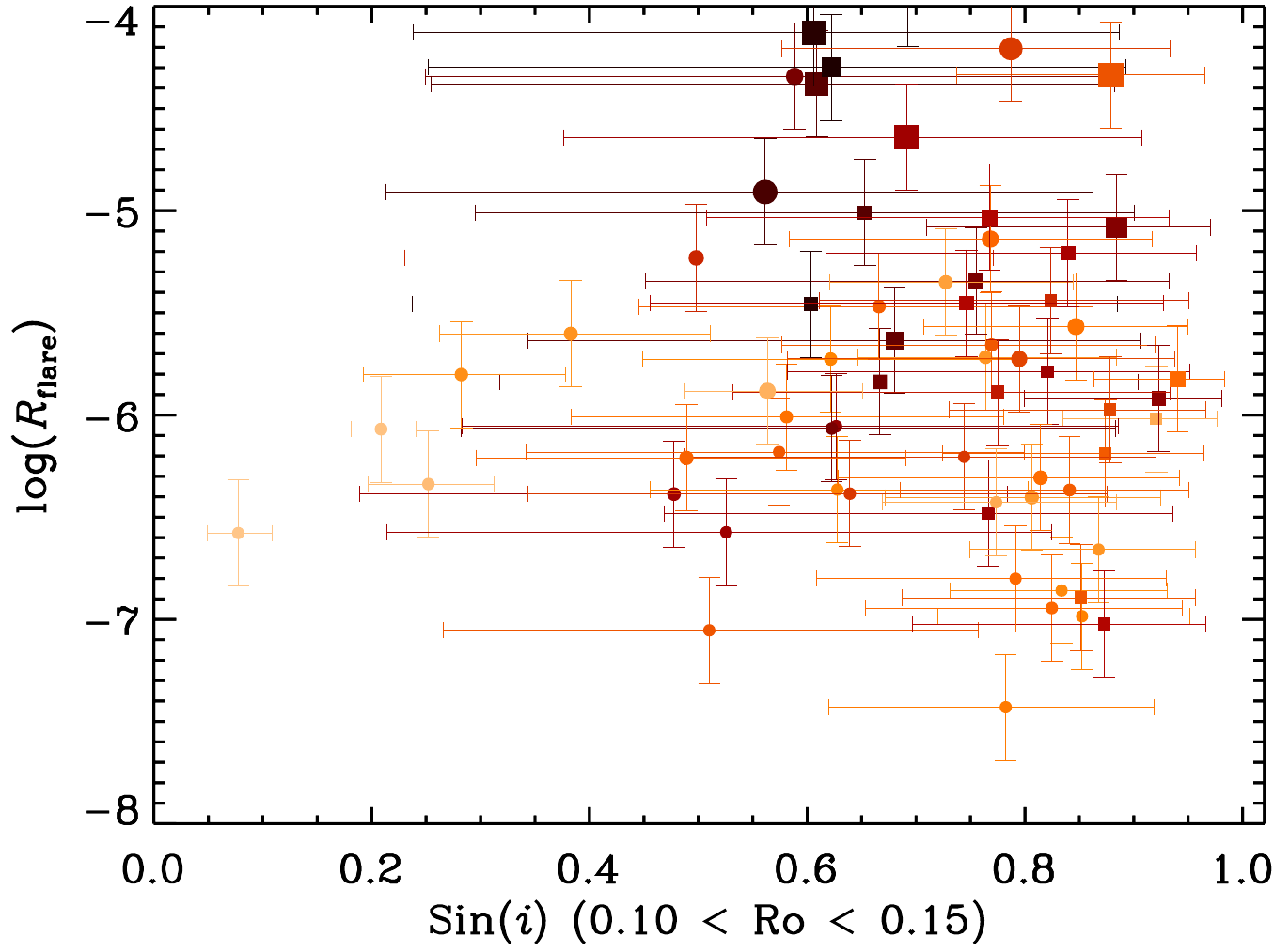}
\includegraphics[width=0.5\textwidth]{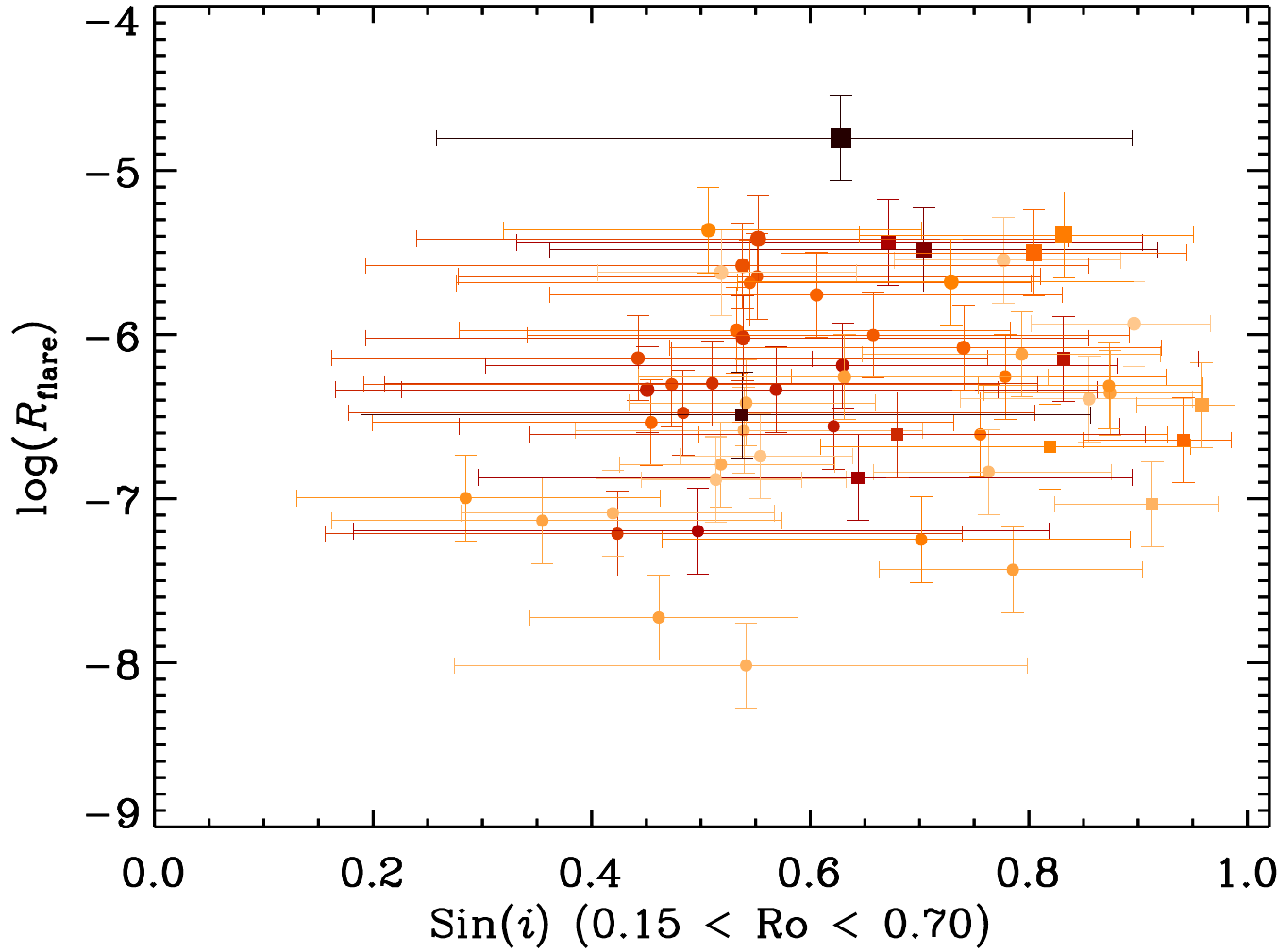}
\includegraphics[width=0.5\textwidth]{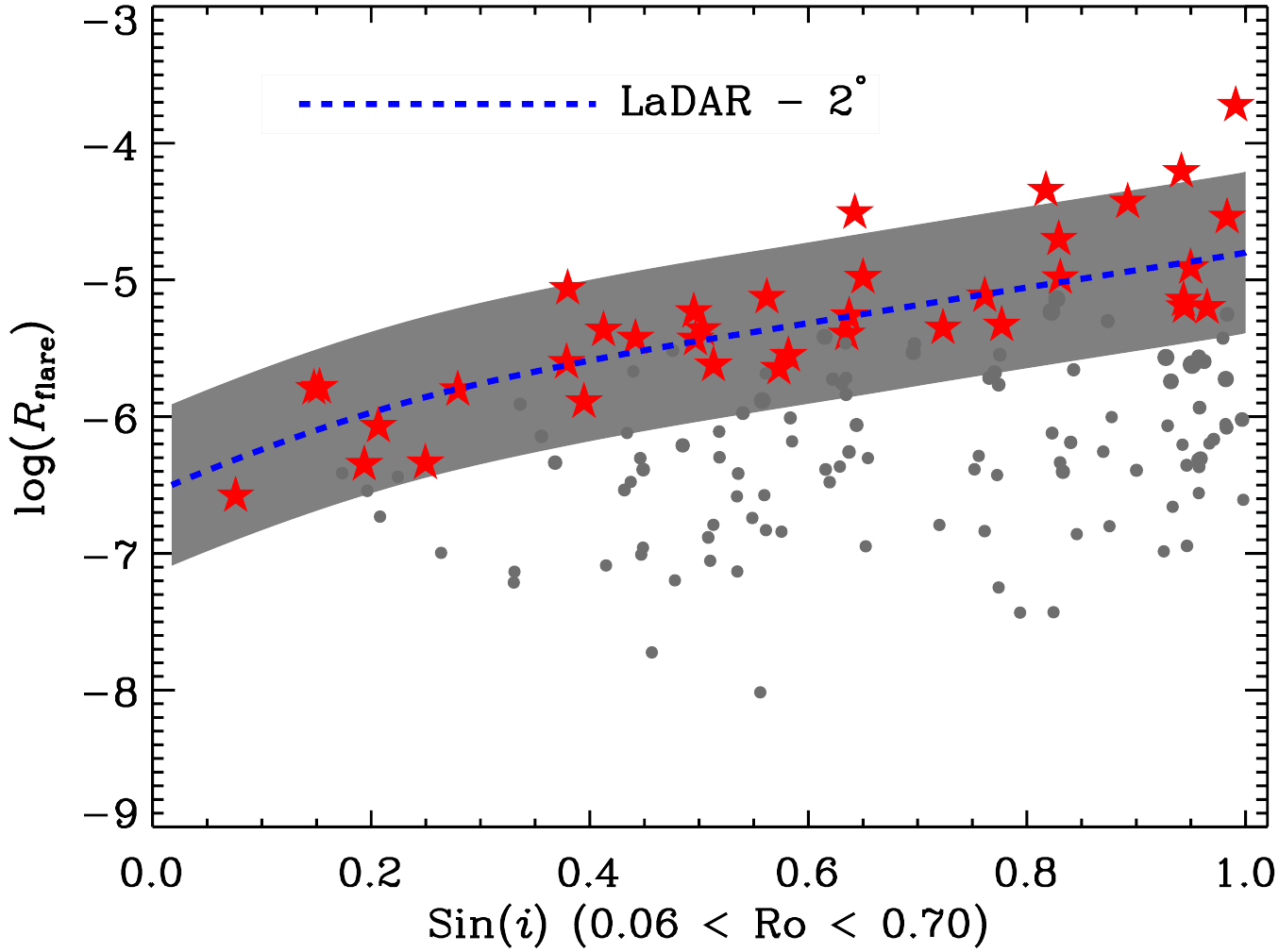}
\caption{Same as Figure~\ref{fig_sini_fa}, but for several larger Ro bins. The bottom right panel shows the relation in the range of $0.06 <$ Ro $< 0.7$. We take stars larger than the 84th percentile (or at least 3 stars) in each sin$i$ bin as the upper envelope of the relation. Stars of the upper envelope are denoted by the symbol of the five-pointed star.}
\label{fig_sini_fa_ext}
\end{figure*}

\begin{figure*}
\includegraphics[width=0.5\textwidth]{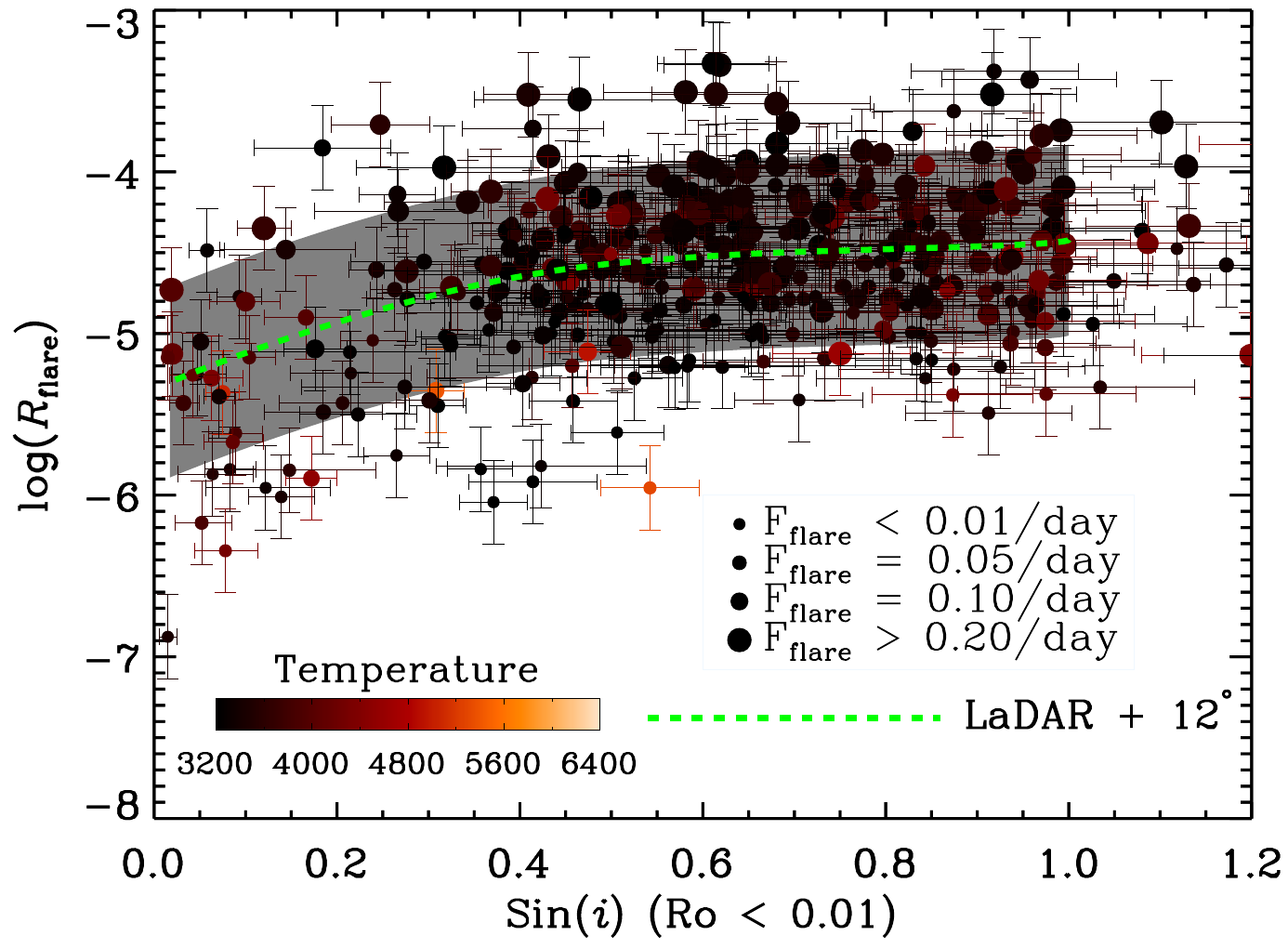}
\includegraphics[width=0.5\textwidth]{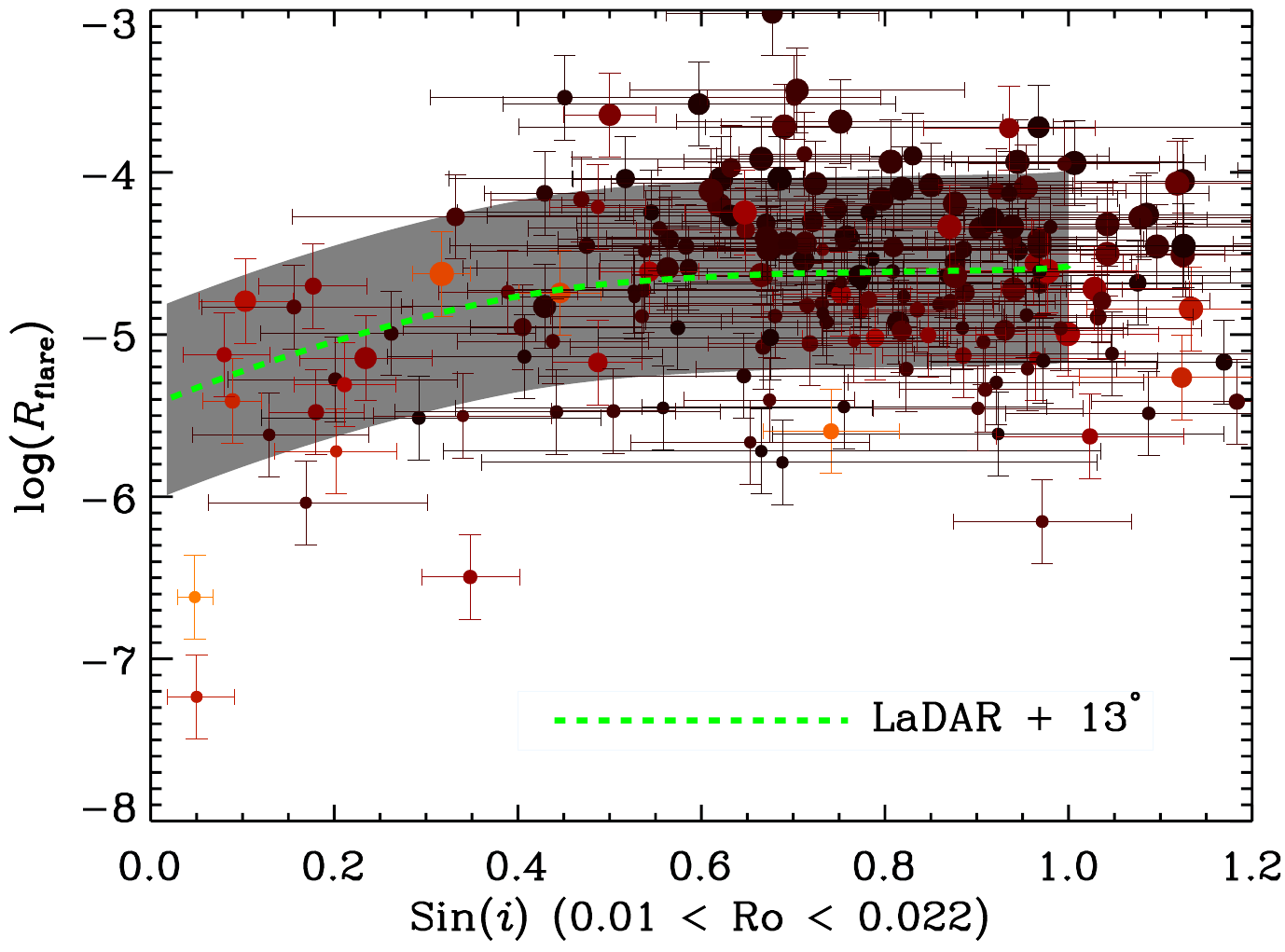}
\includegraphics[width=0.5\textwidth]{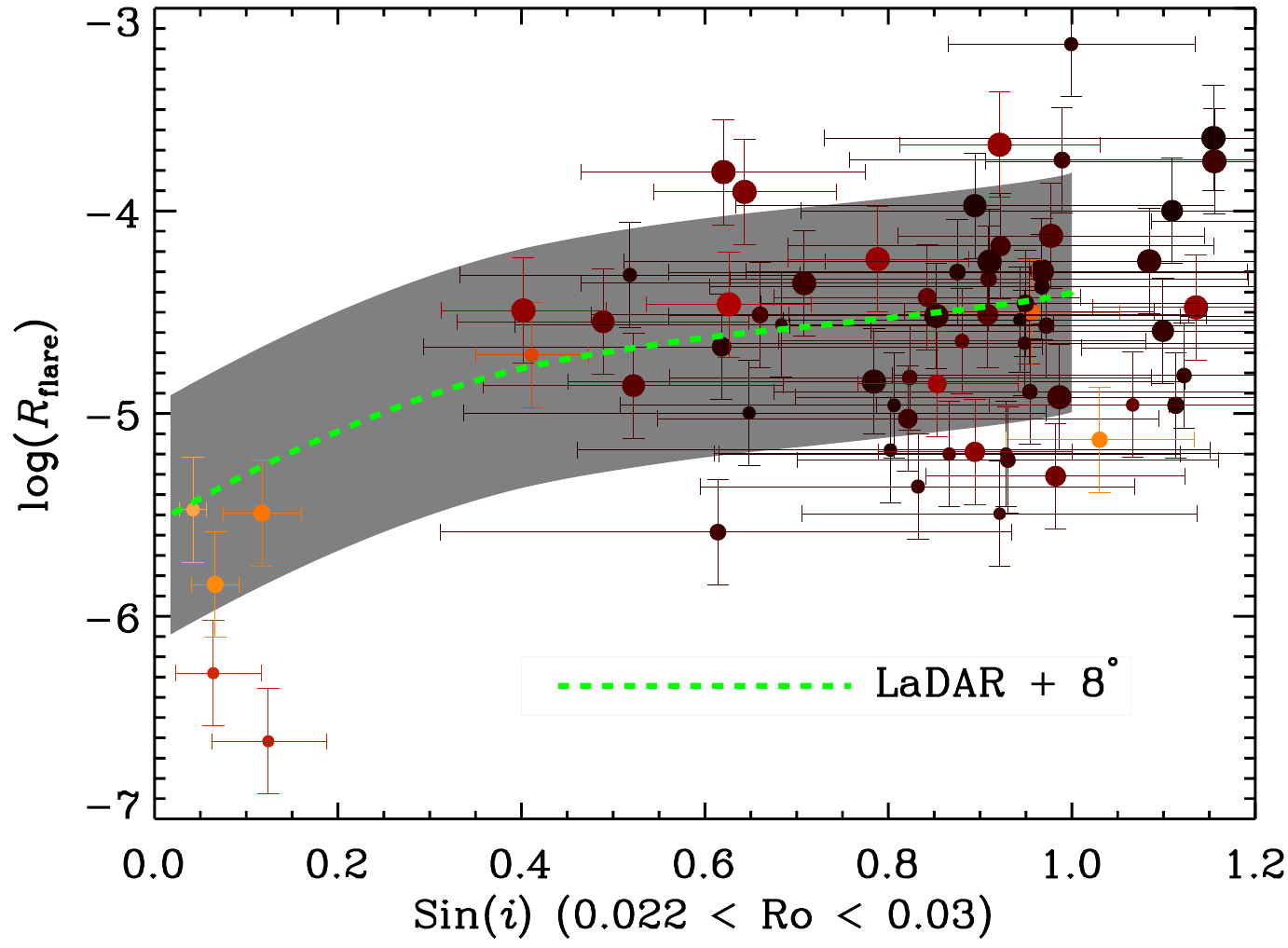}
\includegraphics[width=0.5\textwidth]{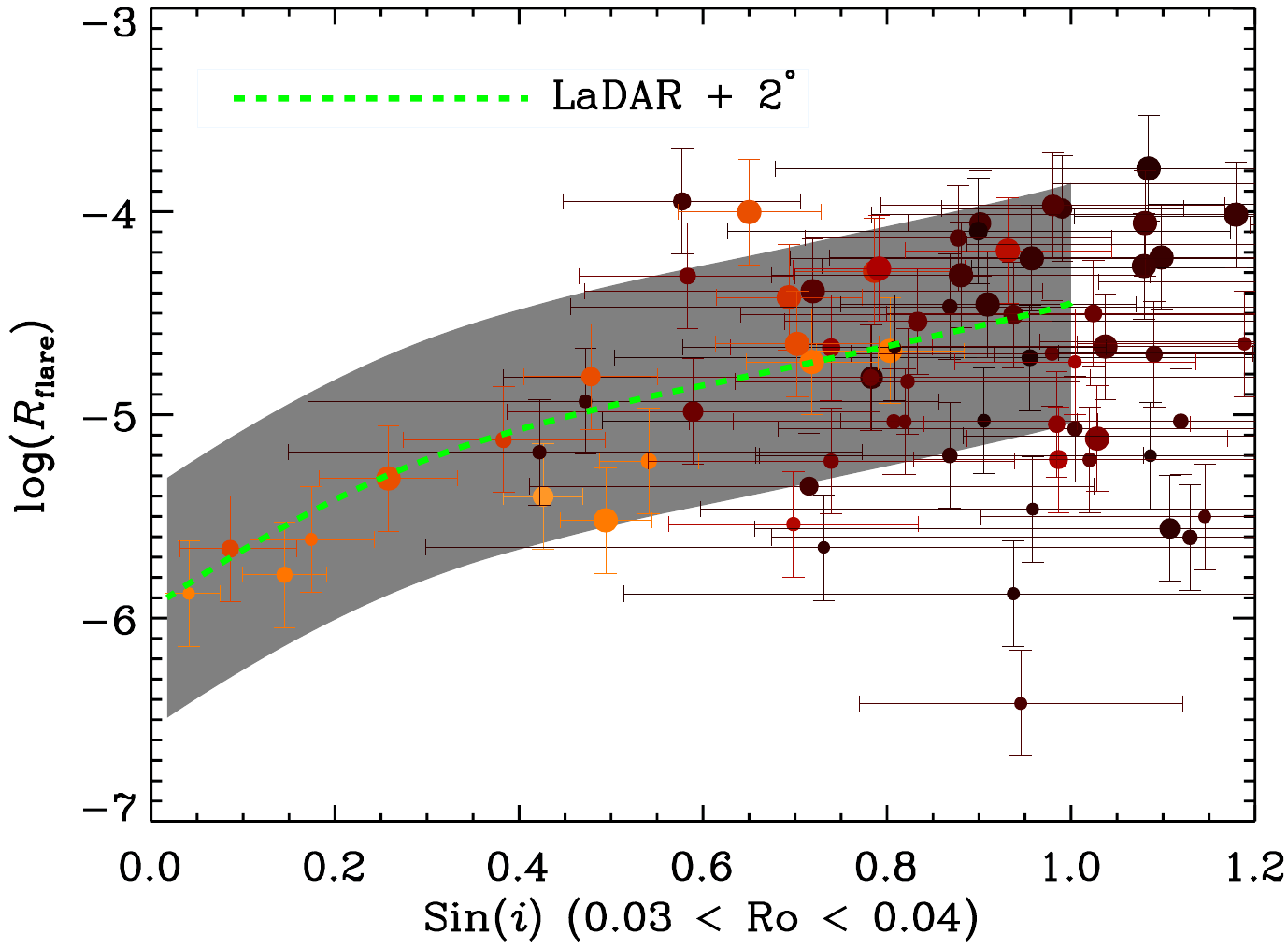}
\includegraphics[width=0.5\textwidth]{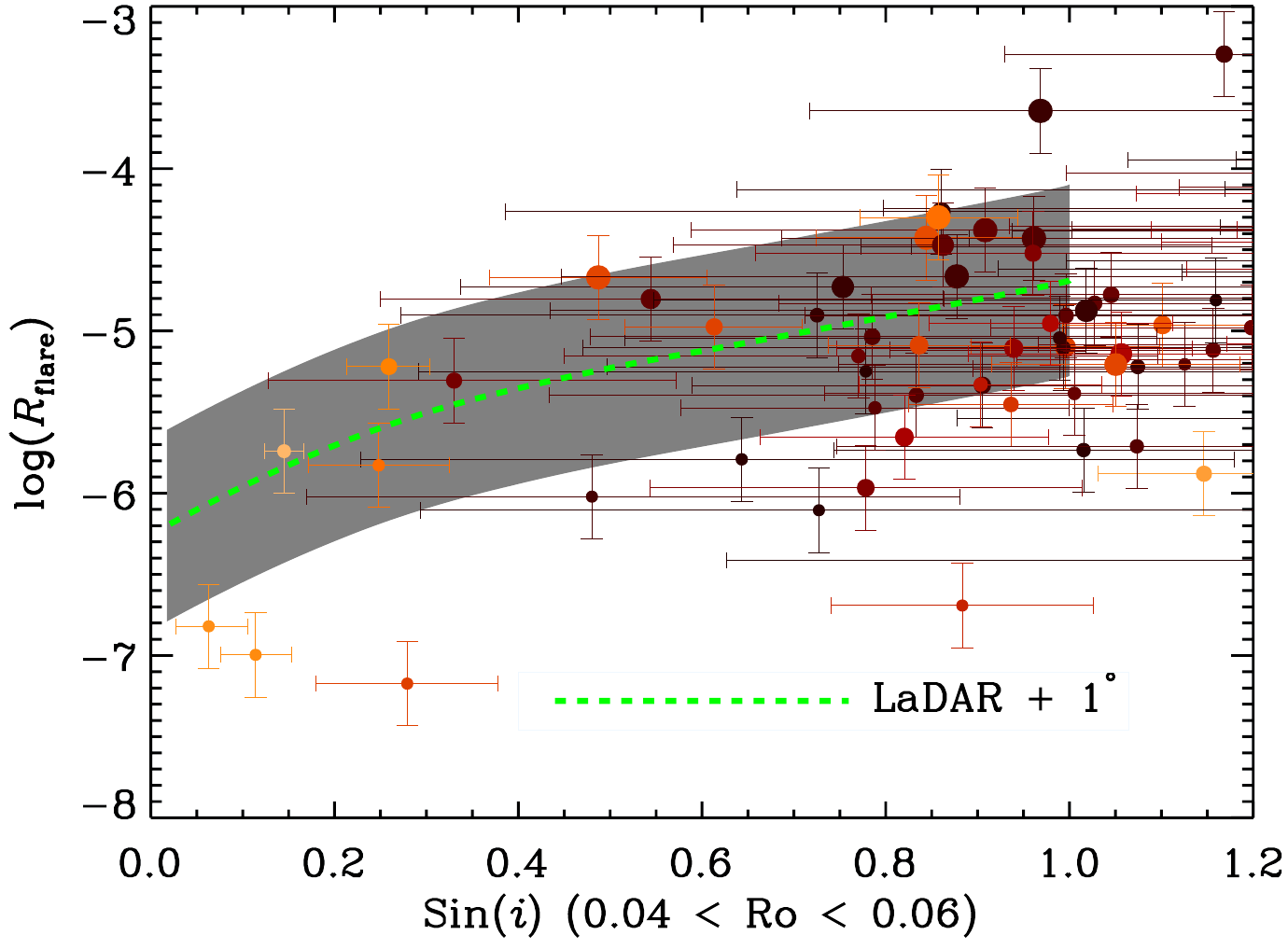}
\includegraphics[width=0.5\textwidth]{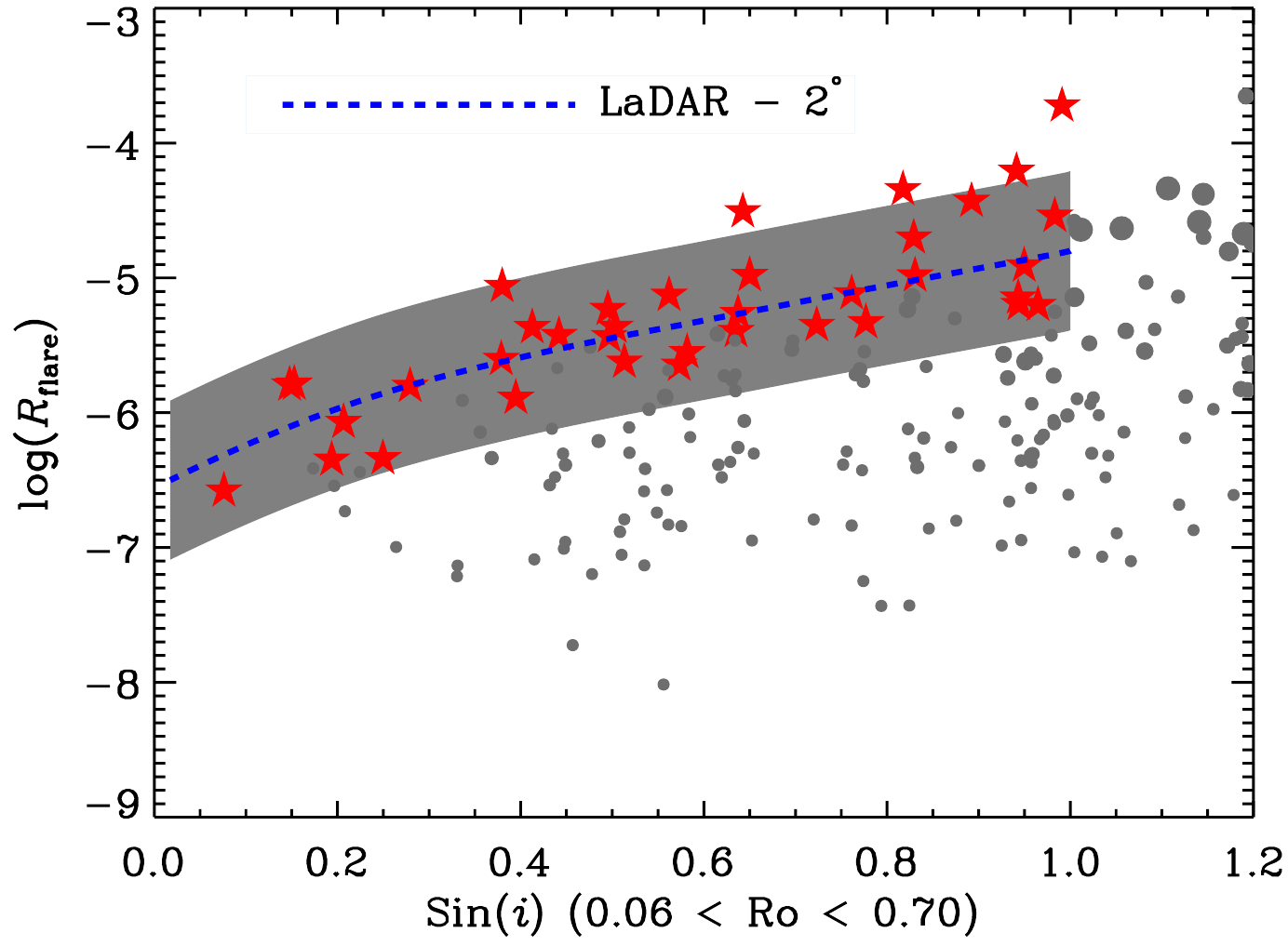}
\caption{Same as Fig.~\ref{fig_sini_fa}, but we allow sin$i$ to exceed 1 by directly integrating $p$(sin$i|D$). This does not change the best match with the simulation data in each Ro bin.}
\label{fig_sini_fa_ext_exceed1}
\end{figure*}
\begin{figure*}
\includegraphics[width=0.5\textwidth]{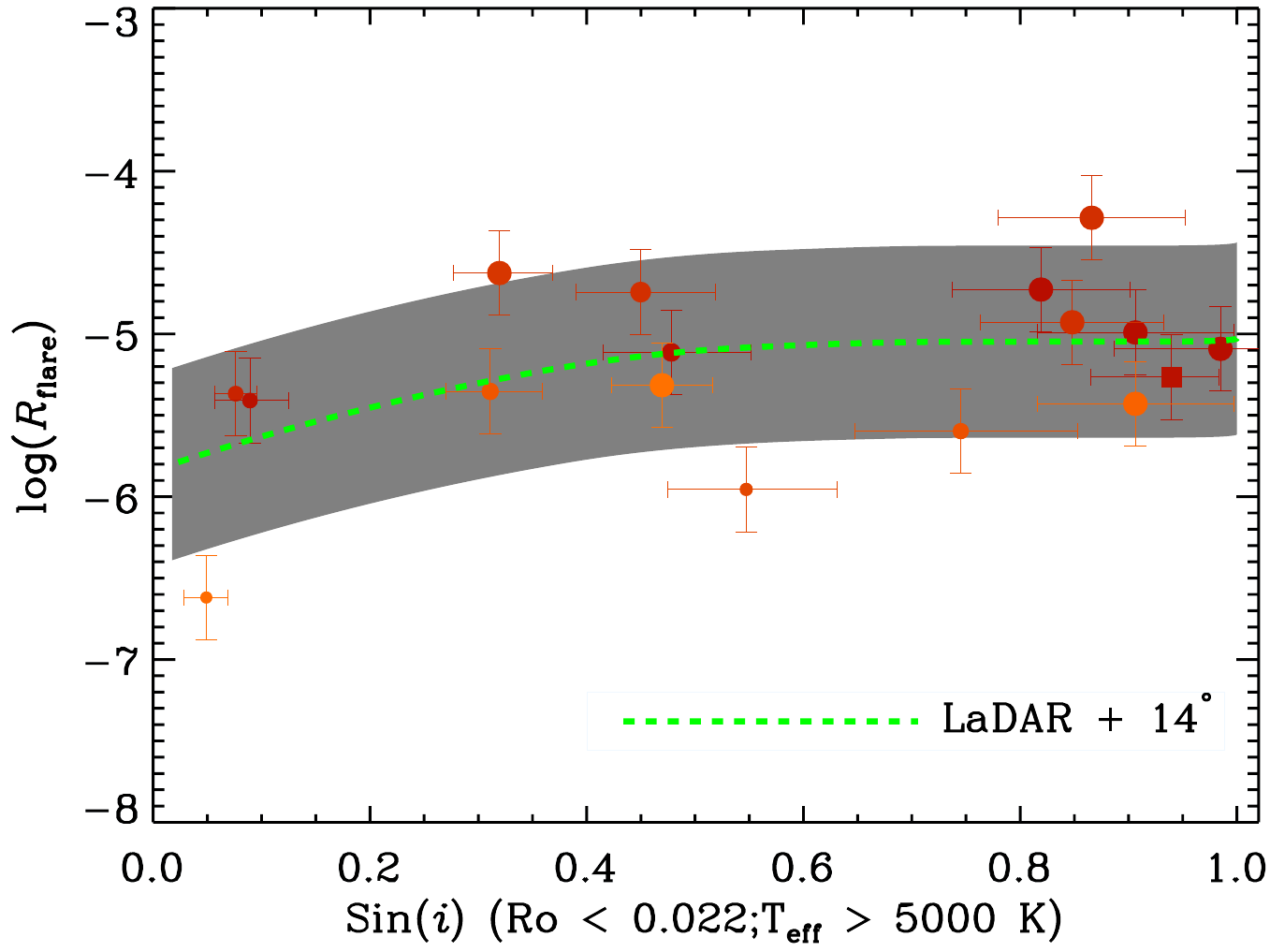}
\includegraphics[width=0.5\textwidth]{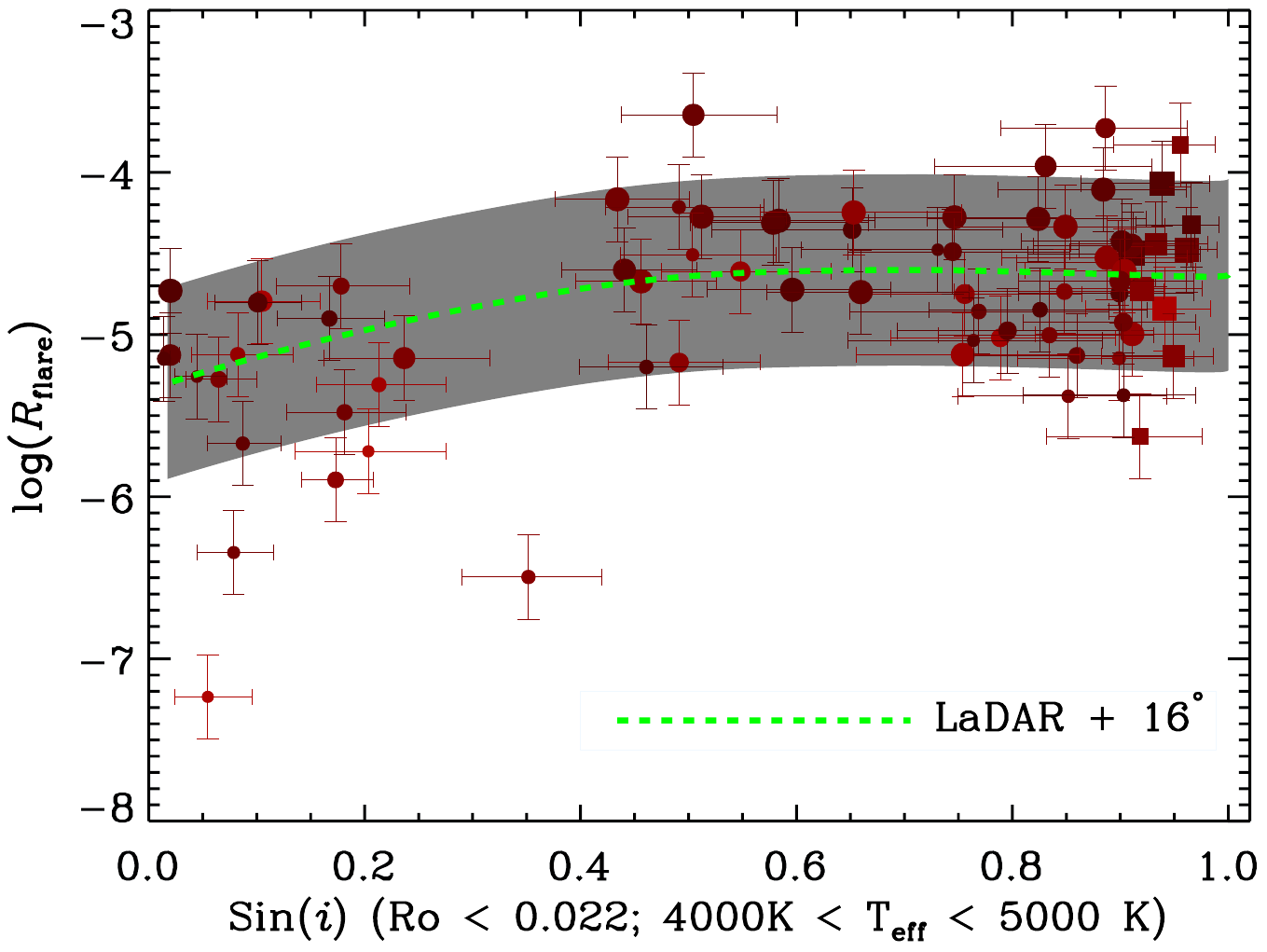}
\includegraphics[width=0.5\textwidth]{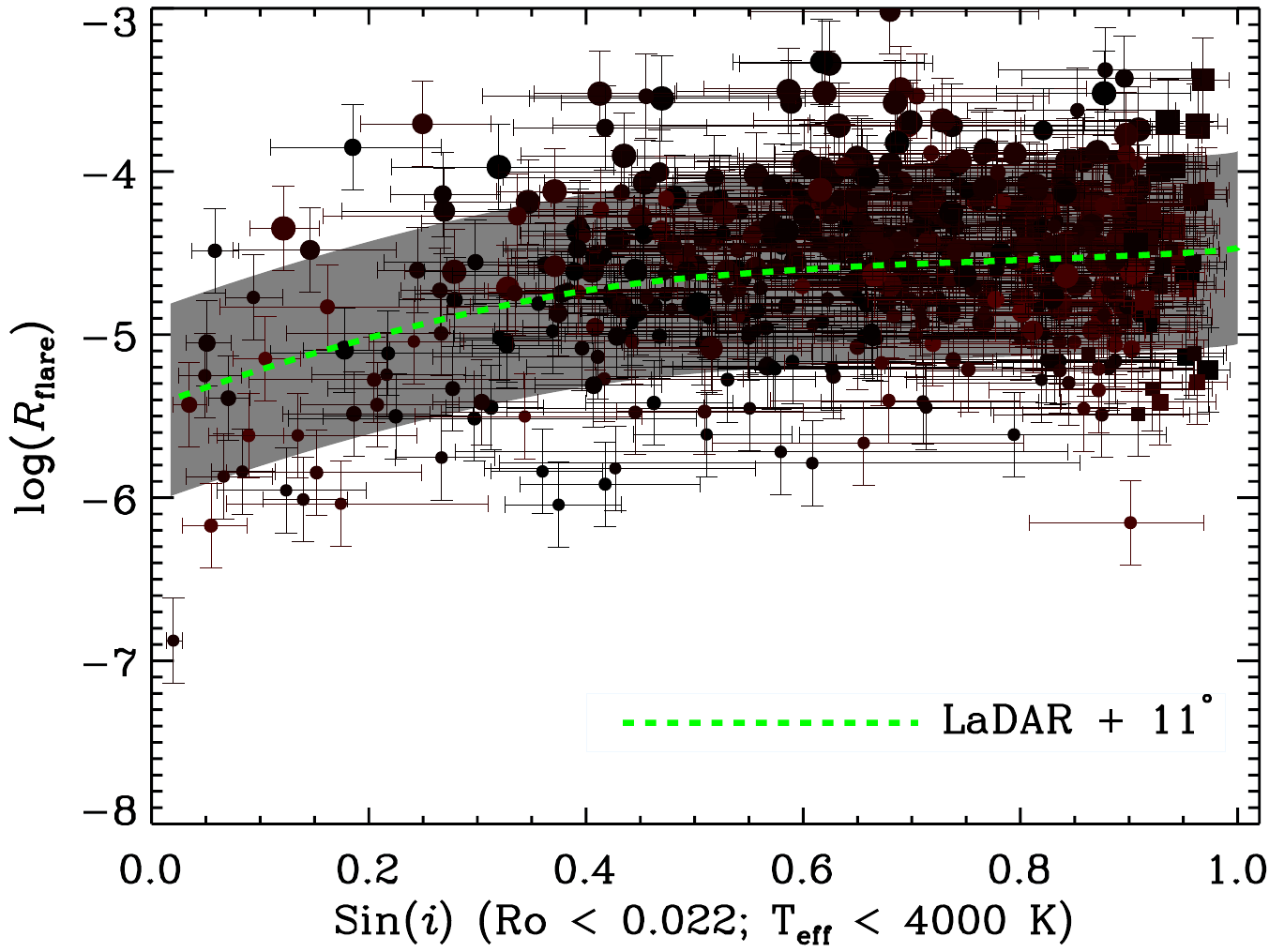}
\caption{Same as Fig.~\ref{fig_sini_fa}, but for three temperature bins of fast rotators. The sample includes polar-spot stars whose inclinations are from \citet{Strass2009} (Fig.~\ref{fig_polarcap}). }
\label{fig_sini_fa_teffbin}
\end{figure*}

\begin{figure}
\includegraphics[width=0.4\textwidth]{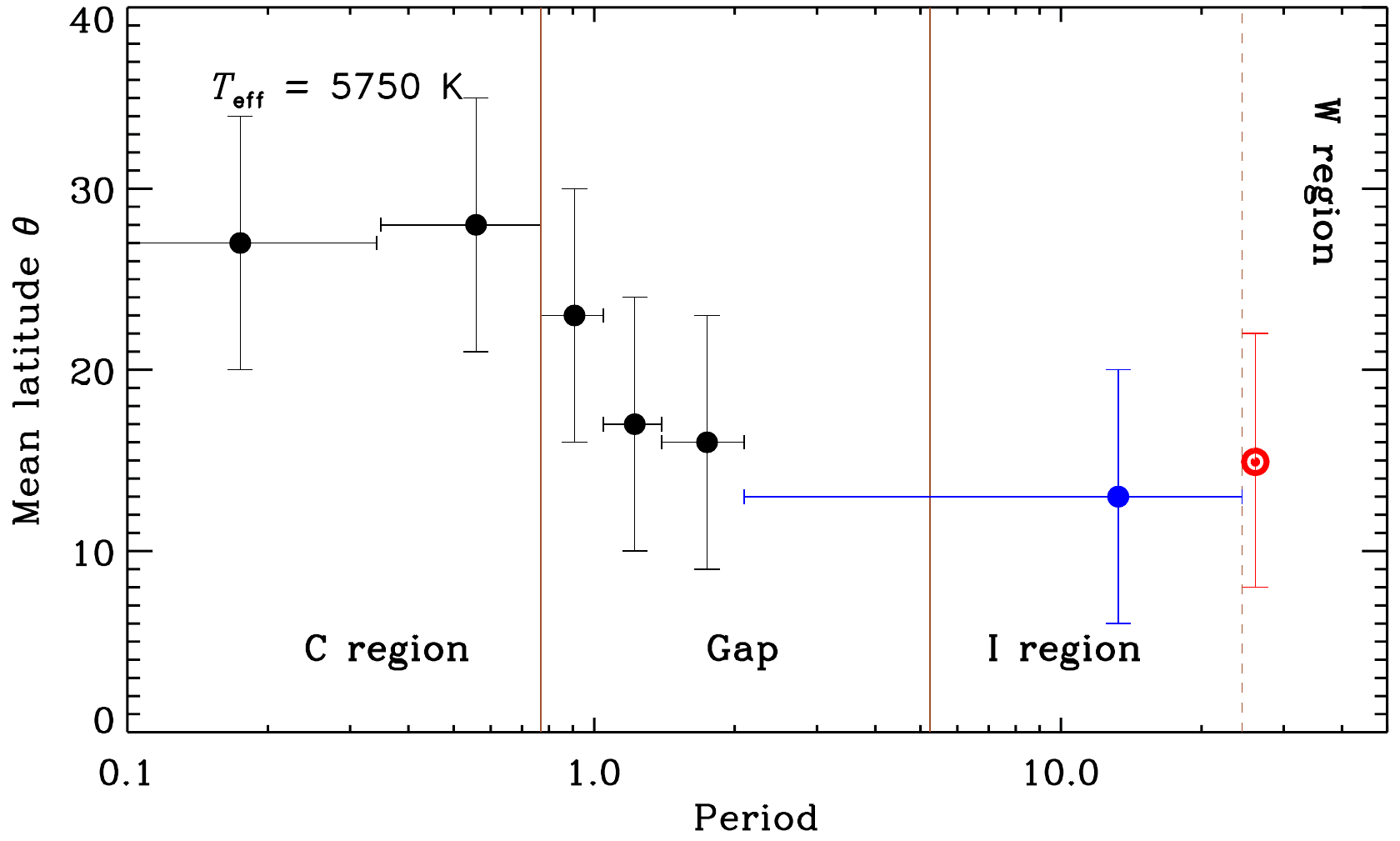}
\includegraphics[width=0.4\textwidth]{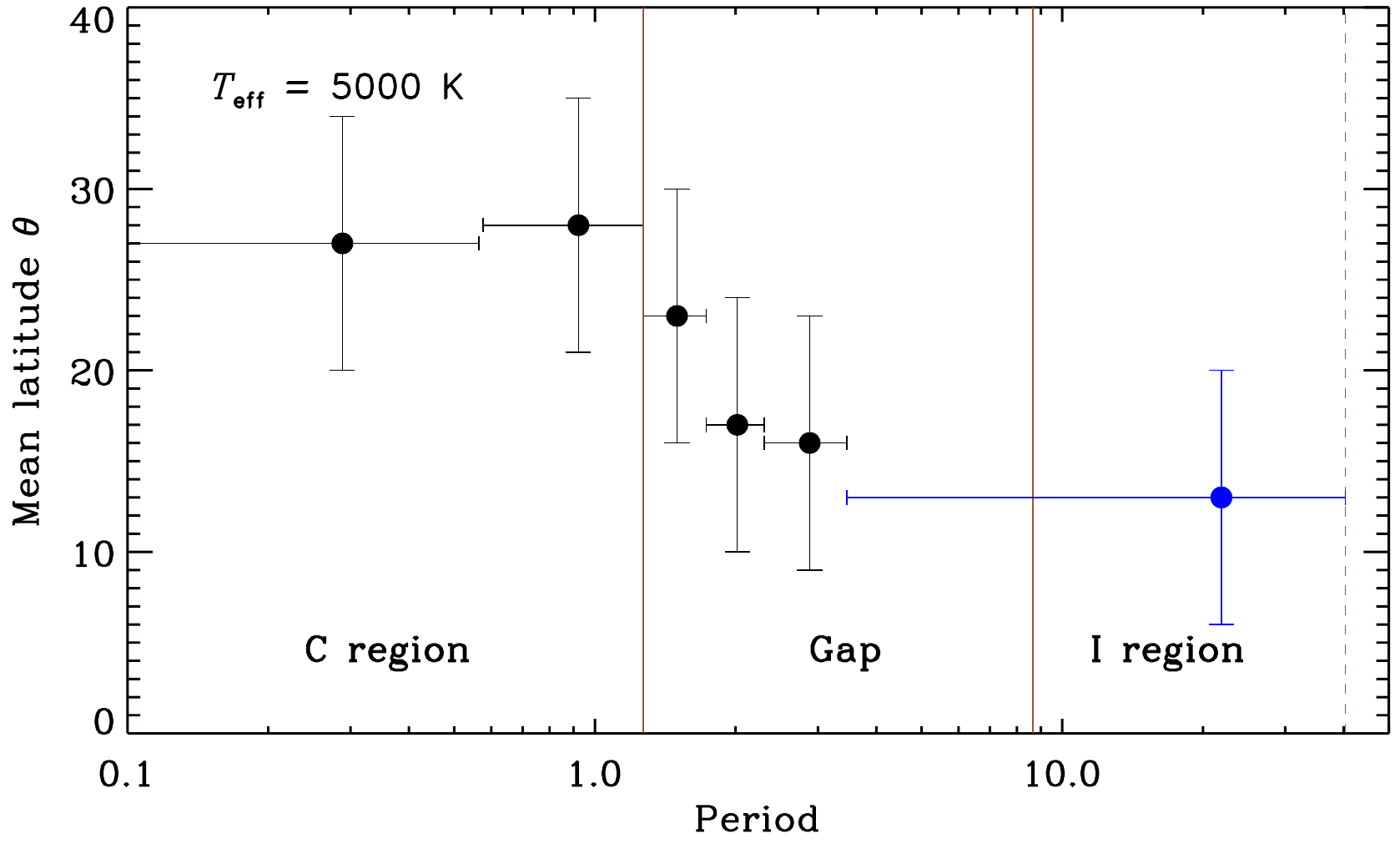}
\includegraphics[width=0.4\textwidth]{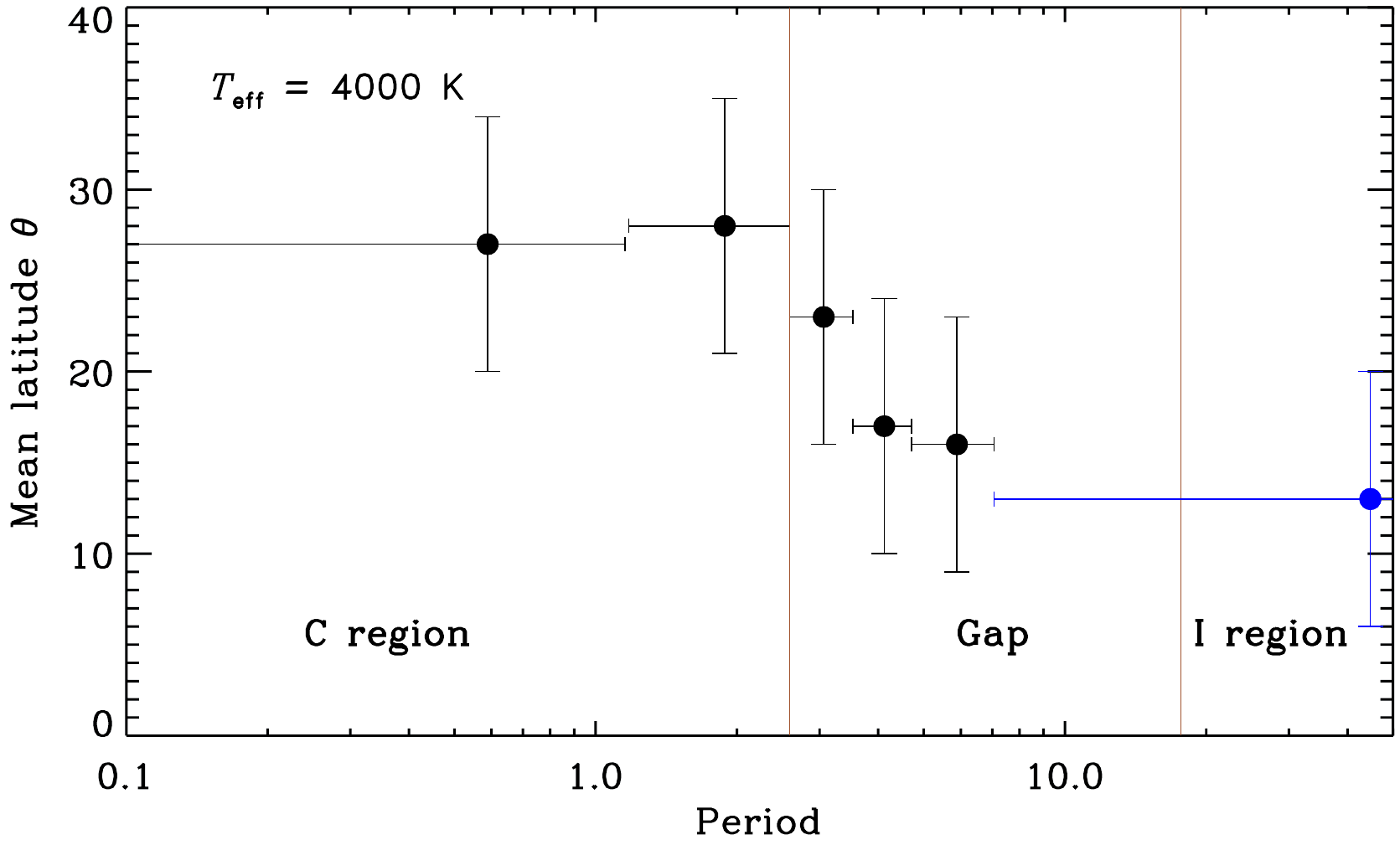}
\caption{Same as Figure~\ref{fig_ro_ladar}, but for rotation period. We convert Ro to rotation period by using the tabulated data of \citet{Barnes2010} for three effective temperatures.}
\label{fig_period_ladar_ind}
\end{figure}

\begin{figure*}
\includegraphics[width=0.5\textwidth]{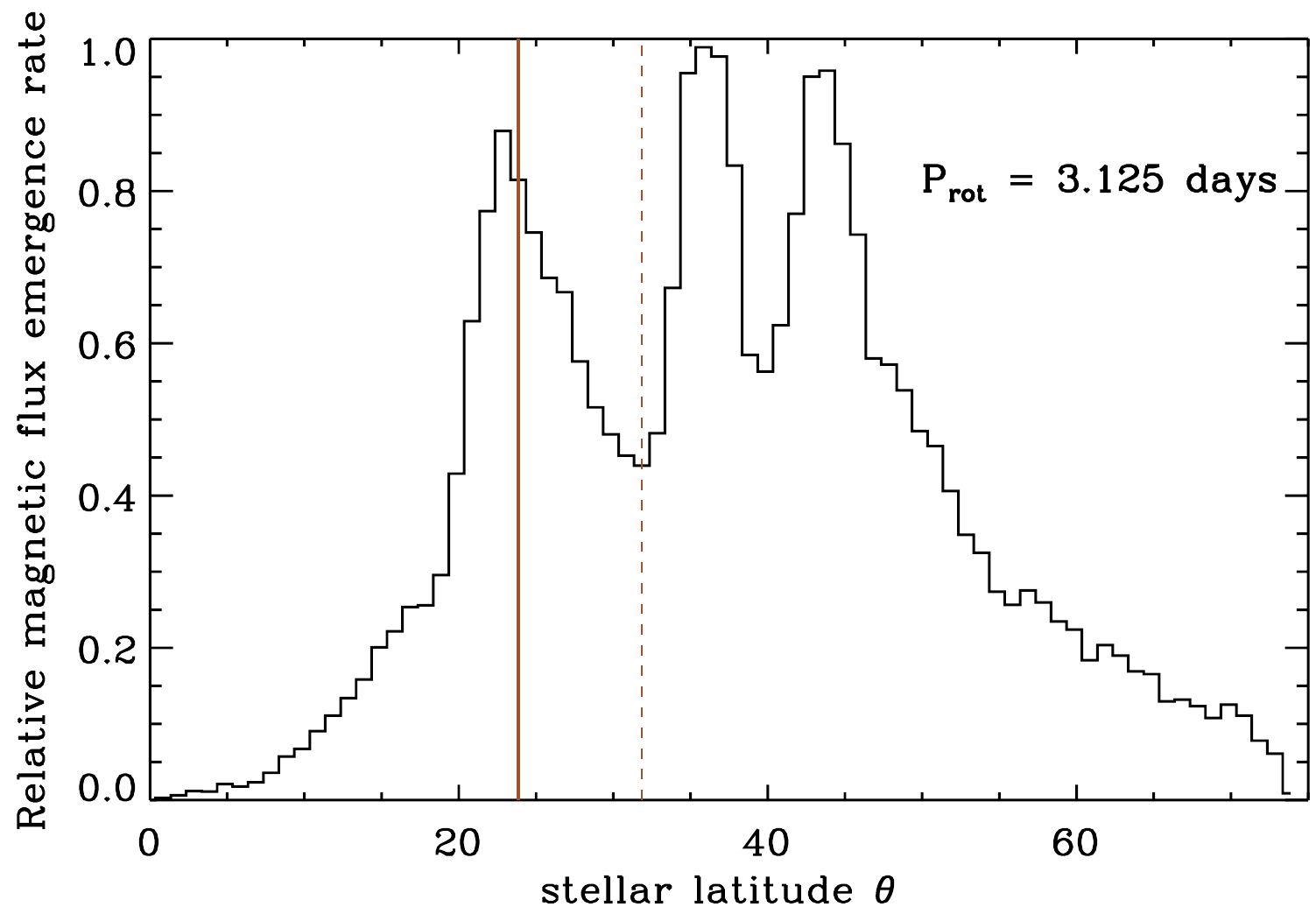}
\includegraphics[width=0.5\textwidth]{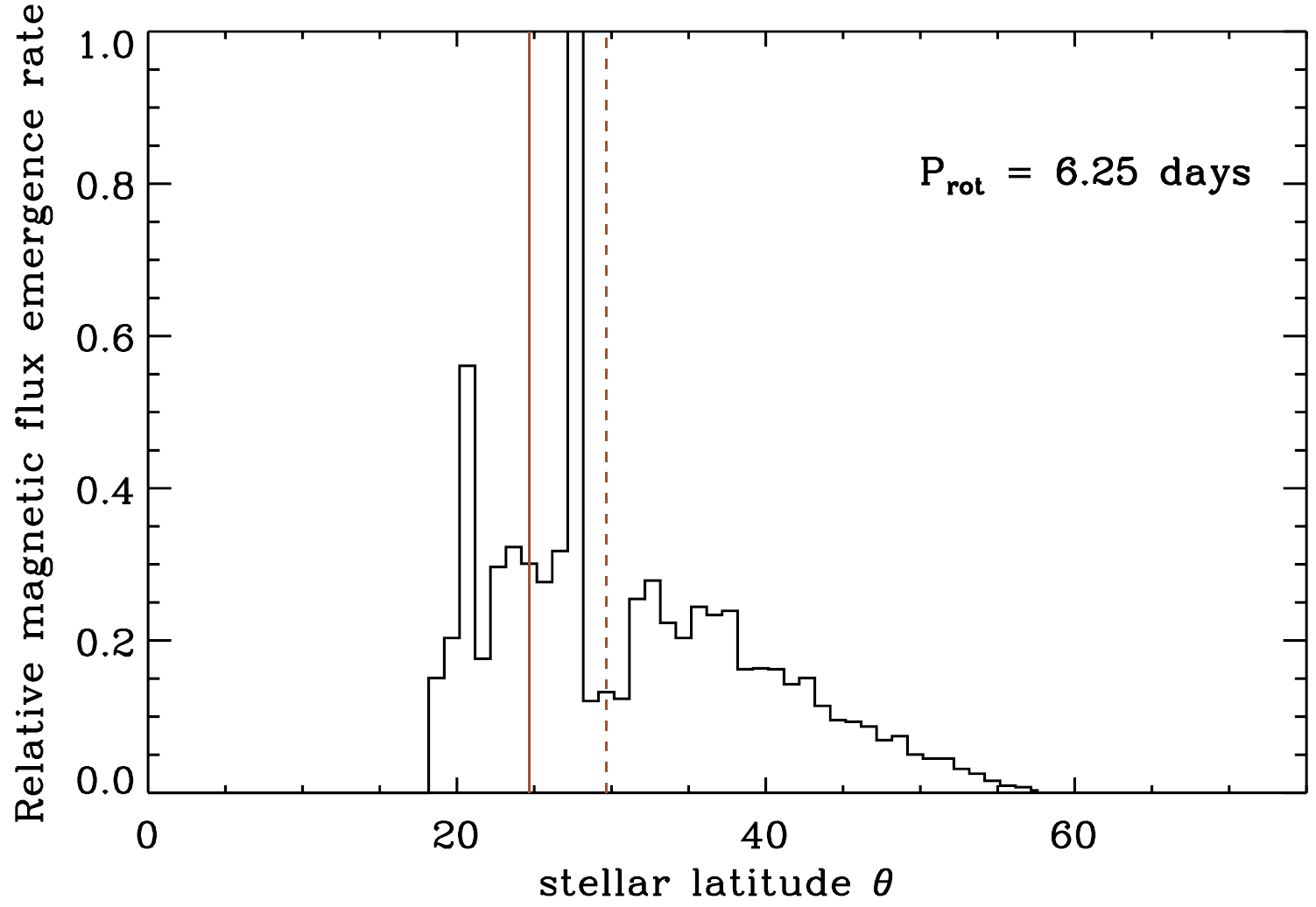}
\includegraphics[width=0.5\textwidth]{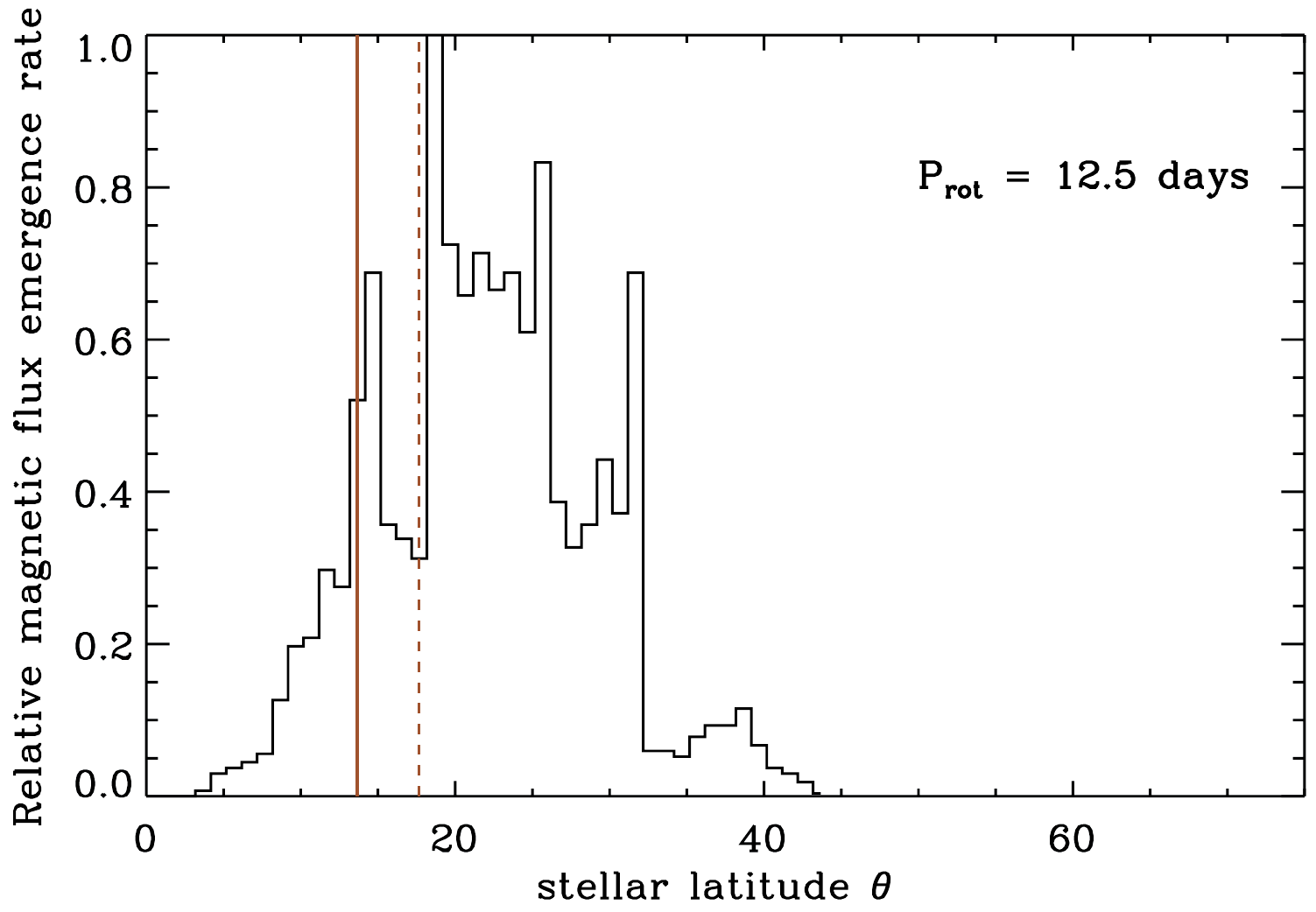}
\includegraphics[width=0.5\textwidth]{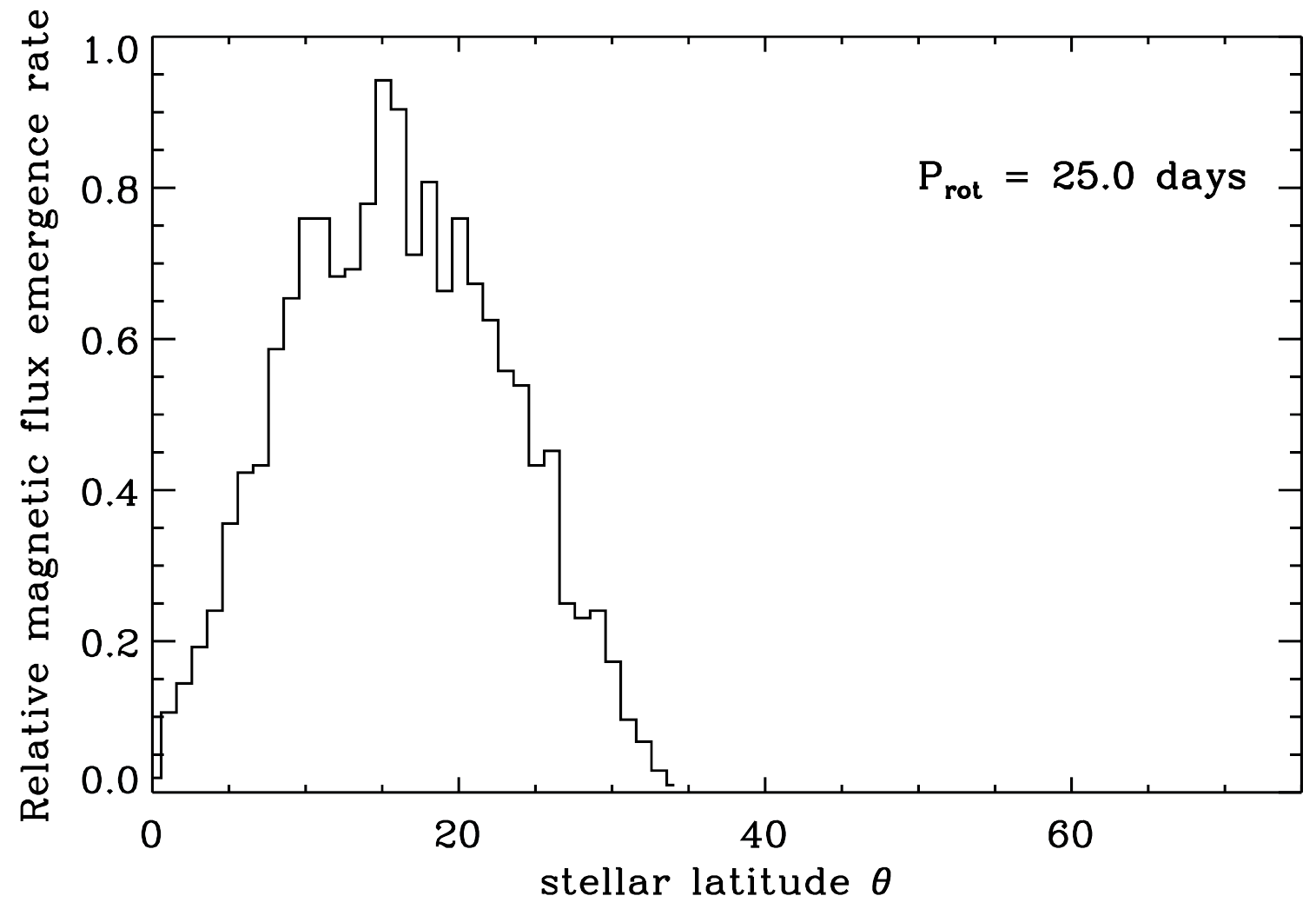}
\caption{A dynamo simulation on the latitudinal distribution of the magnetic flux emergence for different rotation periods. The data of the top left panel is from \citet{Isik2024}, and the others are from \citet{Isik2018}. The vertical dashed lines separate the lower component of the distributions that are associated with low-latitude spots and small-scale fields. The vertical solid lines denote the mean latitudes of the lower components.}
\label{fig_isik_hist}
\end{figure*}

\begin{figure*}
\includegraphics[width=0.5\textwidth]{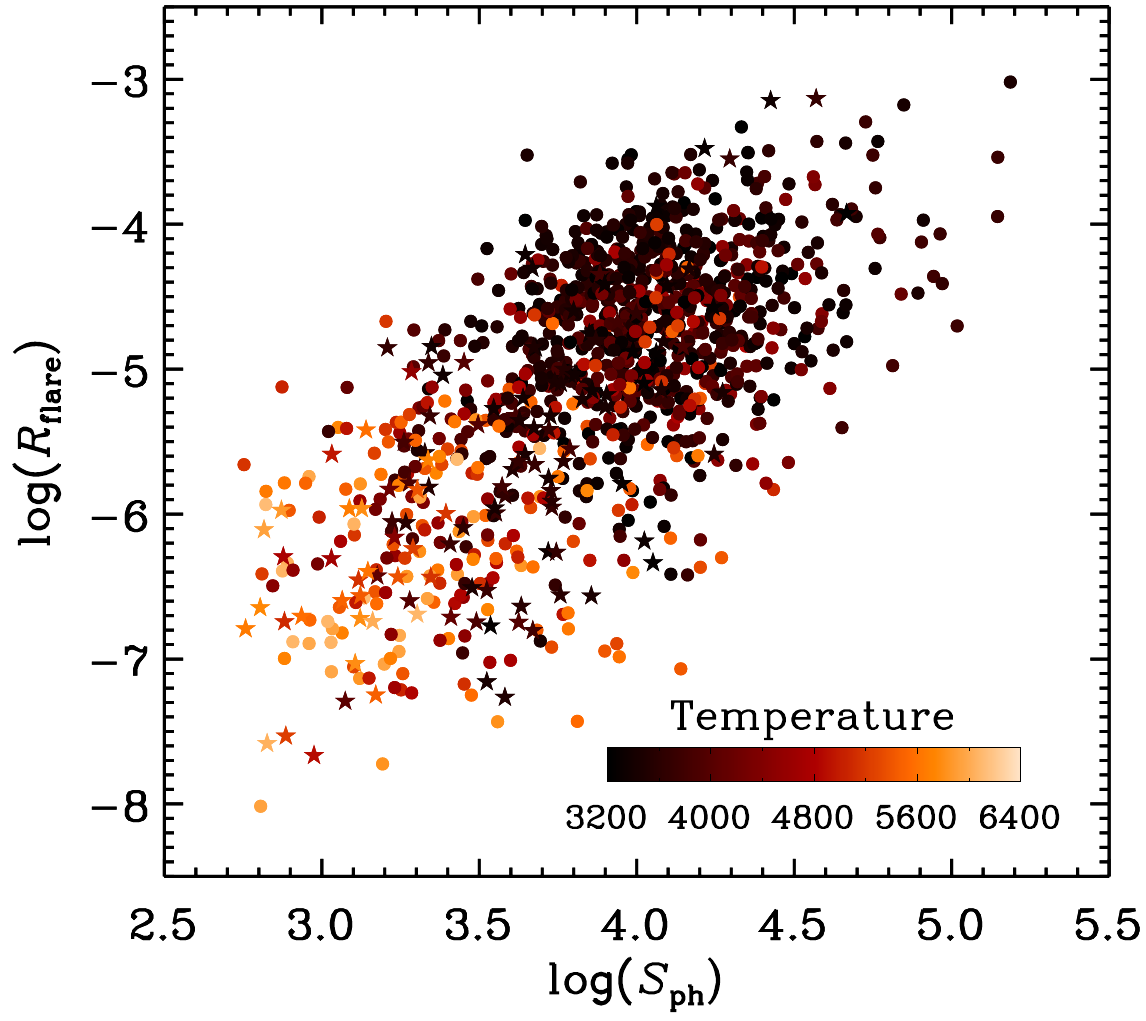}
\includegraphics[width=0.49\textwidth]{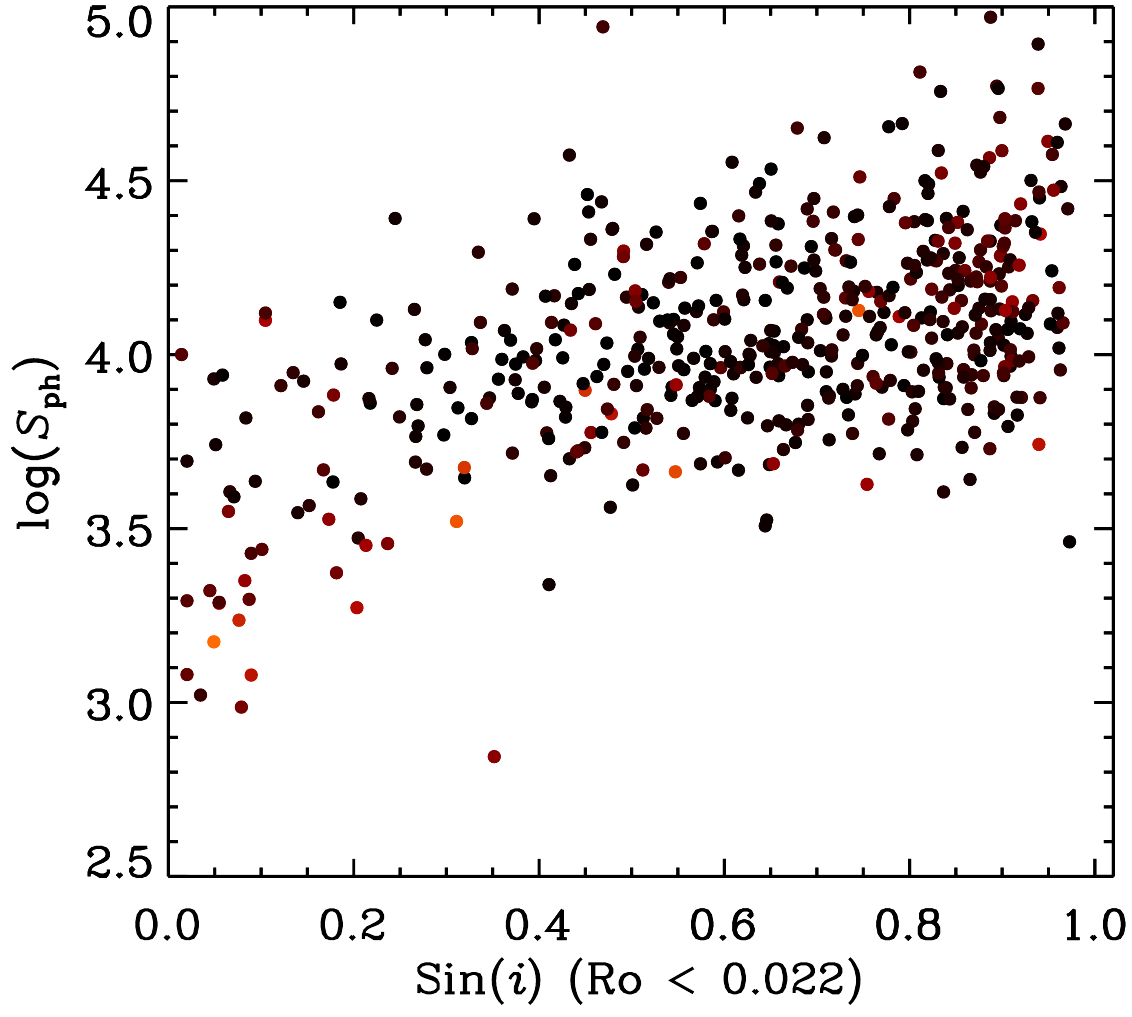}
\includegraphics[width=0.5\textwidth]{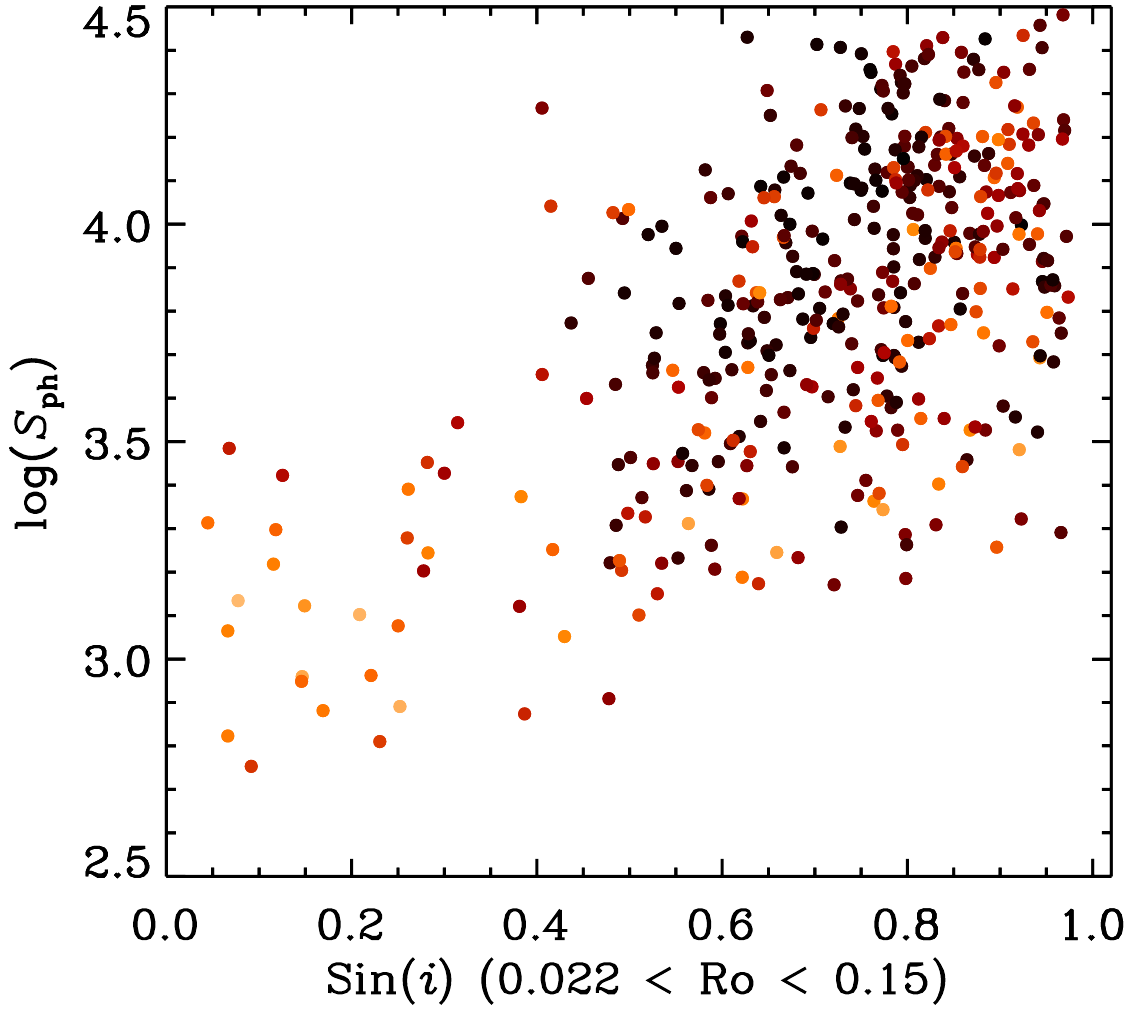}
\includegraphics[width=0.49\textwidth]{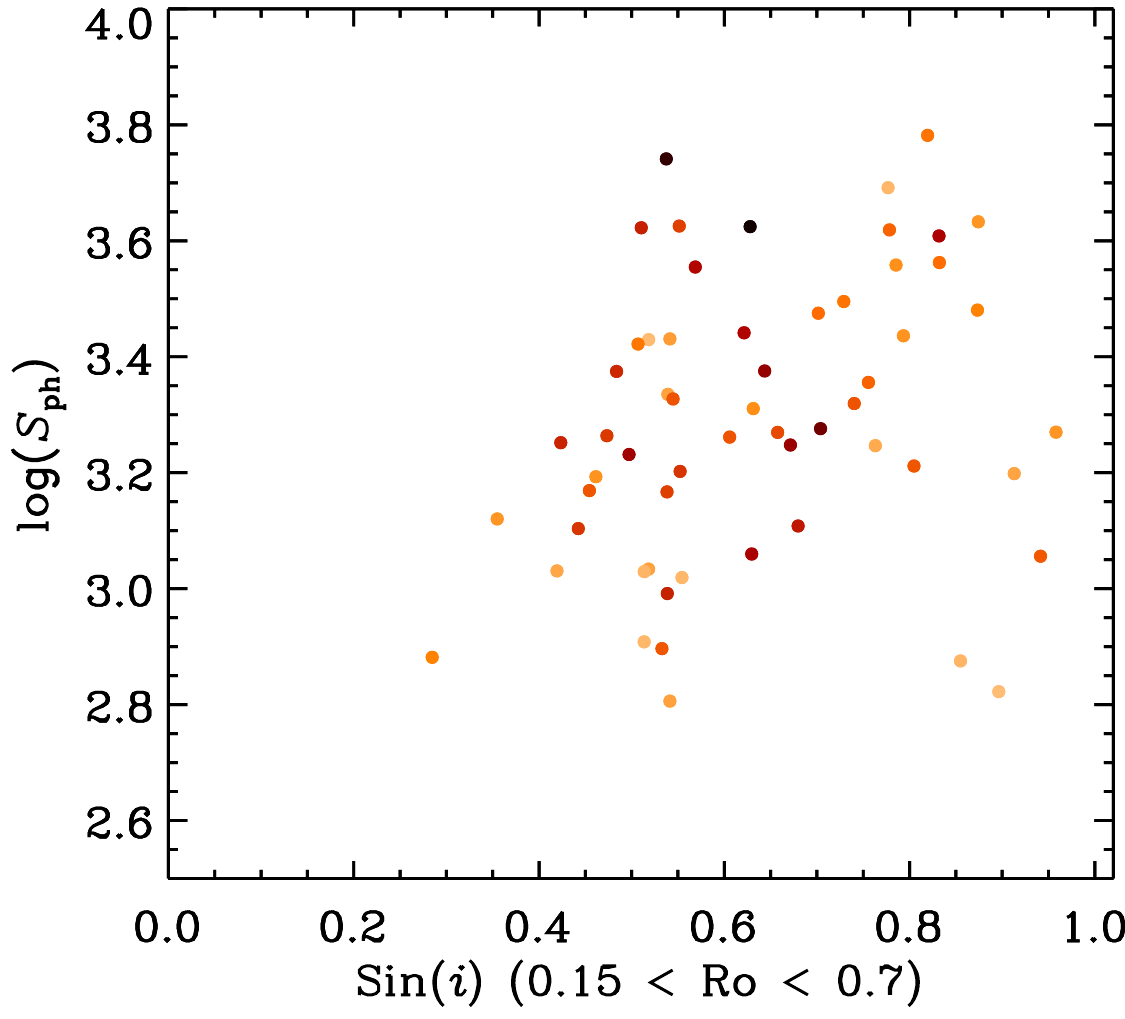}
\caption{ Relations on the proxy of light-curve modulation $S_{\rm ph}$. The top left panel shows its relation to the flaring activity. The other panels show its relations to the inclination in the C, g, and I phase, respectively. }
\label{fig_sini_sph}
\end{figure*}

\begin{figure*}
\includegraphics[width=0.32\textwidth]{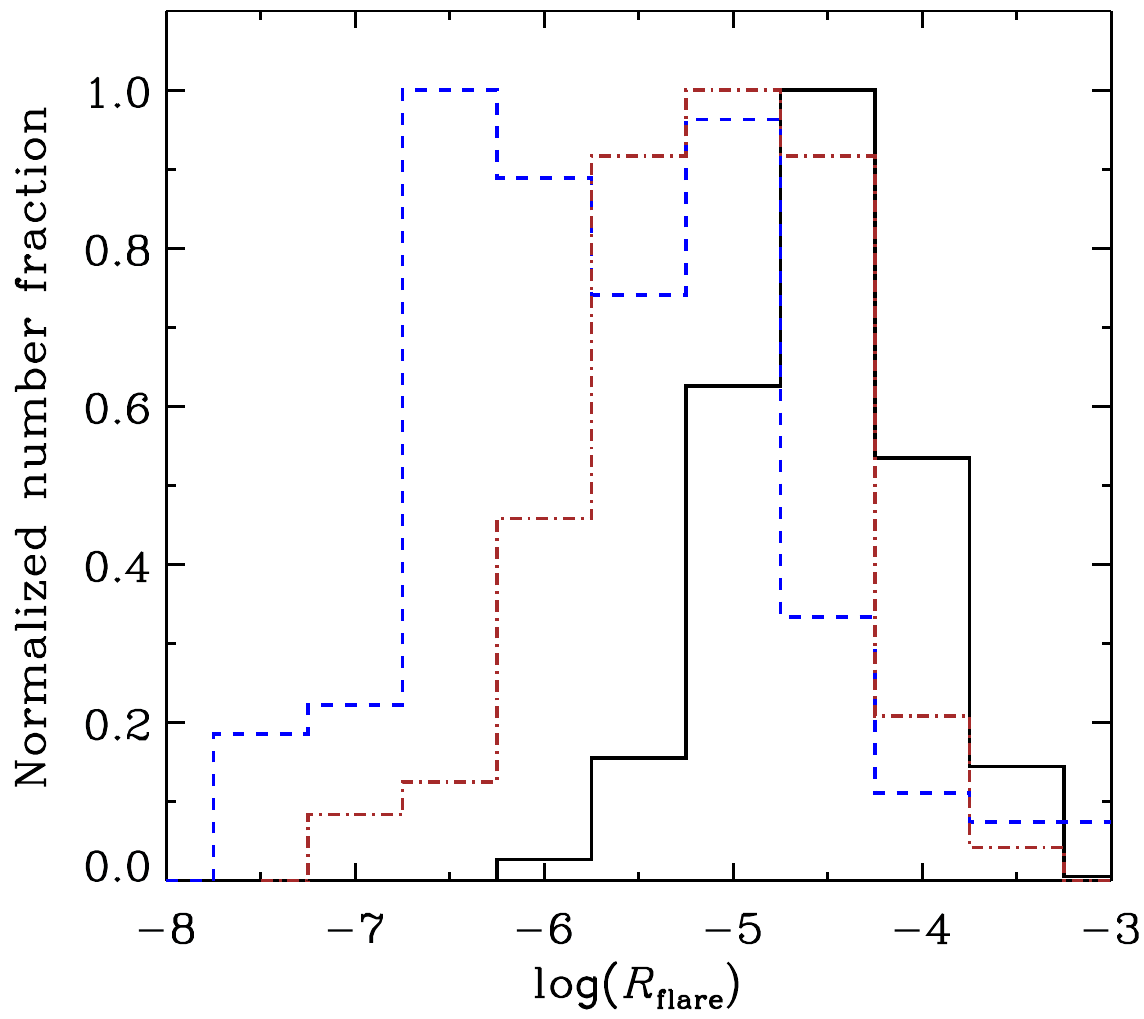}
\includegraphics[width=0.32\textwidth]{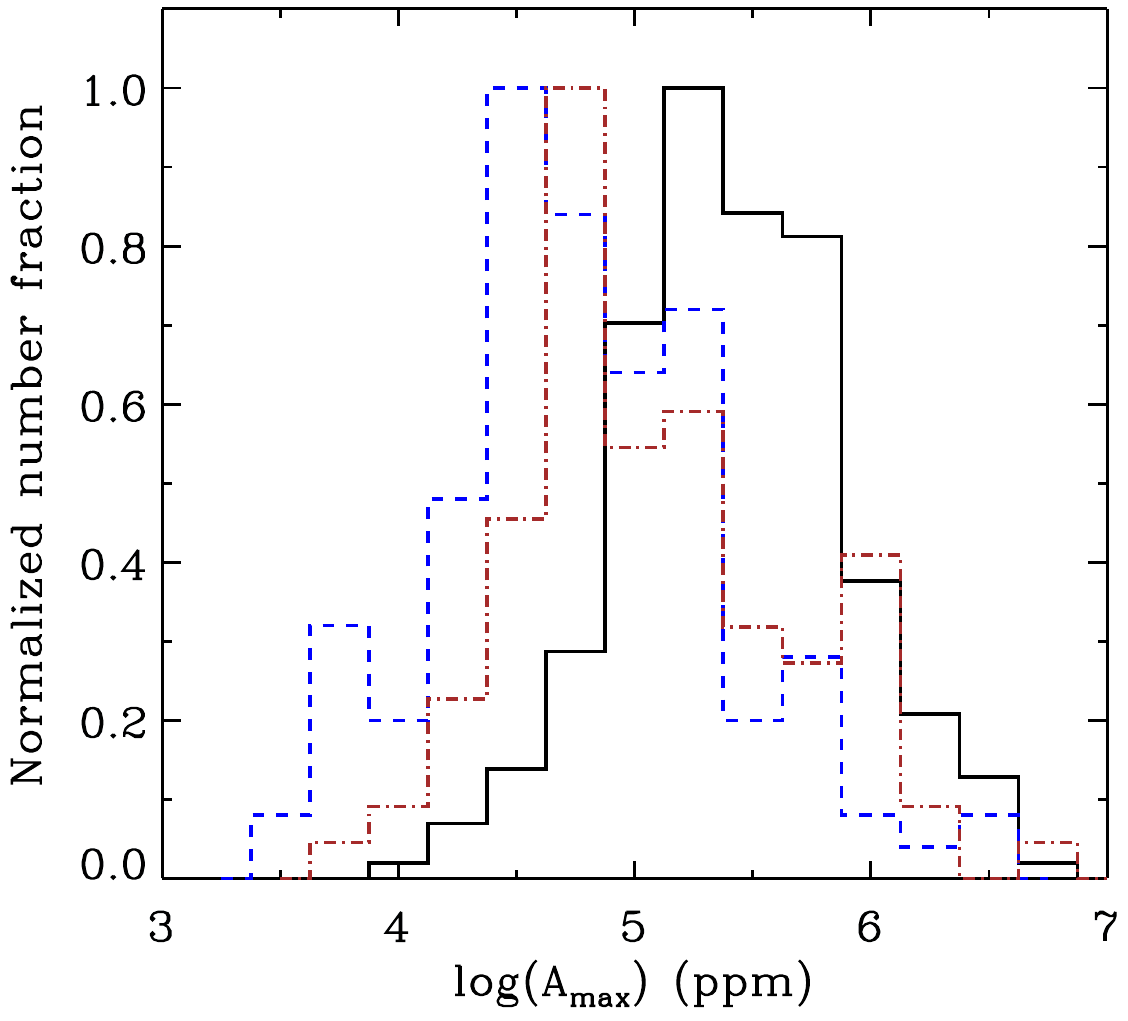}
\includegraphics[width=0.32\textwidth]{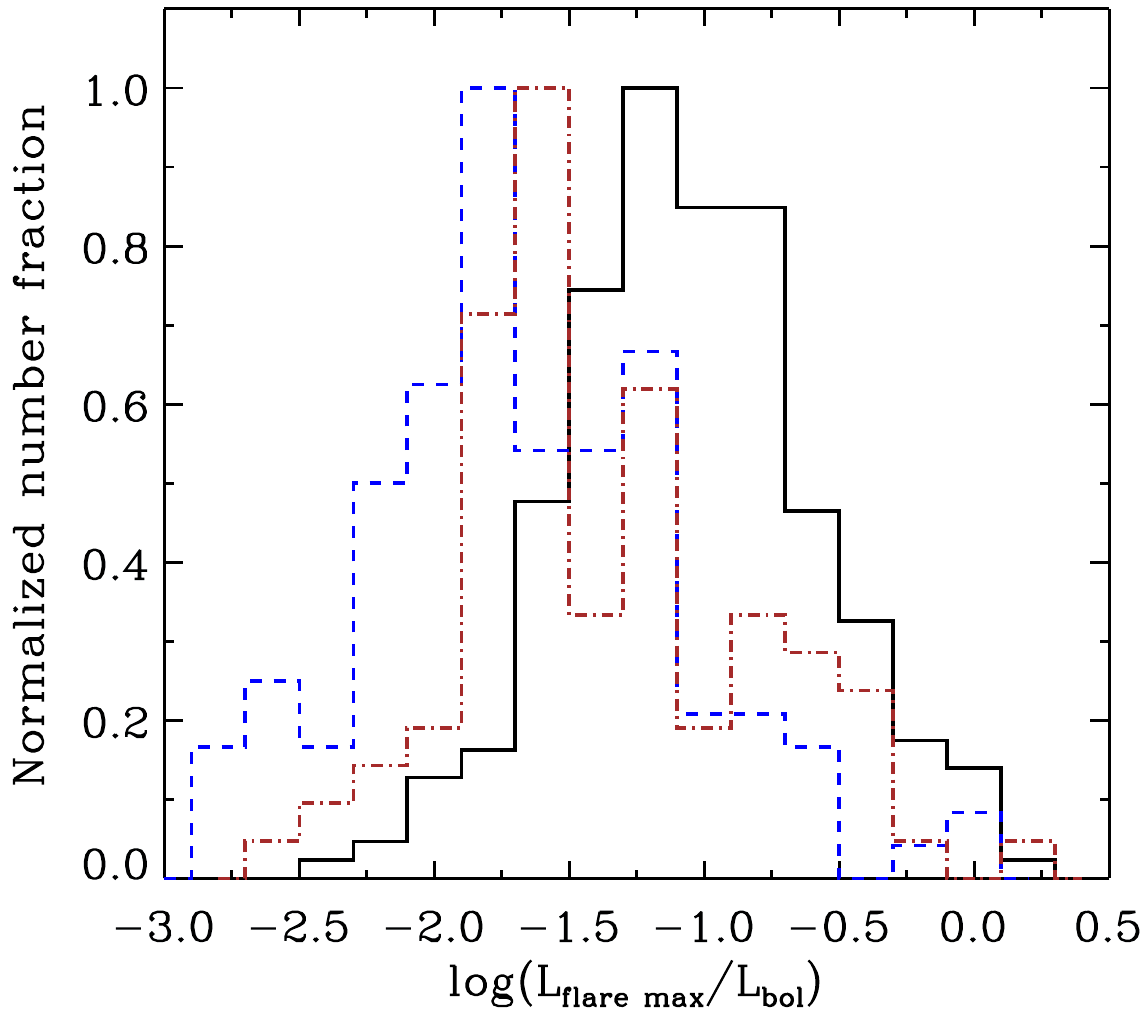}
\caption{Normalized number fractions for stars with different inclinations. The black solid lines represent stars of the C phase which are near equator-on (sin$i > 0.4$ and Ro $< 0.022$). The brown dashed lines represent stars of the C phase which are near pole-on (sin$i < 0.4$ and Ro $< 0.022$). The blue dashed lines represent 130 stars without detectable rotation periods in our sample. From the left to the right panel, the $x$ axis are the flaring activity, the maximum flaring amplitude of a star, and the maximum flaring luminosity of a star, respectively.}
\label{fig_np_hist}
\end{figure*}

\begin{figure*}
\includegraphics[width=0.8\textwidth]{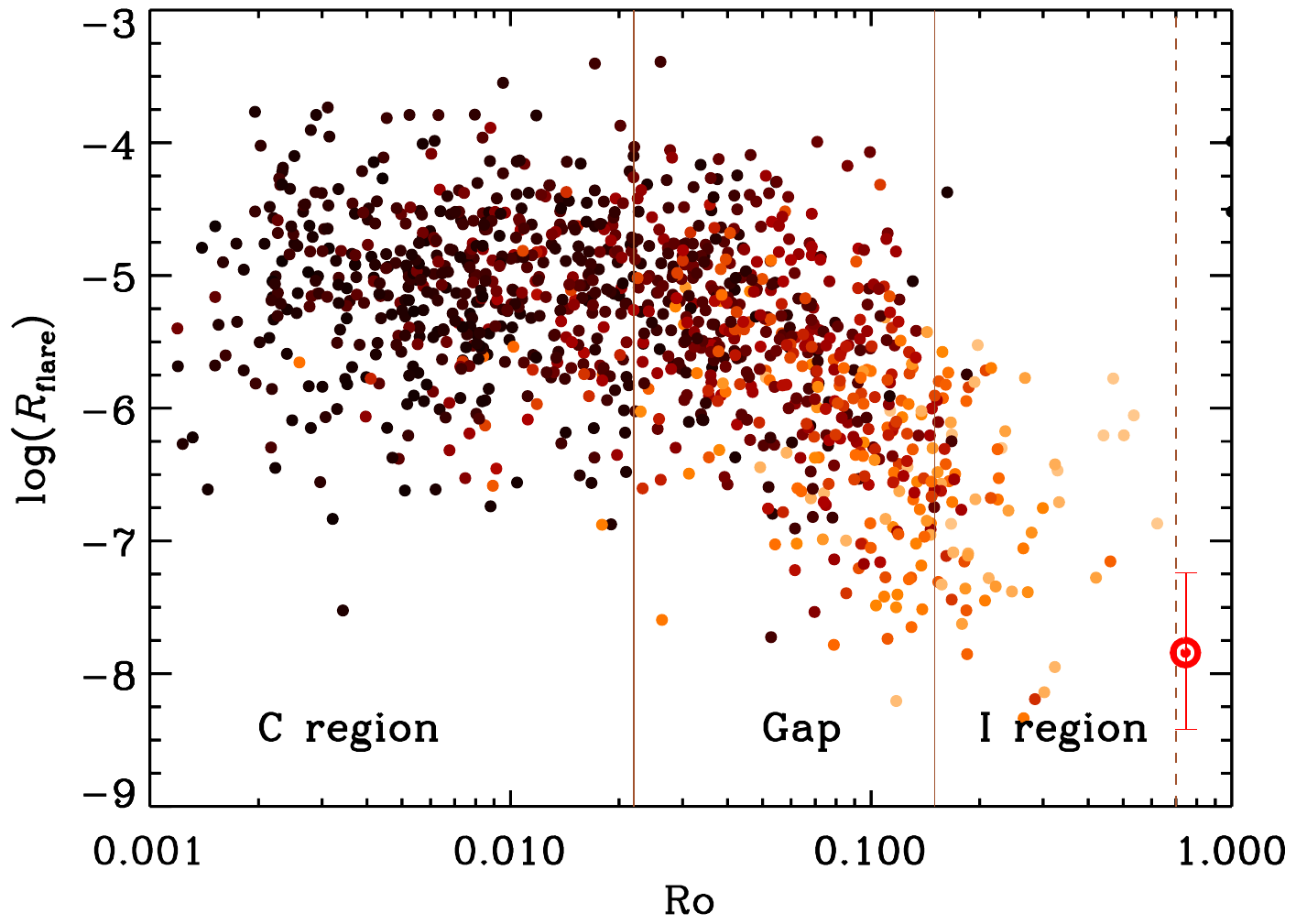}
\includegraphics[width=0.5\textwidth]{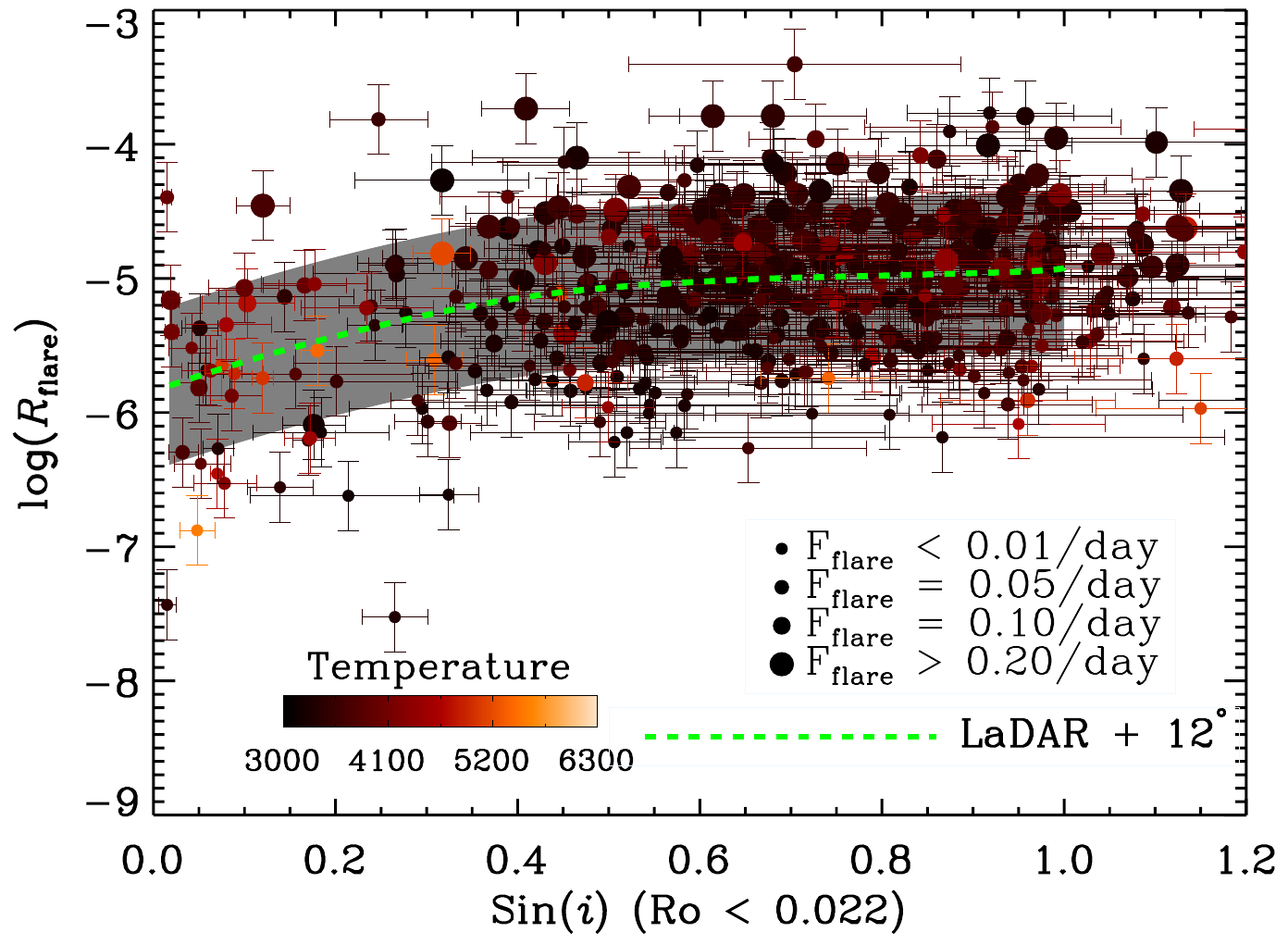}
\includegraphics[width=0.5\textwidth]{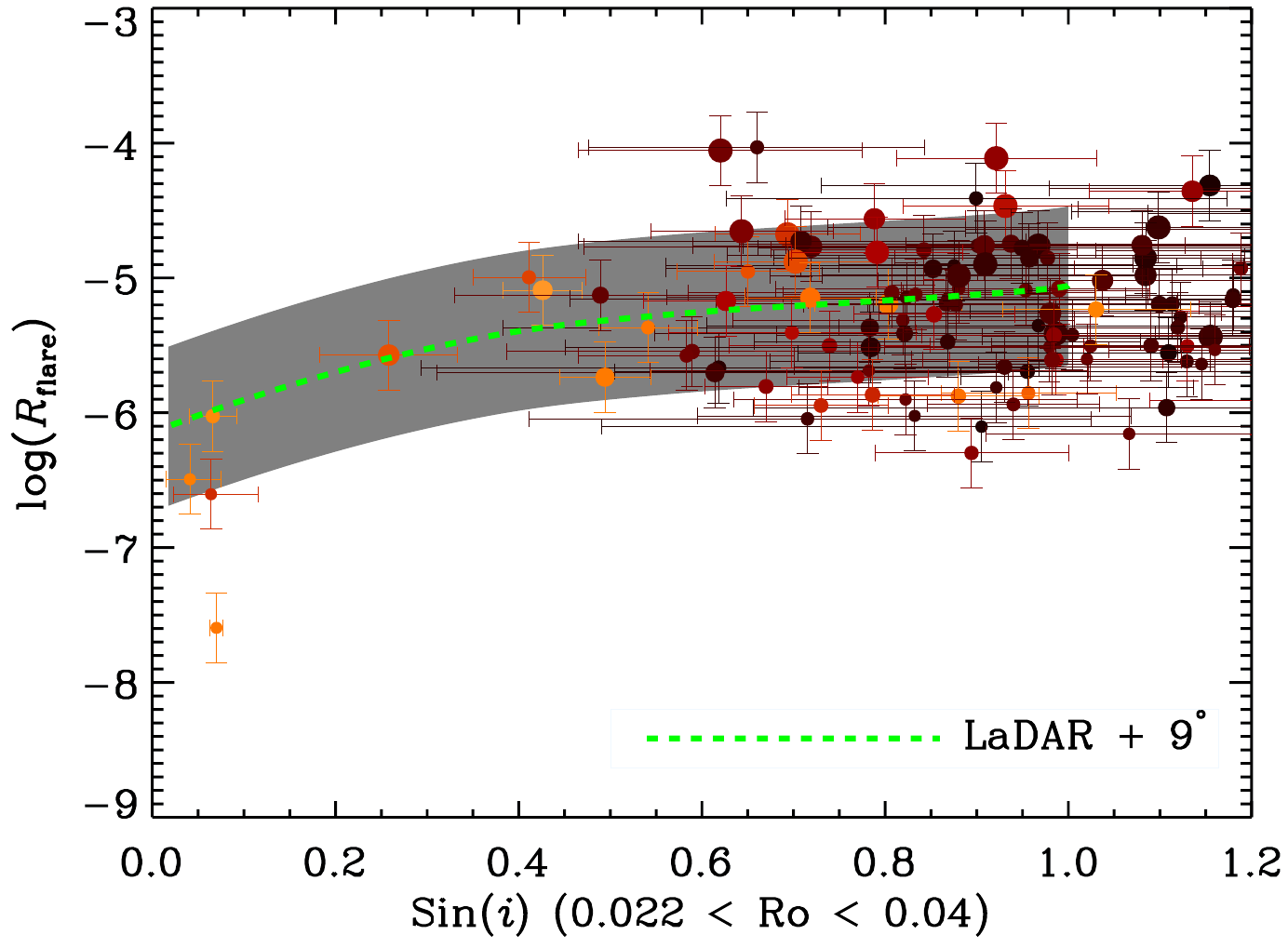}
\includegraphics[width=0.5\textwidth]{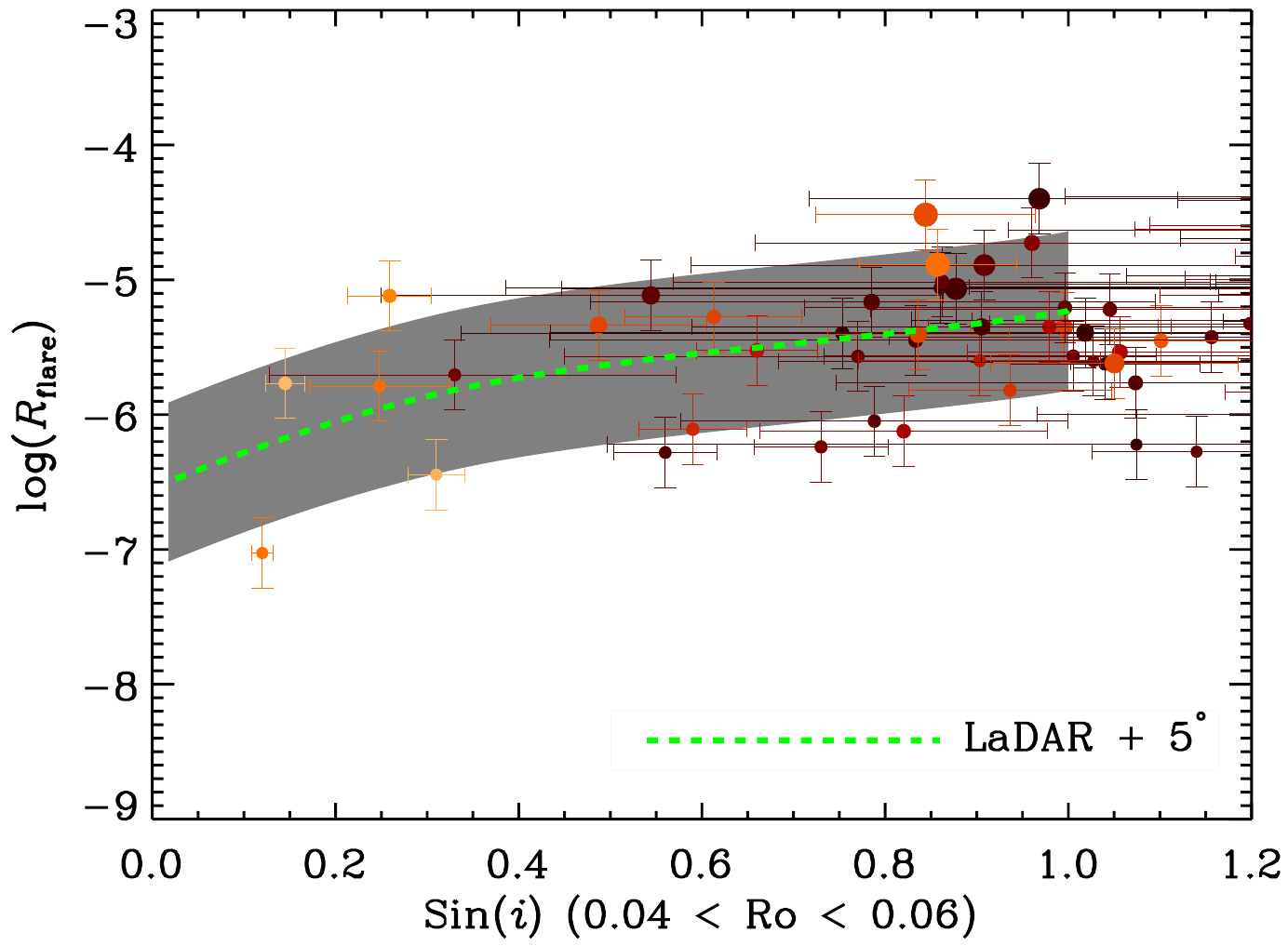}
\includegraphics[width=0.5\textwidth]{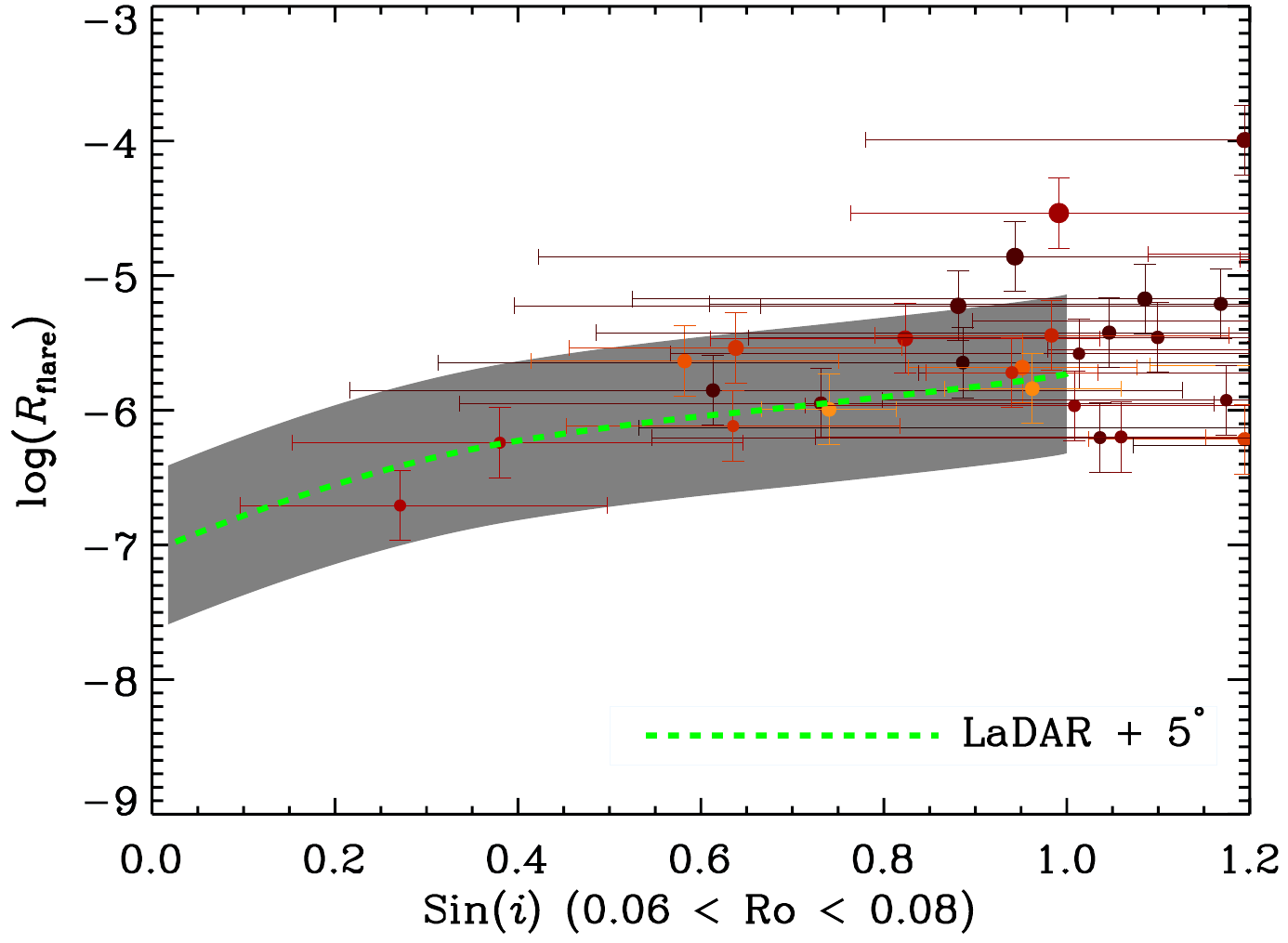}
\caption{ Top panel: Same as Figure~\ref{fig_ro_fa} but for the flare catalog of \citet{Seli2025}. Middle and bottom panel: Same as Figure~\ref{fig_sini_fa} but for the flare catalog of \citet{Seli2025}. }
\label{fig_seli}
\end{figure*}

\begin{table*}
\setlength{\tabcolsep}{3pt}
\caption{\label{table_polarcap} An example of 26 stars with polar spots reported by the (Z)DI technique.}
\centering
\begin{tabular}{ccccccccccccccccccccccccccccc}
\hline\hline\\
Name&TIC&$T_{\rm eff}$&SpT.&$\tau_{g}$& $P_{\rm rot}$& $N_{\rm flare}$& $t_{\rm obs}$&$R_{\rm flare}$&$i_{\rm min}$&$i_{\rm max}$&flagP\\[1ex]
&&(K)&& (Day)& (Day) & & (Day)&&(deg.)&(deg.)&\\[1ex]
\hline

EK Dra &159613900 &5870& G2V &27.9 &2.60& 67 & 201.88 & -5.13  &60 &60&Yes\\[1ex]
DF Tau &268016868 &3750& M0 &165.7 &8.50& 1 & 43.66 & -4.714 &60 &60&Yes\\[1ex]
$\tau$ Boo &72506701 &6360& F7V &5.3 &3.31& 0 & 32.69 & -7.78  &40 &40&Yes\\[1ex]
V830 Tau &268397995 &4000&M0& 117.8 &2.75 & 11 & 43.24 & -4.44  &80 &80&Yes\\[1ex]
Hk Aqr &5656273 &3870&M2V& 137.4 &0.43 & 84 & 58.81 & -3.95  &90 &90&No\\[1ex]
\hline
\end{tabular}
\tablecomments{The entire table is available online as some columns are not shown. Two stars have not been detected flares in the TESS observation. Their flaring activities are the upper limit. The information are from \citet{Strass2009}, where $i_{\rm min}$ and $i_{\rm max}$ represent the minimum and maximum inclination from different measurements, respectively, and flagP denotes if high latitude sports are reported. The flare catalog of the 26 stars is available online.}
\end{table*}

\section{The measurement of the projected rotational velocity $v$sin$i$}
If we adopt the $v$sin$i$ released by the APOGEE and GALAH pipeline, our flaring sample would have $~15\%$ stars with $v$sin$i$/$v > 1$ for fast rotators, which makes no sense in physics. And the fraction increases to $\sim 45\%$ for slow rotators. There are various reasons for this phenomenon , one of which is the uncertainty of $v$sin$i$. The APOGEE and GALAH pipeline is based on template matching through a cross-correlation function. The method of template matching may give more weights to deep and broad absorption lines and overestimate $v$sin$i$ for fast rotators. In order to validate $v$sin$i$ of our flaring stars, we selected several naturally narrow and less blending Fe lines to measure their $v$sin$i$. We first generated a series of synthetic spectra based on the parameters of flaring stars and convolved them with a Gaussian kernel to simulate instrumental broadening. We then convolved the synthetic spectra with a rotational kernel \citep{Gray2005} to find the best match to the observational spectra. For the APOGEE spectra, we applied five Fe lines following \citet{Yang2025}. For GALAH spectra, we applied seven Fe lines (5658.74, 5762.99, 5816.37, 6518.37, 6609.11, 6643.63, 6677.99 {\AA}) in the G and R bands. We took the average $v$sin$i$ of those lines that have good fits as the final $v$sin$i$ for each star. 

We found that most of our results are consistent with those of APOGEE and GALAH except for some fast rotators. Since those fast rotators will have a $v$sin$i$/$v > 1$ in terms of the values of APOGEE and GALAH, we adopted our results for them. After correction, the fraction of stars with $v$sin$i$/$v > 1$ would drop to $\sim 9\%$ and $\sim 42\%$  for fast rotators and slow rotators, respectively (the fraction for the whole sample would drop from $\sim 35\%$ to $\sim 25\%$, and the fraction of $v$sin$i$/$v > 1.2$ will drop from  $\sim 23\%$ to $\sim 10\%$).

\section{The stellar radius, convective turnover time and flare energy}

We used the stellar parameters of tracks with starspots \citep[SPOTS;][]{Somers2020} to estimate stellar radius by finding the best match between the effective temperature and log$g$ of the APOGEE and GALAH parameters and the trackis of SPOTS. We refer to \citet{Yang2025} for details of the determination of the stellar radius. For comparison, we also used the Gaia magnitude, stellar distance \citep{Bailer2021}, extinction \citep{Green2019}, and stellar physical parameters to estimate the stellar bolometric luminosity and stellar radius\citep{Yang2025b}. We note that the two methods obtained similar results.

We used the Yale-Postsdam Stellar Isochrones \citep[YaPSI;][]{Spada2017} to obtain the global convective turnover time by finding the best match between the effective temperature and log$g$ and the tracks of YaPSI. We refer to \citet{Yang2025b} for details.

We estimated the flare energy by assuming a flare radiated as a blackbody with an effective temperature of 9000 K per unit area, and the flare area was estimated based on the flare amplitude, the stellar radius, the stellar effective temperature, and the TESS response function \citep{Shibayama2013}. The uncertainty of the flare energy is 60\% \citep{Shibayama2013}. In the method of energy estimate, the flare energy depends on the effective temperature of the flare $T_{\rm flare}$ and the flare area $A_{\rm flare}$. Since $A_{\rm flare}$ has a negative correlation with $T_{\rm flare}$, this results in that the energy estimate is not very sensitive to $T_{\rm flare}$ (e.g, $E_{\rm 9K}/E_{\rm 15K} \approx 38\%, E_{\rm 5K}/E_{\rm 9K} \approx 60\% $). Our assumption is that $T_{\rm flare}$ is always a 9000 K blackbody spectrum. However, several studies have shown that near the peak of a flare, there is an excess of energy in the UV band, indicating that the bolometric flare energy is underestimated using $T_{\rm flare} =9000$ K \citep{Kowalski2019}. It should be noted that, although $T_{\rm flare}$ varies from $\sim$ 15000 K to $\sim$ 5000 K during the lifetime of a flare \citep{Howard2020,Howard2025}, the average $T_{\rm flare}$ is more suitable to be described using 9000 K \citep{Howard2020}.

\section{The determination of sin$i$}

We calculated the posterior probability distribution (PPD) of sin$i$ for each star according to the method of \citet{Masuda2020}. We then adopted the median value as sin$i$ and the 16th and 84th percentiles as the lower and upper uncertainties. We briefly introduce the method and discuss the uncertainty here.

First, there are two assumptions: 

(1) As the rotation velocity $v$ and the projected velocity $u$ = $v$sin$i$ are from different measurements, their likelihood functions are assumed to be independent, so that

\begin{equation}
 \mathcal{L}_{vu}(v,u) = \mathcal{L}_v\mathcal{L}_u
\label{eq:ass1}
\end{equation}
Here, the likelihood functions of $v$ and $u$ are assumed to be a Gaussian distribution whose $\sigma$ is determined by their uncertainties, respectively.

(2) The prior probability density functions (PDF) of $v$ and $i$ are assumed to be independent, so that

\begin{equation}
\mathcal{P}_{vi}(v,i) = \mathcal{P}_v(v)\mathcal{P}_i(i)
\label{eq:ass2}
\end{equation}
Here, $\mathcal{P}_v(v)$ is assumed to be uniform between 0 and $v_{\rm max}$, and $\mathcal{P}_i(i)$ is transformed to $\mathcal{P}_{\cos i}(\cos i)$ in the following derivation because $\mathcal{P}_{\cos i}(\cos i)$ =1 if we assume that the orientations of the spin axis are uniformly (or randomly). This transformation makes it more convenient for the numerical computation.

Based on the above assumptions and the Bayes theorem, we can obtain the posterior PDF for $\cos i$:
\begin{equation}
p(\cos i \mid D) \propto \mathcal{P}_{\cos i}(\cos i) \int \mathcal{L}_v(v) \, \mathcal{L}_u(v\sqrt{1 - \cos^2 i}) \, \mathcal{P}_v(v)  dv
\label{eq:masuda_10}
\end{equation}

After normalization, we can obtain the posterior PDF for $\sin i$ through the Jacobian determinant:

\begin{equation}
p(\sin i \mid D) = p(\cos i \mid D) \left| \frac{d \cos i}{d \sin i} \right| 
\label{eq:sini}
\end{equation}

The uncertainty of sin$i$ is from the error propagation of $v$sin$i$, $R$, and $P$, which are influenced by several factors. Some of those factors are independent, and some of them are coupled together and vary with different situations, resulting in that it is difficult to give a precise estimate on the uncertainty of sin$i$. Here, we present those factors and give a rough estimate on the uncertainties of $v$sin$i$ and $v$. We then apply the uncertainties to Eq.~\ref{eq:masuda_10} to determine the likelihood functions of $v$sin$i$ and $v$.

The uncertainties of $v$sin$i$ are mainly due to the resolution capacity of the telescope and the macro-turbulence. The typical value of the macro-turbulent velocity is 2 km$\rm s^{-1}$ for a cool star \citep{Doyle2014} and the typical uncertainty for the APOGEE telescope is 10\% \citep{Tayar2015}. Thus, we assumed that the uncertainty of $v$sin$i$ was 2 km~$\rm s^{-1}$ or 10\%, whichever was greater \citep{Yang2025}.
 
The uncertainty of $R$ comes from the capacity of the telescope to determine the bolometric luminosity. It is also model dependent, which is influenced by the stellar parameter $T_{\rm teff}$ (the so-called radius inflation). The radius inflation could be up to 15\% for fast rotators \citep{Jackson2014}. The uncertainties of $P$ are mainly due to the differential rotation. By assuming that stars have a solar-like differential rotation, the influence from the radius inflation and differential rotation could be partially canceled out. This implies that their combined influence to $v$ could be smaller than their individual influence. Thus, we assumed that the uncertainty of $v$ was 10\%.

\section{Do different flare catalogs influence the results?}
Different flare detection methods may produce different flare catalogs, especially for relatively small flares. We compare our flare catalog with others to check if different catalogs could influence the results.
There are two flaring catalogs for the entire data set of the TESS mission. One is seriously polluted by pulsation stars, so we compare our catalog with the other \citep{Seli2025}. We cross-matched the flare catalog of \citet{Seli2025} with our catalog of rotating stars from sector 1 to sector 60. We find that about 80\% flaring stars of our catalog are also in their catalog, and about 95\% flaring stars of their catalog are in our catalog. The common flares account for about 60\% of our flare catalog (i.e., We have detected an additional 40\% of flares). The discrepancy is due to that their goal is to investigate the profiles of flares rather than to compile a complete catalog. Thus, they have adopted several strict criteria that removed most small flares and some light-curves with low S/N. However, the absence of small flares could systematically reduce the values of $R_{\rm flare}$ but did not change their relation along with inclination. We show the inclination--flaring activity relationships derived from the catalog of \citet{Seli2025} ( Figure~\ref{fig_seli}), which is similar to ours.


\bibliography{ref}{}
\bibliographystyle{aasjournalv7}



\end{document}